\documentclass[trackchanges]{aastex631}

\newcommand{\eg}{e.g., }

\newcommand{\etal}{et al.}
\newcommand{\ie}{i.e., }

\newcommand{\sect}[1]{Section \ref{s:#1}}

\newcommand{\eqn}[1]{Equation (\ref{e:#1})}
\newcommand{\eqntwo}[2]{Equations (\ref{e:#1}) and (\ref{e:#2})}

\newcommand{\figr}[1]{Figure \ref{f:#1}}  

\newcommand{\tbl}[1]{Table \ref{t:#1}}

\usepackage{amsmath}
\usepackage{bm}          
\usepackage{hyperref}    
\usepackage{subfigure}
\usepackage{xcolor}      

\newcommand{\hide}[1]{}  

\turnoffedit  

\received{September 28, 2023}
\revised{April 15, 2024}
\revised{June 21, 2024}  
\accepted{July 3, 2024}

\submitjournal{The Planetary Science Journal}  
\shorttitle{Didymos System Dynamics Before \& After DART}  
\shortauthors{Richardson \etal}
\graphicspath{{./}{figures/}}

\begin{document}

\title{The Dynamical State of the Didymos System Before and After the DART Impact}

\correspondingauthor{Derek C. Richardson}
\email{dcr@umd.edu}



\author[0000-0002-0054-6850]{Derek C. Richardson}  
\affiliation{Department of Astronomy \\
University of Maryland \\
College Park, MD 20742, USA}


\author[0000-0002-3544-298X]{Harrison F. Agrusa}  
\affiliation{Universit\'e C\^ote d'Azur, 
Observatoire de la C\^ote d'Azur, CNRS, Laboratoire Lagrange \\
Bd de l'Observatoire, CS 34229 \\
06304 Nice Cedex 4, France}
\affiliation{Department of Astronomy \\
University of Maryland \\
College Park, MD 20742, USA}

\author[0000-0003-3739-3242]{Brent Barbee}  
\affiliation{NASA Goddard Space Flight Center \\
Greenbelt, MD 20771, USA}

\author[0009-0000-2266-6266]{Rachel H. Cueva}  
\affiliation{Smead Department of Aerospace Engineering Sciences \\
University of Colorado Boulder \\
3775 Discovery Dr, Boulder, CO 80303, USA}

\author[0000-0001-7537-4996]{Fabio Ferrari}  
\affiliation{Department of Aerospace Science and Technology \\
Politecnico di Milano \\ 
Milan, 20159, Italy}

\author[0000-0002-4952-9007]{Seth A. Jacobson}  
\affiliation{Department of Earth and Environmental Sciences \\
Michigan State University \\
East Lansing, MI 48824, USA}

\author[0000-0001-9265-2230]{Rahil Makadia}  
\affiliation{Department of Aerospace Engineering \\
University of Illinois at Urbana-Champaign \\
Urbana, IL 61801, USA}

\author[0000-0001-8437-1076]{Alex J. Meyer}  
\affiliation{Smead Department of Aerospace Engineering Sciences \\
University of Colorado Boulder \\
3775 Discovery Dr, Boulder, CO 80303, USA}

\author[0000-0002-0884-1993]{Patrick Michel}  
\affiliation{Universit\'e C\^ote d'Azur, 
Observatoire de la C\^ote d'Azur, CNRS, Laboratoire Lagrange \\
Bd de l'Observatoire, CS 34229 \\
06304 Nice Cedex 4, France}
\affiliation{The University of Tokyo, Department of Systems Innovation, School of Engineering \\
Hongo 7-3-1, Bunkyo-ku, Tokyo 113-0033, Japan}

\author[0000-0002-9840-2416]{Ryota Nakano}  
\affiliation{Daniel Guggenheim School of Aerospace Engineering \\
Georgia Institute of Technology \\
Atlanta, GA 30332, USA}
\affiliation{Department of Aerospace Engineering \\
Auburn University \\
Auburn, AL 36849, USA}

\author[0000-0003-4045-9046]{Yun Zhang}  
\affiliation{Department of Climate and Space Sciences and Engineering \\
University of Michigan \\
Ann Arbor, MI 48109, USA}


\author[0000-0002-6233-1820]{Paul Abell}  
\affiliation{Johnson Space Center \\
Houston, TX 77058, USA}

\author[0000-0002-5566-0618]{Colby C. Merrill}  
\affiliation{Sibley School of Mechanical and Aerospace Engineering \\
Cornell University \\
Ithaca, NY 14853, USA}


\author[0000-0001-9840-2216]{Adriano Campo Bagatin}  
\affiliation{Instituto de Fisica Aplicada a las Ciencias y las Tecnologias (IUFACyT) \\
Universidad de Alicante \\
Ctra.\ Sant Vicent del Raspeig s/n, 03690 \\
Sant Vicent del Raspeig, Alicante, Spain}

\author[0000-0002-3578-7750]{Olivier Barnouin}  
\affiliation{Johns Hopkins University Applied Physics Laboratory \\
Laurel, MD 20723, USA}

\author[0000-0001-8628-3176]{Nancy L. Chabot}  
\affiliation{Johns Hopkins University Applied Physics Laboratory \\
Laurel, MD 20723, USA}

\author[0000-0001-5375-4250]{Andrew F. Cheng}  
\affiliation{Johns Hopkins University Applied Physics Laboratory \\
Laurel, MD 20723, USA}

\author[0000-0003-3240-6497]{Steven R. Chesley}  
\affiliation{Jet Propulsion Laboratory, California Institute of Technology \\
Pasadena, CA 91109, USA}

\author[0000-0002-1320-2985]{R. Terik Daly}  
\affiliation{Johns Hopkins University Applied Physics Laboratory \\
Laurel, MD 20723, USA}

\author[0000-0002-1398-6302]{Siegfried Eggl}  
\affiliation{Department of Aerospace Engineering / Astronomy \\
University of Illinois at Urbana-Champaign \\
Urbana, IL 61801, USA}
\affiliation{National Center for Supercomputing Applications \\
University of Illinois at Urbana-Champaign \\
Urbana, IL 61801, USA}

\author[0000-0002-9434-7886]{Carolyn M. Ernst}  
\affiliation{Johns Hopkins University Applied Physics Laboratory \\
Laurel, MD 20723, USA}

\author[0000-0003-1391-5851]{Eugene G. Fahnestock}  
\affiliation{Jet Propulsion Laboratory, California Institute of Technology \\
Pasadena, CA 91109, USA}

\author[0000-0002-4767-9861]{Tony L. Farnham}  
\affiliation{Department of Astronomy \\
University of Maryland \\
College Park, MD 20742, USA}

\author[0000-0001-5875-1083]{Oscar Fuentes-Mu\~{n}oz}  
\affiliation{Smead Department of Aerospace Engineering Sciences \\
University of Colorado Boulder \\
3775 Discovery Dr, Boulder, CO 80303, USA}

\author[0000-0001-8776-7922]{Edoardo Gramigna}  
\affiliation{Department of Industrial Engineering \\
Alma Mater Studiorum -- Universit\`a di Bologna \\
Forl\`i (FC) 47121, Italy}

\author[0000-0002-5010-0574]{Douglas P. Hamilton}  
\affiliation{Department of Astronomy \\
University of Maryland \\
College Park, MD 20742, USA}

\author[0000-0002-1821-5689]{Masatoshi Hirabayashi}  
\affiliation{Daniel Guggenheim School of Aerospace Engineering \\ Georgia Institute of Technology \\ Atlanta, GA 30332, USA}
\affiliation{Department of Aerospace Engineering \textbackslash Geosciences \\
Auburn University \\
Auburn, AL 36849, USA}

\author[0000-0002-1800-2974]{Martin Jutzi}  
\affiliation{Space Research and Planetary Sciences, Physics Institute \\
University of Bern \\
Bern, 3012, Switzerland}

\author[0000-0001-6420-8423]{Josh Lyzhoft}  
\affiliation{NASA Goddard Space Flight Center \\
Greenbelt, MD 20771, USA}

\author[0000-0002-5733-2554]{Riccardo Lasagni Manghi}  
\affiliation{Department of Industrial Engineering \\
Alma Mater Studiorum -- Universit\`a di Bologna \\
Forl\`i (FC) 47121, Italy}

\author[0000-0002-1847-4795]{Jay McMahon}  
\affiliation{Smead Department of Aerospace Engineering Sciences \\
University of Colorado Boulder \\
3775 Discovery Dr, Boulder, CO 80303, USA}

\author[0000-0003-0670-356X]{Fernando Moreno}  
\affiliation{Instituto de Astrofisica de Andalucia, CSIC \\
18008 Granada, Spain}

\author[0000-0002-9701-4075]{Naomi Murdoch}  
\affiliation{Institut Sup\'erieur de l'A\'eronautique et de l'Espace (ISAE-SUPAERO) \\
Universit\'e de Toulouse, Toulouse, France}

\author[0000-0003-4439-7014]{Shantanu P. Naidu}  
\affiliation{Jet Propulsion Laboratory, California Institute of Technology \\
Pasadena, CA 91109, USA}

\author[0000-0001-6755-8736]{Eric E. Palmer}  
\affiliation{Planetary Science Institute \\
Tucson, AZ 85719, USA}

\author[0000-0003-3743-6302]{Paolo Panicucci}  
\affiliation{Department of Aerospace Science and Technology \\
Politecnico di Milano \\ 
Milan, 20159, Italy}

\author[0000-0001-9619-7271]{Laurent Pou}  
\affiliation{Jet Propulsion Laboratory, California Institute of Technology \\
Pasadena, CA 91109, USA}

\author[0000-0001-8434-9776]{Petr Pravec}  
\affiliation{Astronomical Institute of the Academy of Sciences of the Czech Republic, Fri\v{c}ova 298 \\
Ond\v{r}ejov, CZ 25165 Czech Republic}

\author[0000-0002-7478-0148]{Sabina D. Raducan}  
\affiliation{Space Research and Planetary Sciences, Physics Institute \\
University of Bern \\
Bern, 3012, Switzerland}

\author[0000-0002-9939-9976]{Andrew S. Rivkin}  
\affiliation{Johns Hopkins University Applied Physics Laboratory \\
Laurel, MD 20723, USA}

\author[0000-0001-9311-2869]{Alessandro Rossi}  
\affiliation{Istituto di Fisica Applicata ``Nello Carrara'' (IFAC-CNR) \\ 
Sesto Fiorentino 50019, Italy}

\author[0000-0003-3610-5480]{Paul S\'anchez}  
\affiliation{Colorado Center for Astrodynamics Research \\
University of Colorado Boulder \\
3775 Discovery Dr, Boulder, CO 80303, USA}

\author[0000-0003-0558-3842]{Daniel J. Scheeres}  
\affiliation{Smead Department of Aerospace Engineering Sciences \\
University of Colorado Boulder \\
3775 Discovery Dr, Boulder, CO 80303, USA}

\author[0000-0001-8518-9532]{Peter Scheirich}  
\affiliation{Astronomical Institute of the Academy of Sciences of the Czech Republic, Fri\v{c}ova 298 \\
Ond\v{r}ejov, CZ 25165 Czech Republic}

\author[0000-0001-5475-9379]{Stephen R. Schwartz}  
\affiliation{Planetary Science Institute \\
Tucson, AZ 85719, USA}
\affiliation{Instituto de Fisica Aplicada a las Ciencias y las Tecnologias (IUFACyT) \\
Universidad de Alicante \\
Ctra. Sant Vicent del Raspeig s/n, 03690 Sant Vicent del Raspeig. Alicante (Spain)}

\author[0000-0003-4058-0815]{Damya Souami}  
\affiliation{LESIA, Observatoire de Paris, Universit\'e PSL, CNRS, Sorbonne Universit\'e, Universit\'e de Paris \\
5 place Jules Janssen \\
F-92195 Meudon, France}
\affiliation{Departments of Astronomy, and of Earth and Planetary Science \\ 
University of California Berkeley \\
501 Campbell Hall, University of California \\
Berkeley, CA 94720-3411, USA}
 \affiliation{Universit\'e C\^ote d'Azur, 
Observatoire de la C\^ote d'Azur, CNRS, Laboratoire Lagrange \\
Bd de l'Observatoire, CS 34229 \\
06304 Nice Cedex 4, France}
\affiliation{naXys, Department of Mathematics \\
University of Namur \\
Rue de Bruxelles 61, 5000 Namur, Belgium}

\author[0000-0002-4943-8623]{Gonzalo Tancredi}  
\affiliation{Departamento de Astronom\'ia \\
Facultad de Ciencias \\
Udelar, Uruguay}

\author[0000-0002-2718-997X]{Paolo Tanga}  
\affiliation{Universit\'e C\^ote d'Azur, 
Observatoire de la C\^ote d'Azur, CNRS, Laboratoire Lagrange \\
Bd de l'Observatoire, CS 34229 \\
06304 Nice Cedex 4, France}

\author[0000-0001-9259-7673]{Paolo Tortora}  
\affiliation{Department of Industrial Engineering \\
Alma Mater Studiorum -- Universit\`a di Bologna \\
Forl\`i (FC) 47121, Italy}
\affiliation{Centro Interdipartimentale di Ricerca Industriale Aerospaziale \\
Alma Mater Studiorum -- Universit\`a di Bologna \\
Forl\`i (FC) 47121, Italy}

\author[0000-0001-8417-702X]{Josep M. Trigo-Rodr\'iguez}  
\affiliation{Institute of Space Sciences (CSIC-IEEC), Campus UAB \\
c/Can Magrans s/n, Cerdanyola del vall\`es \\
Barcelona, Catalonia, Spain}

\author[0000-0003-3334-6190]{Kleomenis Tsiganis}  
\affiliation{Department of Physics \\
Aristotle University of Thessaloniki \\
GR 54124 Thessaloniki, Greece}

\author[0009-0004-6736-309X]{John Wimarsson}  
\affiliation{Space Research and Planetary Sciences, Physics Institute \\
University of Bern \\
Bern, 3012, Switzerland}

\author[0000-0002-4151-9656]{Marco Zannoni}  
\affiliation{Department of Industrial Engineering \\
Alma Mater Studiorum -- Universit\`a di Bologna \\
Forl\`i (FC) 47121, Italy}
\affiliation{Centro Interdipartimentale di Ricerca Industriale Aerospaziale \\
Alma Mater Studiorum -- Universit\`a di Bologna \\
Forl\`i (FC) 47121, Italy}





\begin{abstract}  
    NASA's Double Asteroid Redirection Test (DART) spacecraft impacted Dimorphos, the natural satellite of (65803) Didymos, on 2022 September 26, as a first successful test of kinetic impactor technology for deflecting a potentially hazardous object in space. The experiment resulted in a small change to the dynamical state of the Didymos system consistent with expectations and Level 1 mission requirements. In the pre-encounter paper \citep{Richardson2022}, predictions were put forward regarding the pre- and post-impact dynamical state of the Didymos system. Here we assess these predictions, update preliminary findings published after the impact, report on new findings related to dynamics, and provide implications for ESA's Hera mission to Didymos, scheduled for launch in 2024 with arrival in late December 2026. Pre-encounter predictions tested to date are largely in line with observations, despite the unexpected, flattened appearance of Didymos compared to the radar model and the apparent pre-impact oblate shape of Dimorphos (with implications for the origin of the system that remain under investigation). New findings include that Dimorphos likely became prolate due to the impact \edit1{and may have} entered a tumbling rotation state. A possible detection of a post-impact transient secular decrease in the binary orbital period suggests possible dynamical coupling with persistent ejecta. Timescales for damping of any tumbling and clearing of any debris are uncertain. The largest uncertainty in the momentum transfer enhancement factor of the DART impact remains the mass of Dimorphos, which will be resolved by the Hera mission.
\end{abstract}



\section{Introduction} \label{s:intro}

On 2022 September 26, NASA's Double Asteroid Redirection Test (DART) spacecraft impacted Dimorphos, the natural satellite of (65803) Didymos, as the first full-scale demonstration of a kinetic impact deflection technique \citep{Daly2023}. Prior to intercept, the Didymos system was not a hazard to Earth, and the experiment did not increase the likelihood of collision \citep{Makadia2022}. Rather, the impact reduced the orbital period of Dimorphos around Didymos by $33.0 \pm 1.0$ (3-$\sigma$)~min, from 11.92~h to 11.37~h \citep{Thomas2023}. Preliminary modeling found this corresponded to an along-track orbital speed change of $-2.70 \pm 0.10$ (1-$\sigma$)~mm~s$^{-1}$ for the satellite, implying a momentum transfer enhancement factor (ratio of target momentum change to spacecraft momentum, accounting for any ejecta boost) of $\beta = 3.61^{+0.19}_{-0.25}$ (1-$\sigma$) assuming a bulk density of 2,400~kg~m$^{-3}$ \citep{Cheng2023}. \edit1{These results, together with later analyses \citep{Chabot2024}, indicate DART met all of its Level~1 requirements, namely to impact Dimorphos, to change the orbital period by at least 73~s, to measure the period change to an accuracy of 7.3~s, and to measure $\beta$ \citep{Rivkin2021}.} There was no direct mass measurement of Dimorphos, so $\beta$ ranges plausibly from 2.2 to 4.9, \edit1{assuming} a Dimorphos bulk density range of 1,500 to 3,300~kg~m$^{-3}$. Hera, ESA's follow-on mission to the Didymos system with a planned launch in October 2024 and rendezvous in late December 2026, promises to constrain the post-impact mass of Dimorphos and therefore $\beta$ to much higher precision \citep{Michel2022}. Together, Hera and DART comprise the Asteroid Impact and Deflection Assessment (AIDA) cooperation between ESA and NASA.

Prior to DART's encounter, \citet{Richardson2022} put forward predictions for the dynamical state of the Didymos system before and after the DART impact based on best-available data and modeling. \tbl{predictions} summarizes the main predictions of both pre- and post-impact dynamical states and their evaluation in light of observations during and after encounter. \tbl{params} provides an update on key dynamical parameters from the best-available data and analysis, noting that in some cases different approaches give slightly different values, as indicated in the table footnotes (see the primary sources for more details).

\begin{table}
    \caption{Summary of pre-encounter predictions and their post-encounter evaluation.}
    \label{t:predictions}
    \centering
    \begin{tabular}{p{0.45\textwidth}p{0.45\textwidth}}  
        \tableline \tableline
        Prediction & Evaluation \\ 
        \tableline
        The system at encounter will be in a low-energy state, with small/zero eccentricity, no excited modes, and Dimorphos tidally locked to Didymos; no recent big impacts or encounters. &
        Insufficient observational constraints of the pre-impact state at encounter; low/zero eccentricity consistent with modeled low post-impact eccentricity (\sect{dynamics}); Dimorphos's apparent pre-impact oblate shape consistent with undetectable signature of rotation in lightcurves, implying slow but not necessarily zero tidal damping (\sect{shape}); no indication of excitement from recent perturbation but not ruled out. \\
        \tableline
        Impact yields $\beta$ between 1 and 5, reducing orbit period (for $\beta > 1$) and inducing few-minute orbital-period variations; heliocentric $\beta$ may be measurable. &
        $\beta$ between 2.2 and 4.9 for plausible Dimorphos bulk density; period reduced by ${\sim}$33~min but post-impact non-secular variations unobservable due to high frequency (\sect{dynamics}); successful future heliocentric $\beta$ measurement likely (\sect{beta}). \\
        \tableline
        Impact alters rotation state, inducing libration; may excite instability depending on Dimorphos's inertia moments. &
        Post-impact secondary lightcurve minima offsets from mutual events consistent with libration of a few tens of degrees amplitude
        (\sect{dynamics}); observed varying post-impact secondary lightcurve amplitude and post-impact orbit modeling consistent with possible non-principal-axis rotation state/tumbling (\sect{shape}). \\
        \tableline
        Shape change of either body could change $\beta$ measurably. &
        No Didymos spin change detected to 1~s precision (\sect{shape}); Dimorphos likely reshaped due to detection of secondary lightcurve (\sect{shape}). \\
        \tableline
        Post-impact secular effects (tidal friction, BYORP) may alter system state prior to Hera arrival. &
        Possible post-impact drop(s) in eccentricity and orbital period (\sect{dynamics}) noted in post-impact orbit modeling but likely too soon as of this writing to observe measurable effects of tides/BYORP (\sect{shape}). \\
        \tableline
    \end{tabular}
\end{table}

\begin{table}
    \caption{Selected dynamical parameters of the Didymos system before and after the DART impact.$^\mathrm{a}$}
    \label{t:params}
    \centering
    \begin{tabular}{llll}
        \tableline \tableline
        Parameter & Pre-encounter & Pre-impact & Post-impact \\
        \tableline
        Volume-Equivalent Diameter of Primary$^\mathrm{b}$ [m] & $780 \pm 30$ & $730 \pm 17$ & assumed unchanged \\
        Volume-Equivalent Diameter of Secondary$^\mathrm{c}$ [m] & $164 \pm 18$ & $150.0 \pm 2.5$ & assumed unchanged \\
        Bulk Density of Primary, Secondary$^\mathrm{d}$ [kg m$^{-3}$] & $2170 \pm 350$ (both) & $2790 \pm 140$, $2400 \pm 300$ & assumed unchanged \\
        Mean Separation of Component Centers$^\mathrm{e}$ [km] & $1.20 \pm 0.03$ & $1.189 \pm 0.017$ & $1.152 \pm 0.018$ \\
        Secondary Shape$^\mathrm{f}$ $a_s/b_s$, $b_s/c_s$ & $1.3 \pm 0.2$, $1.2$ & $1.06 \pm 0.03$, $1.47 \pm 0.04$ & $1.300 \pm 0.010$, $1.3 \pm 0.2$ \\
        Total Mass of System$^\mathrm{g}$ [$10^{11}$ kg] & $5.55 \pm 0.42$ & $5.3 \pm 0.2$ & assumed unchanged \\
        Mutual Orbital Period$^\mathrm{h}$ [h] & $11.921629 \pm 0.000003$ & $11.921493 \pm 0.000016$ & $11.3674 \pm 0.0004$ \\
        Mutual Orbital Eccentricity$^\mathrm{i}$ & $< 0.03$ & $< 0.03$ & $0.0274 \pm 0.0015$ \\
        Primary Rotation Period$^\mathrm{j}$ [h] & $2.2600 \pm 0.0001$ & $2.2600 \pm 0.0001$ & $2.260 \pm 0.001$ \\
        Secondary Rotation Period$^\mathrm{k}$ [h] & $11.921629 \pm 0.000003$ & $11.92149 \pm 0.00002$ & pending \\
        Secondary Orbital Inclination [$^\circ$] & $0$ (assumed) & $0$ (assumed) & pending \\
        Apsidal Precession Rate$^\mathrm{l}$ [$^\circ$/d] & --- & --- & $6.7 \pm 0.2$ \\
        \tableline
        \multicolumn{4}{>{\raggedright}p{\textwidth}}{$^\mathrm{a}$``Pre-encounter'' values are from \citet{Richardson2022} before DART's arrival in the system. ``Pre-impact'' (using DART data) and  ``Post-impact'' values are from just prior to and just after the DART impact, respectively, based on the latest measurements and modeling as of this writing. Uncertainties are 1-$\sigma$. $^\mathrm{b}$Pre-impact value from \cite{Barnouin2023}. $^\mathrm{c}$Pre-impact value from \cite{Daly2024}. $^\mathrm{d}$Pre-impact values from \cite{Naidu2024} for Didymos and \cite{Daly2023} for Dimorphos. A larger pre-encounter bulk density estimate of $2370\pm300$~kg~m$^{-3}$ for both is reported by \cite{Scheirich-2022}. $^\mathrm{e}$Pre-impact value from \cite{Naidu2024}; post-impact change in value from \cite{meyer2023b}. $^\mathrm{f}$Pre-impact values from \cite{Daly2024}; post-impact values from \cite{Naidu2024}. Note: the shape of Dimorphos is not directly measured post-impact in this model. The reported axis ratios correspond to the shape of an ellipsoid having the same moments of inertia. \edit1{\citet{Pravec2024} find a post-impact range of 1.1--1.4 for $a_s/b_s$.} $^\mathrm{g}$Pre-impact value from \cite{Naidu2024}. $^\mathrm{h}$Pre- and post-impact values from \cite{Naidu2024}. Using a different model, \cite{Scheirich2024} find a post-impact value of $11.3675 \pm 0.0004$~h. $^\mathrm{i}$Pre-impact value from \cite{Scheirich2009}; post-impact value from \cite{Naidu2024}. \cite{Scheirich2024} find a post-impact value of $0.028 \pm 0.005$, but possibly dropping to zero 70~d after impact gives the best fit in their model (see \sect{eccentricity}). $^\mathrm{j}$Pre-impact value from \cite{Pravec2006}; post-impact value from J. \v{D}urech \& P. Pravec (2024), in preparation. $^\mathrm{k}$Pre-impact value assumed to be same as secondary orbital period (tidal lock). $^\mathrm{l}$Value from \cite{Naidu2024}. \cite{Scheirich2024} find $7.3 \pm 0.7$~$^\circ$/d in their model.}
    \end{tabular}
\end{table}

In the remainder of this paper, we expand on the post-encounter observations and dynamics implications in detail. \sect{dynamics} describes constraints on the perturbed dynamics of the system from observations of the post-impact mutual binary orbit, with implications for the possible new shape and spin state of the secondary, limits on the system mass distribution, and the possible importance of post-impact ejecta momentum exchange. \sect{beta} provides an update on the $\beta$ estimate, including progress toward measuring the heliocentric momentum change, modeling of exchanges between ejecta fragments and the binary components, and the effect of surface curvature on recoil efficiency. \sect{shape} provides implications for the inferred component rubble structure in light of new observations, the possible origin of the oblate pre-impact shape of Dimorphos, effects on $\beta$ and the rotation state arising from possible reshaping of the secondary, and implications for secular effects driven by tides and BYORP over year-long timescales. \sect{hera} presents expectations for Hera given our best assessment of the post-impact dynamical state of the system. \sect{concl} summarizes our conclusions from this work. The reader is referred to companion papers in this focus issue to form a complete picture of the aftermath of the DART mission.


\section{Perturbed Dynamics} \label{s:dynamics}

Spin-orbit coupling is common in binary asteroids due to their generally irregular shapes and close mutual proximity. \edit1{Thus, binary asteroids are best modelled by the Full Two-body Problem (F2BP). We can numerically simulate the dynamics of binary asteroids by integrating the F2BP using software such as the General Use Binary Asteroid Simulator} \edit2{(\textsc{GUBAS}) \citep{Davis2020, Davis2021GUBAS}.} The effects of perturbations are also strongly coupled as explored in detail in \cite{meyer2023a}. The observed changes in the orbit of Dimorphos allow us to constrain the properties of the system in the following way. The impact decreased the semi-major axis of the system and increased its eccentricity (at least initially; see \sect{eccentricity}). As a result, the system was perturbed out of equilibrium and the argument of periapsis precessed at the rate given in \tbl{params}. Simultaneously, the rotation state of Dimorphos was probably excited in the impact (\sect{Dimo_rotation}), and its evolution was coupled to the evolution of the binary eccentricity and semi-major axis. The tangential component of the orbital velocity change ($\Delta V_T$) is one of the main measurements that allows us to bridge pre-encounter and post-encounter states, giving insight into the dynamical parameters.

Based on the initial results published by \cite{Thomas2023}, \cite{Li-Nature-2023}, and \cite{Daly2023}, \cite{Cheng2023} calculated $\Delta V_T=2.7\pm0.1$ mm~s$^{-1}$ due to the DART impact. This calculation was corroborated by \cite{meyer2023a}, who go on to calculate the change in eccentricity and semimajor axis of the orbit.  \edit1{Using so-called `observable elements', defined using only the physical separation between the two asteroids,} they report a post-impact \edit1{observable} eccentricity of $0.027\pm0.001$ and a post-impact \edit1{observable} semimajor axis of $1189\pm17$~m.

Since these publications, the post-impact measurements of the system have been refined. Using the iterative algorithm defined in \cite{Agrusa2021Excited} and \cite{meyer2023a}, the nominal parameters of the pre- and post-impact Didymos-Dimorphos system (listed in Table \ref{t:params}) lead to an estimate of the system's bulk density equal to $2.79$~g~cm$^{-3}$. This value of the bulk density is in line with the bulk densities measured for other S-type asteroids: somewhat higher than Itokawa ($1.95$~g~cm$^{-3}$), but similar to Eros and Ida ($2.6$~g~cm$^{-3}$) \citep{Abe2006,Wilkison2002,Belton1995}. This is also between the density of (66391) Moshup ($1.97$~g~cm$^{-3}$) and its secondary Squannit ($2.81$~g~cm$^{-3}$) \citep{Ostro2006}.

Using the calculated change in semimajor axis from \cite{meyer2023a}, we find a post-impact \edit1{observable} semimajor axis of $1152\pm18$~m. The measured post-impact \edit1{eccentricity} is consistent with a physically circular pre-impact orbit, as described in \cite{meyer2023a}.

\edit1{Similar to the approach in} \cite{meyer2023b}, \edit1{we iterate the system mass to match the pre-impact orbit period, then iterate the tangential $\Delta V_T$ until we match the post-impact orbit period. Using this approach with the} updated nominal post-impact system, we calculate a nominal $\Delta V_T$ \edit1{of} 2.42~mm~s$^{-1}$. Interestingly, \citep{Naidu2024} calculate a $\Delta V_T$ of 2.62~mm~s$^{-1}$, slightly larger than our value. Our smaller value is the result of the change in orbit period caused by secondary reshaping \citep{meyer2023a,nakano2022nasa}. However, these values are largely consistent when considering the system uncertainties.

\subsection{Orbit Precession}
\label{s:orbit_precession}

The measured post-impact precession rate of around $6.7^\circ$ per day warrants discussion \citep{Naidu2024, Scheirich2024}. The $J_2$ value reported in \cite{Naidu2024}, \edit2{estimated using an orbit fit matching the photometry data}, is about 0.092. \edit2{For a detailed discussion of the methods used for this estimation, see \cite{naidu2022anticipating}} \edit1{ This is consistent with a uniform-density distribution in Didymos and} suggests a much-faster precession rate, around $13^\circ$/d \edit1{using classical perturbation theory \citep{MurrayDermott}}. One interpretation of this discrepancy is that Didymos \edit1{in reality} has an over-dense interior that decreases its effective $J_2$. However, a reduction in $J_2$ of around 50\% \edit1{that is required to achieve the observed precession rate} is physically unrealistic, \edit1{requiring an inner core with a radius equal to the polar radius of Didymos and a density of 4--5~g/cm$^3$, which is unrealistic for a rubble-pile S-type asteroid such as Didymos}.

Another consideration is how the secondary's elongation and libration affect the system's precession rate, \edit1{as shown by} \citet{cuk2010orbital}. \cite{meyer2022} demonstrated \edit1{that} both the $\beta$ value of the impact and the secondary's elongation affect the apsidal precession of the orbit. Once \cite{Cheng2023} estimated the impact's $\beta$, \cite{meyer2023a} calculated that increasing the post-impact elongation of Dimorphos reduced the system's precession rate. \edit2{We note the apsidal precession rate is given by the expression \citep{cuk2010orbital}}

\begin{equation}
    \dot{\varpi} = \frac{3}{2}\frac{J_2 n}{a^2(1-e^2)^2} + \frac{C_{22}a\sqrt{1-e^2}}{e}\bigg[9\cos2\phi\cos f+6\sin2\phi\sin f \frac{2+e\cos f}{1+e\sin f}\bigg] ,
\end{equation}
\edit2{where $f$ is the true anomaly, $\phi$ is the secondary's physical libration angle, $a$ is the semimajor axis, $e$ is the eccentricity, $n$ is the mean motion, and $J_2$ and $C_{22}$ are the second-degree gravity terms. Changing the elongation of the secondary changes its $C_{22}$ gravity term, which affects the precession rate. We also note how both the libration angle and true anomaly appear in this expression. By changing the $\beta$ value of the impact, the relationship between these two angles will also change, again affecting the precession rate. Furthermore, larger impact $\beta$ values also result in a smaller semimajor axis and larger eccentricity. This demonstrates the complex nature of orbital precession in perturbed binary asteroid systems.}

\begin{figure}[ht!]
   \centering
   \includegraphics[width = 3in]{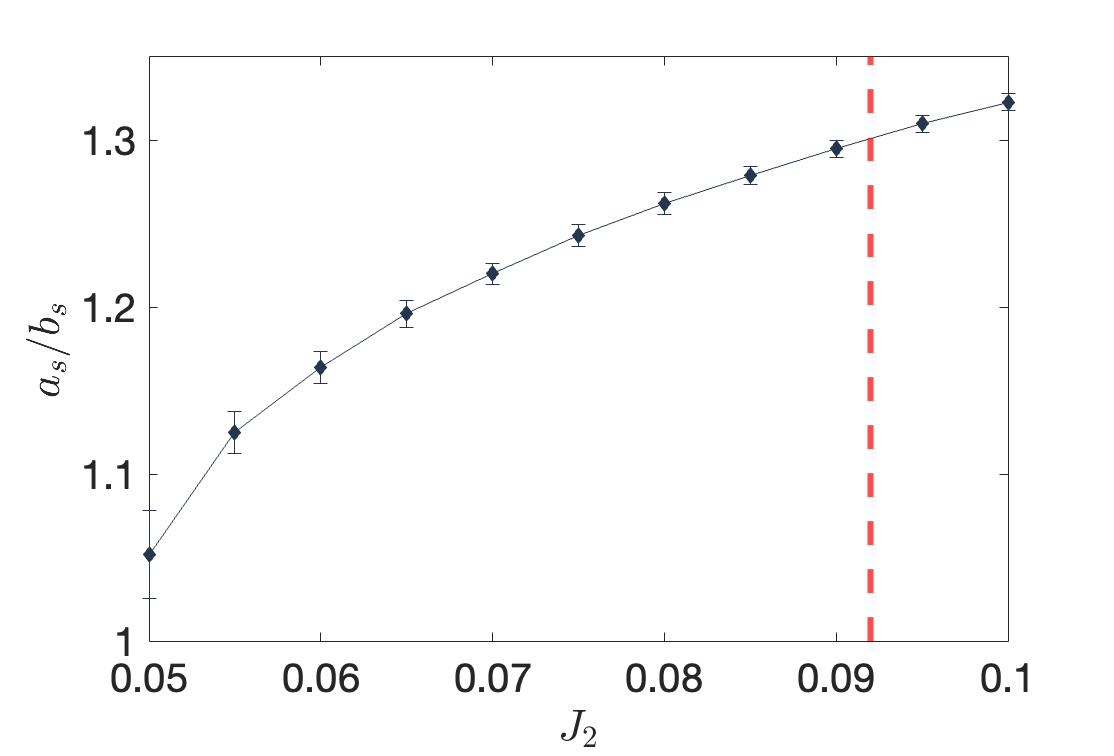} 
   \caption{For the observed precession rate of the mutual orbit, the curve indicates the necessary Dimorphos elongation for a given $J_2$ value for Didymos, \edit1{with error bars corresponding to the uncertainty on the precession rate measurement}. The \edit1{homogeneous density} $J_2$ is shown as a red dashed line, \edit1{which corresponds to a uniform density distribution in Didymos}.}
   \label{f:ab_v_j2}
\end{figure}

Thus, we find a degeneracy in explaining the orbit precession rate, as both the primary's interior structure and the elongation \edit1{and libration} of Dimorphos can contribute to the observed value. To illustrate, \edit1{using fully coupled numerical simulations} we plot the necessary Dimorphos elongation, $a_s/b_s$, as a function of the primary's $J_2$, in \figr{ab_v_j2}. This curve represents possible combinations of primary oblateness and secondary elongation that can account for the observed precession rate. When accounting for the secondary's libration, the estimate in \cite{Naidu2024} suggests a $J_2$ value largely consistent with a homogeneous density distribution in Didymos, \edit1{shown as a dashed red line in \figr{ab_v_j2}}. 

Furthermore, lightcurves suggest a more elongated secondary \edit1{with $a_s/b_s$ between 1.1 and 1.4 \citep{Pravec2024}}, also indicating a homogeneous density distribution within Didymos. To go along with their estimate of $J_2$, \cite{Naidu2024} also estimate an elongated secondary around $a_s/b_s=1.3$. \edit1{These results are consistent with impact simulations suggesting a global reshaping of Dimorphos \citep{raducan2024physical}}. The reshaping of Dimorphos could also substantially affect its post-impact orbital period change and rotational state (see \sect{Dimorphos_reshaping} for details).

\subsection{Dimorphos's Rotation}
\label{s:Dimo_rotation}

The rotation state of Dimorphos is of particular interest to Hera (\sect{hera}), which will deploy two CubeSats to perform proximity operations around the secondary. The likely circular pre-impact orbit indicates a lack of substantial \edit1{forced} libration in Dimorphos prior to the impact, \edit1{meaning the secondary's rotation rate is nearly identical to the orbit rate over the full orbit}. Thus, the typical assumption of a perfectly synchronous pre-impact secondary still holds \citep{Richardson2022}. We can therefore determine the minimum amount of libration in the post-impact system, which occurs if DART impacted in line with the center of mass of Dimorphos.

\begin{figure}[ht!]
   \centering
   \includegraphics[width = 3in]{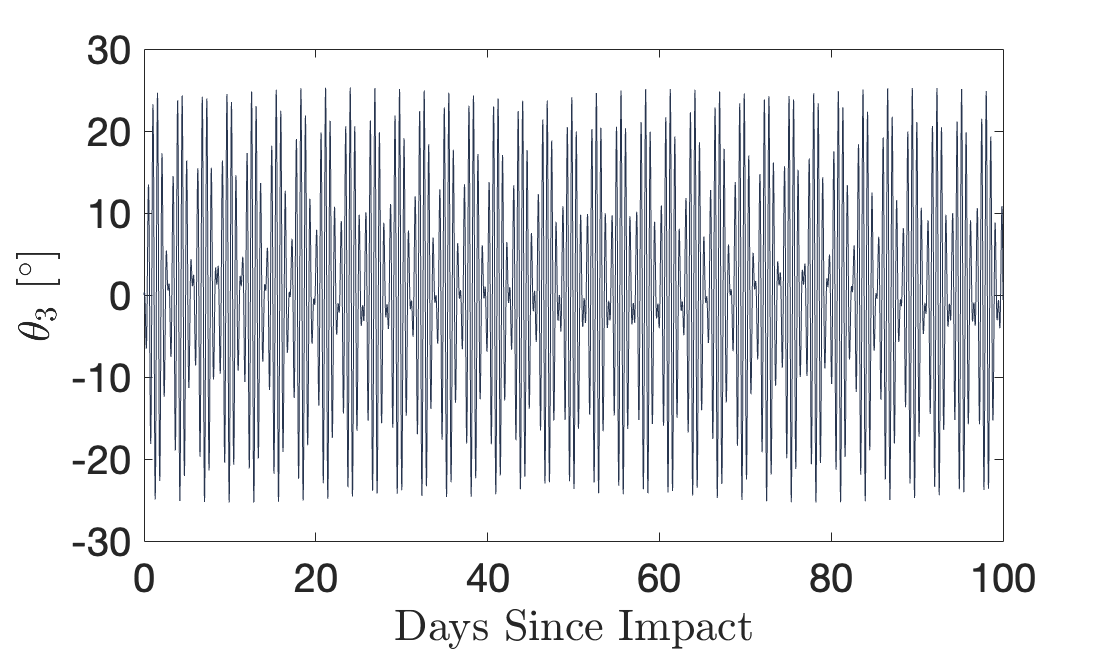} 
   \caption{Libration angle of Dimorphos for the nominal system if the impact vector passed through its center of mass.}
   \label{f:min_lib}
\end{figure}

\edit1{To investigate the post-impact rotation of Dimorphos, we perform \edit2{high-fidelity} numerical simulations using \textsc{GUBAS}, \citep{Davis2021GUBAS} following} the methodology from \cite{Agrusa2021Excited}: \edit1{we apply a $\Delta V_T$ to \edit2{Dimorphos when it is rotating synchronously with the pre-impact orbit period, then} simply track the \edit2{post-impact} attitude of the perturbed secondary over time.} We use the nominal system parameters \edit1{without uncertainties as inputs for a \textsc{GUBAS} simulation, and integrate for 100 days.}  \edit2{This simulation is used to predict the libration amplitude in the system, which has not previously been studied in high-fidelity simulations.} \figr{min_lib} shows the libration angle, \edit1{measuring the angle between the secondary's long axis and the line connecting the centers of the two bodies,} over time for the case where the impact vector passes through the secondary's center of mass \edit1{meaning there is no torque imparted by the impact}. Thus, this is the \textit{minimum} amount of libration present in the system, corresponding to an amplitude of $\pm25^\circ$. We note that this nominal system remains attitude stable after the impact, where rotation only occurs about the secondary's major principal inertia axis. There are several underlying periods driving the behavior of the libration angle, all with periods on the order of hours. \edit1{Using a Fast Fourier Transform decomposition, we find the} main frequency \edit1{of libration} has a period around 14~h, with a secondary period around 11.4~h \edit1{(equal to the mean motion)}, and several other minor frequencies. These commensurate periods result in beating in the libration angle.

\begin{figure}[ht!]
   \centering
   \includegraphics[width = 3in]{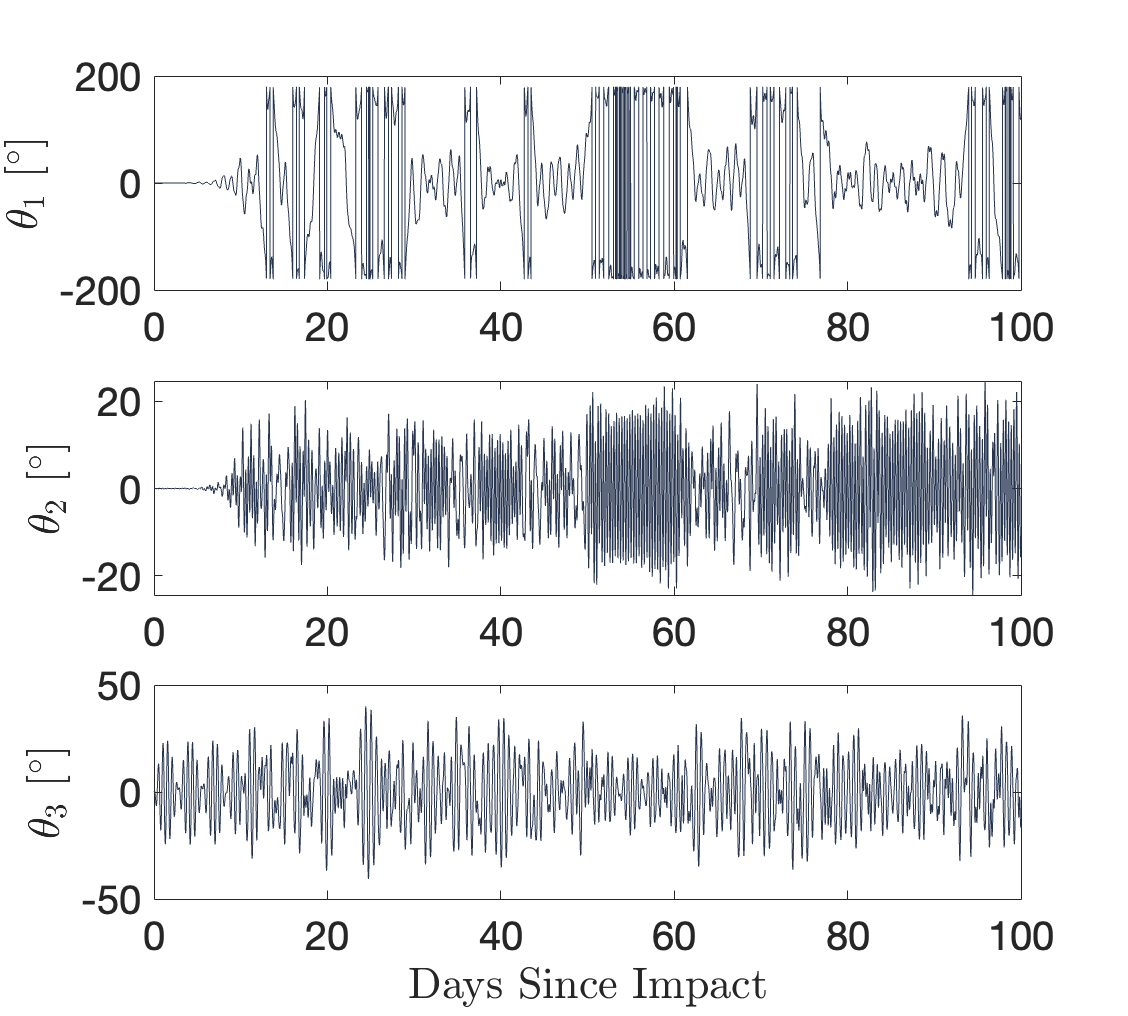} 
   \caption{1-2-3 roll, pitch, yaw Euler angles for the secondary for a tumbling case. On average, the secondary is still tidally locked with the primary.}
   \label{f:tumble_euler}
\end{figure}

\edit1{However, in reality there} is still considerable uncertainty surrounding the post-impact secondary shape. As a result, not only are larger libration amplitudes possible, Dimorphos may be in a non-principal-axis rotation state (\ie tumbling).\footnote{\edit1{\citet{Pravec2024} indicate that post-impact tumbling is expected for the range of shapes that fit their models while \citet{Naidu2024} find a non-tumbling spin state is consistent with their data and cannot be ruled out.}} The nominal secondary shape given in Table \ref{t:params}, $a_s/b_s=1.3$ $b_s/c_s=\edit1{1.3}$, is near a large region \edit1{of attitude instability} \citep{Agrusa2021Excited}. Adjusting the shape slightly can induce attitude instabilities. For example, the 1-2-3 roll, pitch, yaw Euler angles, \edit1{measuring the secondary's attitude relative to a rotating Hill frame,} for the shape $a_s/b_s=1.2$, $b_s/c_s=1.4$ are plotted in \figr{tumble_euler} \citep[for methods, see][]{Agrusa2021Excited}. This demonstrates that not only are larger libration amplitudes possible, the secondary can begin tumbling after the impact. However, \edit1{this} tumbling state of Dimorphos is still on-average synchronous with the orbit rate, so its long axis is still generally pointed toward Didymos. In \figr{tumble_euler}, this is illustrated by the yaw angle $\theta_3$ remaining $< 90^\circ$.

The impact and corresponding ejecta almost certainly imparted some amount of torque to Dimorphos, \edit1{so we now relax our torque-less impact assumption and include a perturbation to the secondary's spin rate}. The spin angular momentum imparted to Dimorphos is
\begin{equation}
	\Delta \vec{L} = \vec{r}_\mathrm{imp}\times\Delta \vec{p},
\end{equation}
where $\vec{r}_\mathrm{imp}$ is the impact location of DART relative to the center of mass of Dimorphos, and $\Delta \vec{p}$ is the momentum transfer caused by DART. The spacecraft impacted roughly 25~m off from the center of figure \citep[under our uniform density assumption this is also the center of mass; ][]{Daly2023}. The change to Dimorphos's spin vector is then
\begin{equation}
    \label{eq:deltaOmega}
	\Delta\vec{\omega} = \mathbf{I}^{-1}\big(\vec{r}_\mathrm{imp}\times\Delta \vec{p}\big),
\end{equation}
where $\mathbf{I}$ is the (unmeasured) inertia tensor of Dimorphos. 

The true impact geometry, taken from \cite{Daly2023}, does not pass through Dimorphos's center of mass. Importantly, the resultant torque of the impact and ejecta acts to increase the secondary's spin period. On the other hand, the post-impact orbit period is reduced. This means the new Didymos system has a larger difference between the secondary's rotation period and the orbit period than if the impact did not change the spin period, leading to a larger libration amplitude.

\begin{figure}
\gridline{\fig{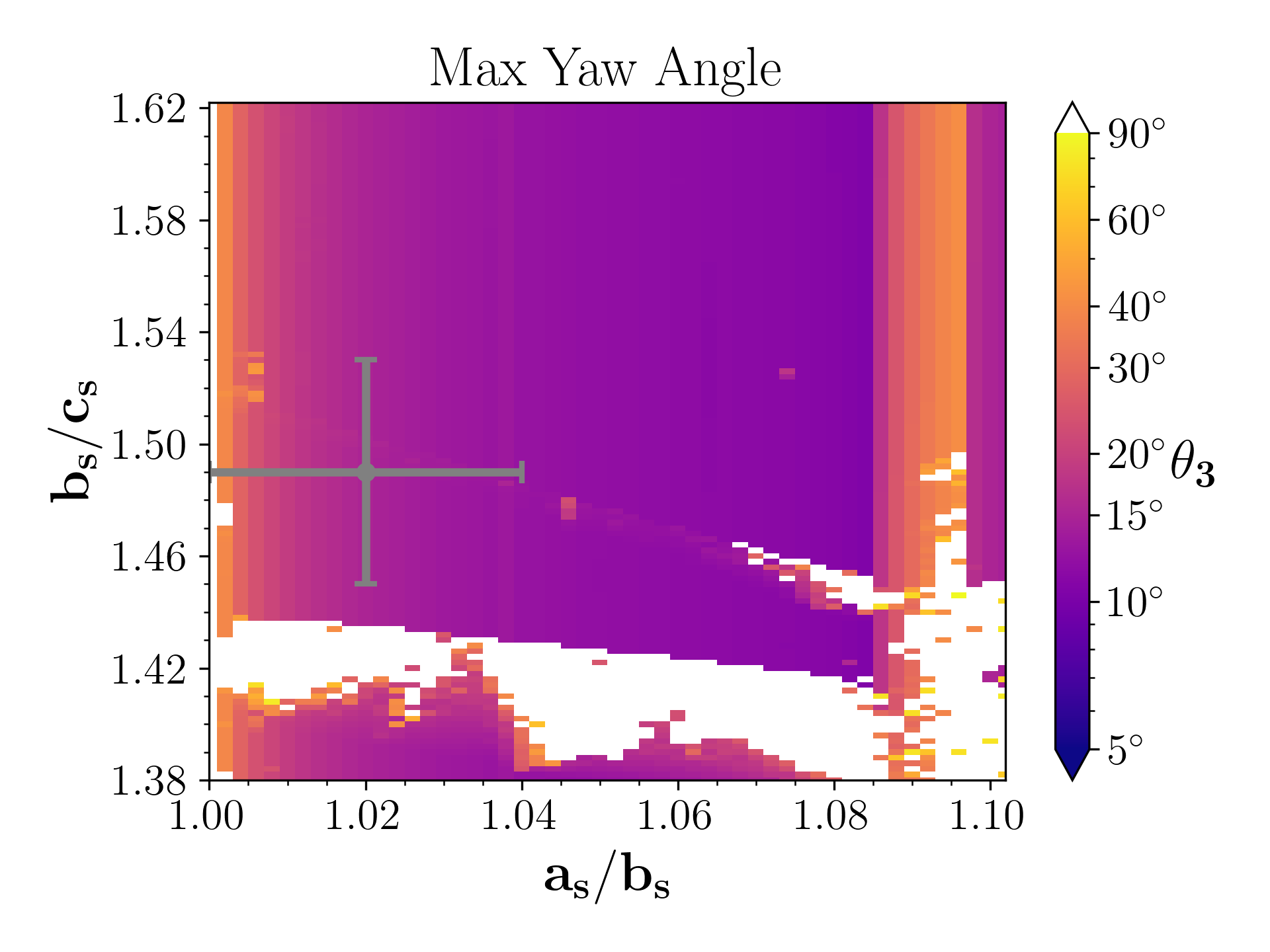}{0.5\textwidth}{(a) Only including $\Delta\vec{V}$}
            \fig{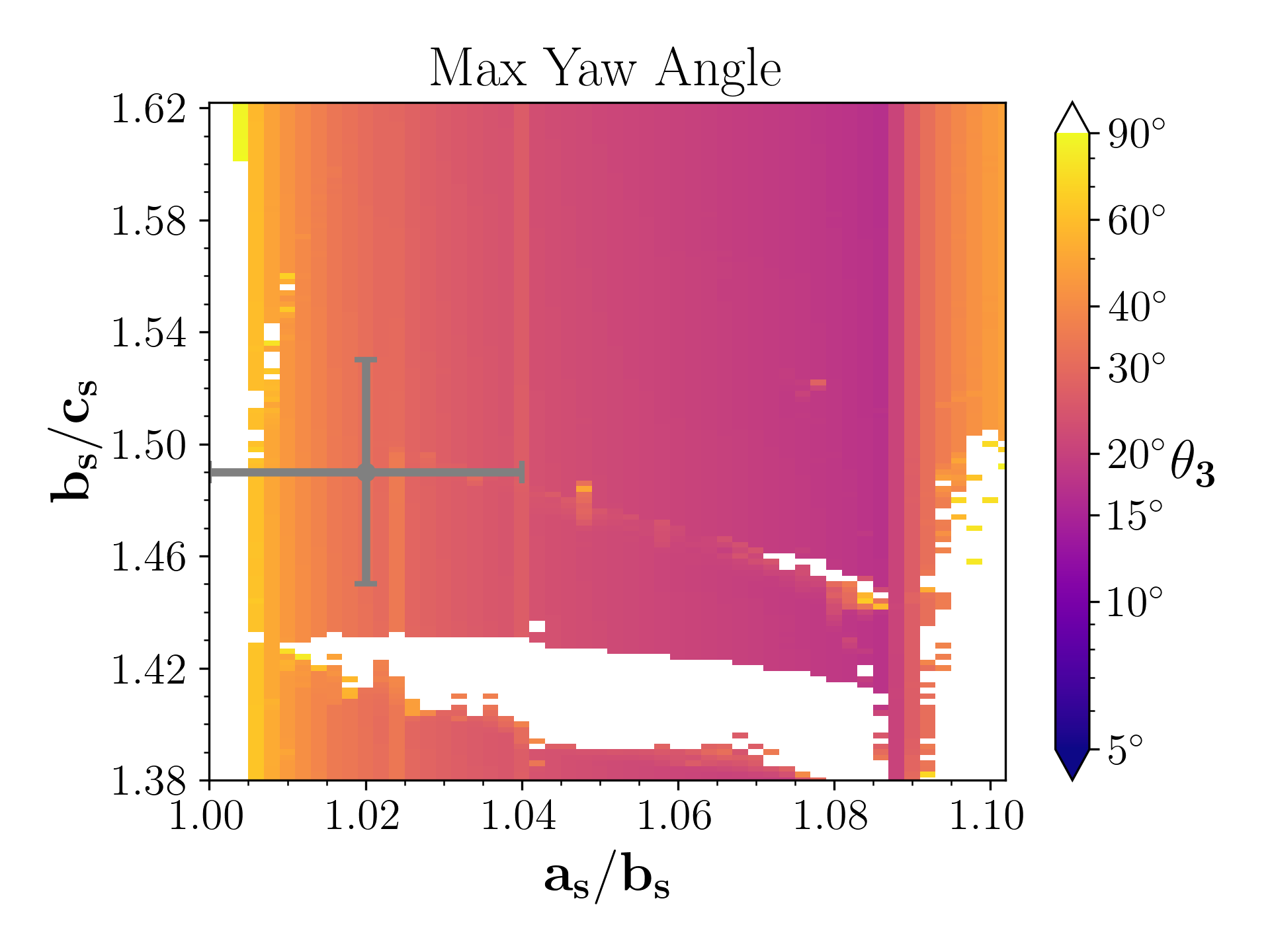}{0.5\textwidth}{(b) Including both $\Delta\vec{V}$ and $\Delta\vec{\omega}$}
            }
   \caption{Maximum yaw angle of Dimorphos, comparing the effects of including the impact torque. When this torque is included, the libration amplitude of the post-impact system generally increases on the order of $10^\circ$. When $\Delta \vec{\omega}$ is included and Dimorphos has a dynamically equivalent $a_s/b_s\lesssim1.005$, the impact can cause Dimorphos to immediately break from synchronous rotation. The shape of the best-fit ellipsoid is shown by the gray errorbars \citep{Daly2023}. The structure of these plots is explained by resonances among Dimorphos's natural frequencies that can trigger attitude instabilities \citep[see][and references therein]{Agrusa2021Excited}}.
   \label{f:yaw_torque}
\end{figure}

The importance of including the impact torque is illustrated in \figr{yaw_torque}. One case includes the effects of the impact's imparted $\Delta\vec{V}$, whereas the other includes both $\Delta\vec{V}$ and $\Delta\vec{\omega}$. We see larger libration amplitudes when we include the change in rotation rate caused by the impact torque. \edit1{This demonstrates the importance of the torque imparted by the impact, and shows the predictions by \cite{Agrusa2021Excited} are under-estimates.}

\edit1{In a formal analysis of the post-DART rotation state of Dimorphos, \cite{Pravec2024} leverage synthetic lightcurves of Dimorphos to assess the rotational stability of Dimorphos. They find significant NPA rotation is required to reproduce observations of the secondary lightcurves.}

\subsection{Orbit Period} \label{s:orbit_period}

\begin{figure}[ht!]
   \centering
   \includegraphics[width = 3in]{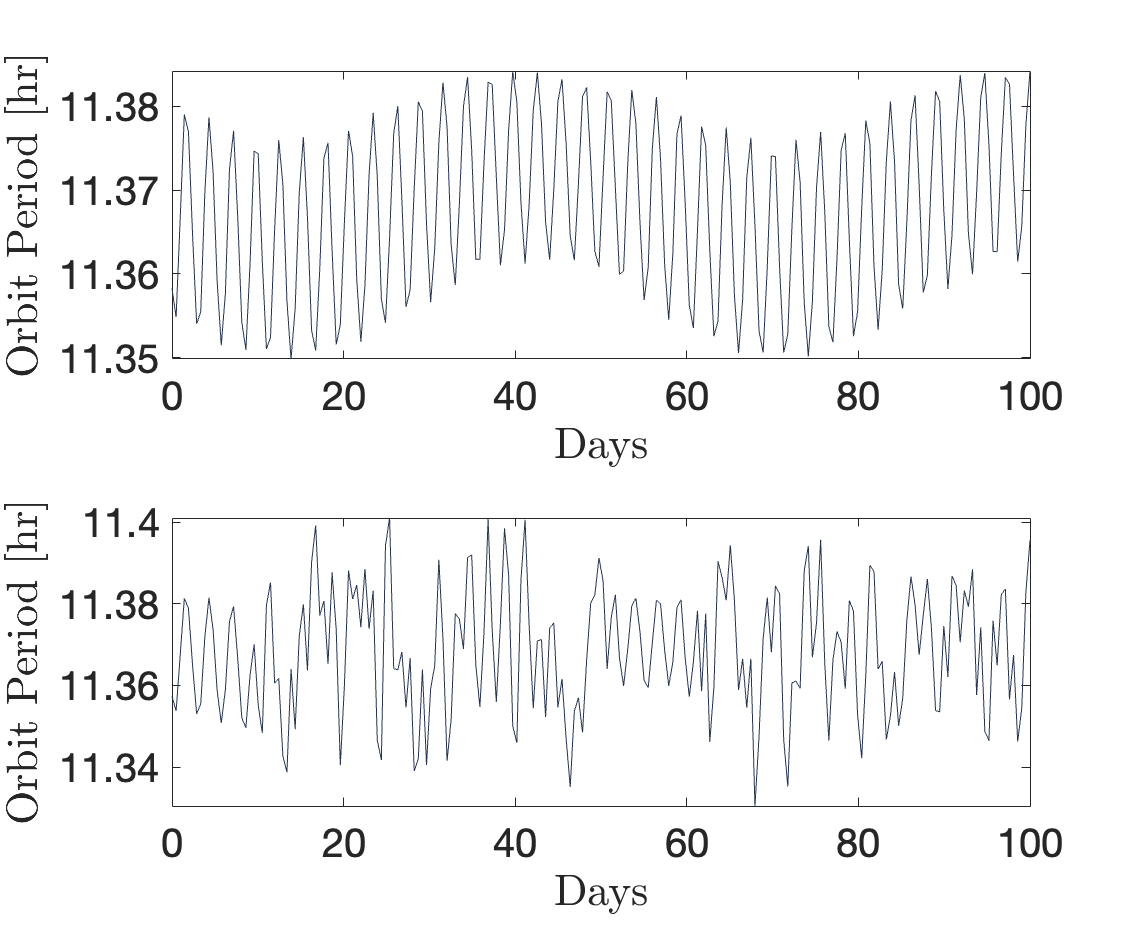} 
   \caption{(Top) Sidereal orbit period for the system with secondary $a_s/b_s=1.3$, $b_s/c_s=1.1$. (Bottom) Sidereal orbit period for the system with secondary $a_s/b_s=1.2$, $b_s/c_s=1.4$.}
   \label{f:orbit_periods}
\end{figure}

While the average orbit period has been measured to high accuracy, the instantaneous orbit period will experience non-secular variations due to precession and angular momentum exchange within the system \citep{meyer2022}. In calculating the system's orbit period, it is useful to use the sidereal orbit period, calculated as the time required for the secondary to move $360^\circ$ around the primary in inertial space \citep{meyer2022,meyer2023b}. From \edit1{the simulations we presented above showing the secondary's attitude, we now also} plot the sidereal orbit period in \figr{orbit_periods}, for the stable libration system ($a_s/b_s=1.3$, $b_s/c_s=1.1$) and the unstable tumbling system ($a_s/b_s=1.2$, $b_s/c_s=1.4$). 

For the stable system, we see two distinct signals in the orbit period. The long-period signal is the result of the orbit's precession, which was measured with observations. The short-period signal is caused by libration within the system, but \edit1{is likely} at too high a frequency and too low an amplitude \edit1{so that the mean anomaly modulation is less than 1 degree and is too small} to be detected in the data \citep{meyer2022}. \edit1{This signal could potentially be found by calculating the variation in mutual event timings over many orbit periods.}

In the unstable system, the onset of tumbling removes the periodic behavior from the variations in favor of chaotic behavior. Again, the variations are \edit1{likely} too high in frequency and too low in amplitude for detection, but otherwise they would indicate an unstable rotation state in the secondary. Also note the long-period signal is lost, suggesting that the loss of the precession signal in the observations also indicates tumbling in the secondary. This is consistent with a relatively large reduction in the orbit's eccentricity caused by the onset of secondary tumbling \citep{meyer2023b}.

\edit2{The detection of these orbit-period variations would require high-quality data over an extended period of time. This is only possible using space-based measurements. Ground-based measurements are limited to sparse coverage and typically combine several nights' worth of data \citep{Thomas2023}. This approach would average out any small modulations of the orbit period. However, space-based measurements could provide full coverage of several orbit periods. Thus, these measurements may be possible from the Hera spacecraft. A variation of the orbit period of 0.03~h to an average orbit period of 11.3674~h is a modulation of about 0.26\%. This corresponds to a correction to the mean anomaly of about $1^\circ$. Extended, high-quality observations have a chance at detecting this error in mean anomaly given sufficient coverage in time. Thus, while orbit-period variations may be hidden from ground-based data, Hera could feasibly detect this phenomenon.}

While secular damping is possible in the near future, it is unlikely to have major effects on the system when Hera arrives \citep{meyer2023a}. Thus, Hera may encounter a tumbling Dimorphos, complicating proximity operations. Monitoring of the rotation state with Hera will allow estimates of $\Delta \omega$, particularly if the system is in stable libration, but much more uncertain if Dimorphos is tumbling.

There is also a possible detection of a secular trend in orbit period for a short interval after impact that may be due to momentum exchange between the binary and the ejecta. This is discussed further in \sect{hardening}.

\subsection{Eccentricity Drop} \label{s:eccentricity}

If Dimorphos entered a tumbling state after impact, this would be accompanied by a relatively rapid decrease in the mutual orbit eccentricity \citep{meyer2023a}. This is driven by an exchange of angular momentum, where the onset of tumbling causes the secondary's angular momentum to decrease on average, resulting in an increase in the orbit's angular momentum and corresponding decrease in eccentricity. The magnitude of this decrease depends on the change of libration amplitude in Dimorphos: a large libration amplitude when Dimorphos starts tumbling corresponds to a large drop in eccentricity, whereas a smaller change in libration amplitude corresponds to a smaller drop in eccentricity. So if the libration amplitude after the onset of tumbling is nearly the same as the amplitude prior to tumbling (\ie there is a small change in the secondary's angular momentum), there will not be a large reduction in eccentricity. But if the libration amplitude is much larger after tumbling begins (\ie a large change in the secondary's angular momentum), there will be a large reduction in eccentricity. Thus, while the eccentricity of 0.027 seen shortly after the impact is consistent with a circular pre-impact orbit, this value may change over the course of several days, indicating non-principal-axis rotation in Dimorphos. Indeed, \cite{Scheirich2024} find in their dynamical model that after a transient period of about 70~d post-impact, during which the average eccentricity is $0.028 \pm 0.005$, either a drop in eccentricity to essentially zero or the start of chaotic orbital behavior with onset of tumbling in Dimorphos, is needed to explain their lightcurve data.


\section{The Momentum Transfer Enhancement Factor} \label{s:beta}

\subsection{Mutual Orbit Change} \label{s:beta_mutual}

A key objective of DART as a planetary defense test mission was to determine the momentum transfer to the target body relative to the incident momentum of the spacecraft, which is quantified by the momentum transfer enhancement factor, $\beta$~\citep{Cheng2023}, defined by the momentum balance of the kinetic impact,

\begin{equation}
    \label{eq:MomBalance}
	M\Delta\vec{V} = m\vec{U} + m \left( \beta-1 \right) \left(\hat{E} \cdot \vec{U} \right) \hat{E} .
\end{equation}

Here, $M$ is the mass of Dimorphos, $\Delta\vec{V}$ is the impact-induced change in Dimorphos's orbital velocity, $m$ is DART's mass at impact, $\vec{U}$ is DART's velocity relative to Dimorphos at impact, and $\hat{E}$ is the net ejecta momentum direction. $M\Delta\vec{V}$ is the momentum transferred to Dimorphos, $m\vec{U}$ is DART's incident momentum, and the final term in the equation is the ejecta's net momentum written in terms of the spacecraft incident momentum. This definition of $\beta$ can be re-expressed as the ratio of the components along $\hat{E}$ of both the momentum transfer and the incident momentum vectors, or $\beta = \left. M\left(\hat{E}\cdot\Delta\vec{V}\right) \middle/ m\left(\hat{E}\cdot\vec{U}\right) \right.$. Since the along-track component of $\Delta\vec{V}$, which is the component along Dimorphos's orbital velocity direction, $\hat{e}_{_{T}}$, can be estimated from Dimorphos's orbit-period change, the momentum transfer enhancement factor $\beta$ is re-expressed in terms of $\Delta\vec{V}\cdot\hat{e}_T = \Delta V_T$ as

\begin{equation}
    \label{eq:betaEqn}
	\beta = 1 + \frac{\frac{M}{m}\left(\Delta V_T\right)-\left(\vec{U}\cdot\hat{e}_T\right)}{\left(\hat{E}\cdot\vec{U}\right)\left(\hat{E}\cdot\hat{e}_T\right)} .
\end{equation}

The determination of $\beta$ for the DART impact~\citep{Cheng2023} required estimation of $\Delta V_T$, $M$, and $\hat{E}$. A Monte Carlo method was used to find a distribution for $\Delta V_T$ consistent with the measured period change incorporating the uncertainties in Didymos system parameters, such as the ellipsoid axial lengths of the asteroids, the pre-impact orbit separation distance between the asteroids, the pre- and post-impact orbit periods, and the net ejecta momentum direction $\hat{E}$. Full two-body numerical simulations \citep{Davis2021GUBAS} of the coupled rotational and orbital dynamics were used to determine $\Delta V_T$ for each sampled combination of input parameters, finding $\Delta V_T = -2.70 \pm 0.10$ $\left(1\sigma\right)$~mm~s$^{-1}$. The mass $M$ of Dimorphos was estimated from the volumes of the ellipsoid shape models~\citep{Daly2023} with assumed values for Dimorphos's density, which was not directly measured by DART. In the Monte Carlo analysis, Dimorphos's density was uniformly sampled between 1,500 and 3,300~kg~m$^{-3}$. The ejecta momentum direction $\hat{E}$ was found from Hubble Space Telescope and LICIACube observations of the ejecta, which yielded an estimate of the ejecta cone axis direction. This direction is identical to $\hat{E}$ assuming the ejecta plume holds the momentum uniformly, and $\hat{E}$ points toward a right ascension (RA) of $138^{\circ}$ and a declination (Dec) of $+13^{\circ}$ with an uncertainty of $15^{\circ}$ around this direction. Combining the inputs for $\Delta V_T$, $M$, and $\hat{E}$, the dynamical Monte Carlo analysis~\citep{Cheng2023} found $\beta$ as a function of Dimorphos's bulk density $\rho_s$,

\begin{equation}
    \label{e:betaEstimate}
	\beta = \left(3.61 \pm 0.2\right)\frac{\rho_s}{\text{2,400~kg~m}^{-3}} - 0.03 \pm 0.02\text{~}\left(1\sigma\right) .
\end{equation}

For a Dimorphos bulk density range of 1,500 to 3,300~kg m$^{-3}$, the expected value of the momentum transfer enhancement factor, $\beta$, ranges between 2.2 and 4.9. These $\beta$ values indicate that significantly more momentum was transferred to Dimorphos from the escaping impact ejecta than was incident with DART.

While additional observational data on the Didymos system have been collected in the months following the results shown in~\citep{Cheng2023}, particularly \eqn{betaEstimate}, none of the updates to the aforementioned input parameters for the $\beta$ analysis have changed sufficiently from their previous values to warrant recalculation of the $\beta$ estimate. Thus, the next anticipated update to the $\beta$ result for DART's impact on Dimorphos is expected to come from Hera mission data.

\subsection{Heliocentric Orbit Change} \label{s:beta_helio}

The DART impact changed the heliocentric orbit of the Didymos system in addition to changing the mutual orbit of Dimorphos and Didymos. The initial impulse delivered to the system's barycenter was augmented by the momentum carried by the ejecta that escaped the system \citep{Jewitt-ApJ-2023}. We define the escape criterion as ejecta crossing the Hill sphere of the binary asteroid. The combination of the DART impulse and the momentum transported out of the system by ejecta can be encapsulated in the heliocentric momentum transfer enhancement factor, $\beta_\odot$. The $\beta_\odot$ value describes the total change in the heliocentric momentum, and therefore the changes to the heliocentric orbit of the Didymos-Dimorphos system caused by the DART mission. Without measuring $\beta_\odot$, the total momentum transfer on the entire system can only be constrained through its natural upper bound, namely the local $\beta$ measured from the change in Dimorphos's orbit. Since some ejecta may stay trapped in the binary system, $\beta_\odot$ cannot exceed $\beta$. Given the short post-impact observation arc at the time of this writing, a measurement of $\beta_\odot$ is not yet possible \citep{Makadia2024}. However, high-quality astrometry derived from stellar occultations between October 2022 and March 2023 has helped to make strides toward a statistically significant $\beta_\odot$ estimate. 

Observing stellar occultations is one of the most accurate ground-based methods for measuring the on-sky position and directly determining the size and shape of solar system objects \citep{Ferreira2022}. However, stellar occultations of $<$1~km-diameter asteroids present their own inherent challenges \citep{Souami2022}. The Asteroid Collaborative Occultation Research via Occultation Systematic Survey (ACROSS) project, funded by ESA in its initial exploratory phase, supports the DART and Hera missions. ACROSS observation campaigns on different continents between June 2022 and March 2023 led to 20 successful stellar occultations by Didymos being recorded, four of which allowed for the detection of Dimorphos. This makes Dimorphos the smallest object ever observed during an occultation campaign \citep{Souami2022}. It is worth noting that at the end of the Didymos campaigns, occultation astrometry was fundamental to reach an exceptional orbit quality (50~m uncertainty on the sky plane in March 2023).

\citet{Makadia2024} investigated whether more occultation data could help speed up the process of determining $\beta_\odot$. \edit1{This was done by directly adding $\beta_\odot$ to the list of estimated parameters during the orbit determination process for the Didymos system barycenter}. \edit2{The least-squares orbit determination process naturally gives estimates and corresponding 1-$\sigma$ uncertainties for $\beta_\odot$ after solving the normal equations.} \tbl{dart_results} shows the summary of future observation scenarios considered for the estimation of $\beta_\odot$. The first row corresponds to a case that assumes five new monthly occultation measurements between June--October 2024, before the launch of the Hera spacecraft. The second row corresponds to five additional occultations in February--June 2027, while the Hera spacecraft is at the Didymos system. The third row considers just five additional pseudorange measurements of the Didymos system barycenter. These pseudorange measurements are radar delay measurements of the system barycenter taken from tracking data of the Hera spacecraft. Finally, the last row is the combination of all previous scenarios. For additional justification of the number and types of these observations, as well as the \edit2{full} methodology behind estimating $\beta_\odot$, the reader is referred to \citet{Makadia2024}.

The highest signal-to-noise ratio (SNR) for $\beta_\odot$ naturally comes from the case with the highest number of future high-accuracy observations. However, the more interesting result is that a statistically significant estimate of $\beta_\odot$ could be generated through additional occultation observations even before the launch of the Hera spacecraft. This scenario is contingent on successful observations from the stellar occultation measurement campaigns in the second half of 2024.

\begin{table}[!ht]
    \centering
    \caption{Estimated accuracy for retrieved $\beta_\odot$ for various future observation campaigns.}
    \begin{tabular}{ccc}
    \hline
        Scenario                                        & $\beta_\odot$ & SNR \\ \hline
        5 Occ.\ (2024)                                  & 3.212         &  4.850 \\
        5 Occ.\ (2024) + 5 Occ.\ (2027)                 & 3.025         &  7.835 \\
        5 Hera (2027)                                   & 3.011         & 16.747 \\
        5 Occ.\ (2024) + 5 Occ.\ (2027) + 5 Hera (2027) & 3.031         & 17.014 \\ \hline
    \end{tabular}
    \tablecomments{The target value to be retrieved in these simulations is $\beta_\odot$=3;\\Occ: Stellar occultation measurements; Hera: Hera pseudorange measurements.}
    \label{t:dart_results}
\end{table}

There is, therefore, a clear benefit from follow-up stellar occultation observations, certainly in 2024. The ACROSS collaboration plans to organize dedicated occultation campaigns to observe those events. We predict that the uncertainty of the current orbit of Didymos propagated to the beginning of its observability period in 2024 is only $\approx$1 Didymos body radius. This means that a small number of observers on the ground can lead to a successful occultation event. Following similar plans to the 2022--2023 campaigns, the ACROSS collaboration will be deploying some of its instruments for the 2024 campaigns and will be interacting with the amateur community to ensure the best coverage of the best events. A first positive detection in 2024 will then provide the accuracy needed for successful stellar occultation campaigns that capture the events that will follow. As mentioned above, if successful, these 2024 occultations should already provide a SNR of $\approx$5 by the end of next year. Given additional measurements from the Hera spacecraft, this SNR can then jump to $\approx$17 in 2027.

\subsection{The dynamical effects of DART-produced ejecta on the Didymos system}

Here we discuss and quantify the dynamical effects that DART-produced ejecta have on the dynamics of the Didymos system.

\subsubsection{Ejecta mass estimates}

Post-impact observations have provided estimates about the mass of the ejecta produced by the kinetic impact of DART on Dimorphos. As reported in \tbl{ejecta_mass}, the observations are in agreement within an order of magnitude around $\sim 10^7$ kg. Studies based on observed ejecta quantify the mass of a subgroup of ejecta, \ie leaving the Didymos system up to 2--3 weeks after the DART impact, or within a certain size range. Also, some studies \citep[i.e.,][]{Roth-PSJ-2023,Jewitt-ApJ-2023,Gudebski2023} assume a lower system density compared to the latest estimates (\tbl{params}). In addition, many observations have preferred sensitivity to specific particle sizes (\eg $\sim$~mm in diameter), likely missing ejecta in the full particle size range. This implies that numbers in \tbl{ejecta_mass} likely represent lower-bound values.

The total quantity and mass of ejecta can be estimated by comparing observations to numerical models of dust dynamics and dispersal taking into account a distribution of ejection velocities and particle sizes as well as both gravity and solar-radiation pressure. Such estimates for the total quantity and mass of ejecta vary widely, as summarized in \tbl{ejecta_mass}. The particle sizes of this material probably extend from micrometers into the boulder size regime \citep{Cheng-Icarus-2020,Cheng-PSJ-2022}.

\begin{table*}[ht]
\centering
    \footnotesize
	\caption{Estimates of total ejecta mass from observations and constrained numerical simulations. The former are limited to observed ejecta (\ie within a certain range of sizes, or within a certain time span), while the latter are limited by model assumptions (\ie size range and distribution of ejecta).}
	\begin{tabular}{lp{8cm}l}
    \hline
    Total mass [kg] & Notes & Reference \\
    \hline
    $4.2\times10^{6}$ & numerical simulations (pre-impact) & \cite{Moreno-MNRAS-2022} \\
    $6.8\times10^6$ & observations, visible, w/HST & \cite{Li-Nature-2023} \\         
        & (integrating under differential size-frequency distribution fit to tail profiles) \\
    $1.3$--$2.2\times10^7$ & observations, visible, w/Unistellar network & \cite{Graykowski-Nature-2023} \\
    $0.9$--$5.2\times10^7$ & observations, mm-wave, w/ALMA & \cite{Roth-PSJ-2023} \\
    $5.2\times10^6$ & observations, visible, w/HST & \cite{Jewitt-ApJ-2023} \\
        & (mass of slow-boulder population alone) \\
    $9.4\times10^6$ & numerical simulations & \cite{Moreno-PSJ-2023} \\
        & \citep[constrained by HST observations in][]{Li-Nature-2023} \\
    $1.1$--$5.5\times10^7$ & numerical simulations & \cite{Ferrari2024} \\
        & \citep[constrained by HST observations in][]{Li-Nature-2023} \\
    $1.7$--$4.3\times10^7$ & numerical simulations & \cite{raducan2024physical} \\
        & \citep[constrained by LICIACube observations in][]{Dotto-Nature-2024} \\
    $0.65$--$4.1\times10^7$ & point source scaling from numerical simulations & \cite{Cheng2023b} \\
        & \citep[constrained by LiciaCube observations in][]{Dotto-Nature-2024} \\
    $1.7$--$2.2\times10^7$ & numerical simulations & \cite{Kim2023} \\       
        & \citep[constrained by HST observations in][]{Li-Nature-2023} \\ 
    $4.5\pm3.7\times 10^6$ & analytic arguments from observed post-impact period change  & \sect{hardening} of this work \\
        & \citep[constrained by ground-based observations in][]{Naidu2024, Scheirich2024} \\
    \hline
    \end{tabular} \\
	\label{t:ejecta_mass}
\end{table*}

\begin{figure}[ht]
    \centering
    \includegraphics[width=0.48\linewidth]{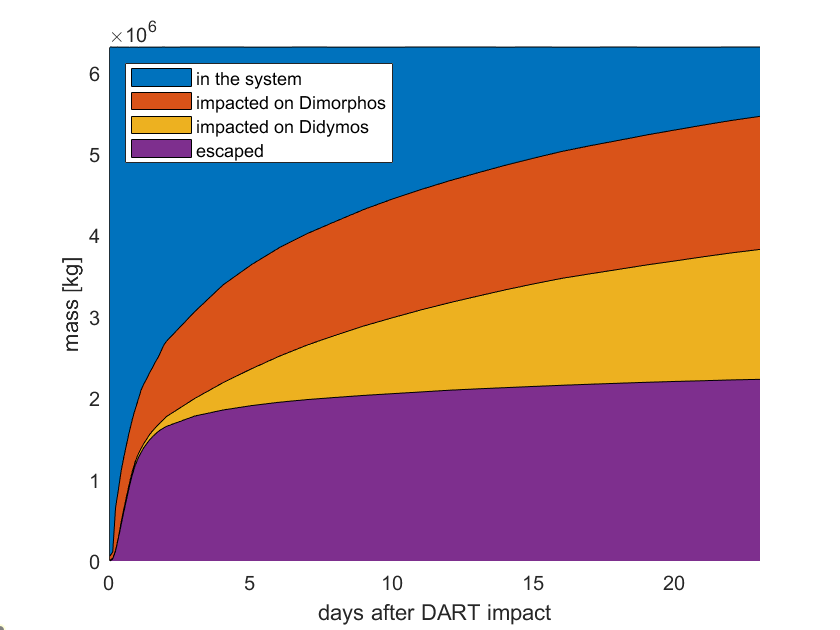}
    \includegraphics[width=0.48\linewidth]{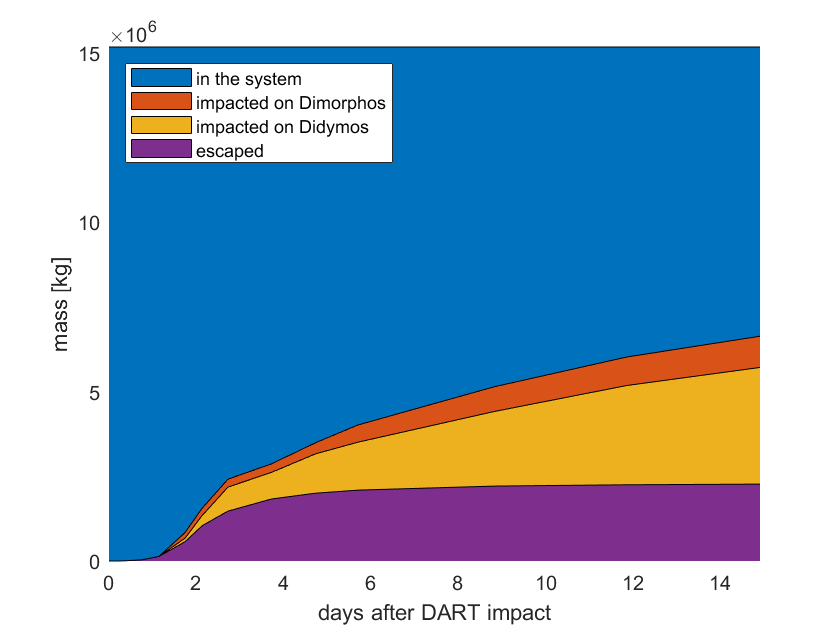}
    \caption{Evolution of ejecta mass in time showing mass escaped from the system vs.\ mass that remains in the system, either by orbiting about it (labeled ``in the system'') or after re-impacting on Didymos/Dimorphos. \edit1{The colored regions indicate the amount of mass that belongs to each subgroup at each time: \ie 100\% of the mass at $t=0$ is ``in the system,'' while some mass is being lost to impact on Didymos/Dimorphos, or to escape the system, as time progresses.} Left: \cite{Moreno-PSJ-2023}. Right: \cite{Ferrari2024}.}
    \label{f:plot_mass}
\end{figure}

\figr{plot_mass} shows two examples of ejecta evolution in time. In these cases, the total ejecta mass is $\sim 6 \times 10^6$ kg \citep{Moreno-PSJ-2023} and $\sim 1.5 \times 10^7$ kg \citep{Ferrari2024}, respectively. Both models resolve the dynamics of the ejecta fragments as they evolve after the DART impact but starting from different assumptions regarding their initial state\edit1{. In more detail, both works consider particles in the range $\mu$m--cm, but \cite{Moreno-PSJ-2023} use a broken power law with index $–2.5$ for particles smaller than 3 mm, and $-3.7$ for larger ones, while \cite{Ferrari2024} use a single power law with index $-2.7$. Also, \cite{Moreno-PSJ-2023} initialize the velocity of ejecta particles using a combination of an isotropic distribution (equal to the escape speed from Didymos) and a power law with index $-0.5$, while \citep{Ferrari2024} investigate velocity power law distributions with index between $-0.4$ and $-1$}. Both cases show the amount of mass contributing to the early tail (escaped) and the ejecta mass that remains in the Didymos system, either orbiting about it (labeled ``in the system'') or impacted on Didymos/Dimorphos. Both cases provide a similar estimate of escaped mass after 15 days ($\sim$2.15--2.27$\times 10^6$~kg), whereas the largest difference is given by the mass that remains in the system by that time.

The estimates above provide a range of values for impacted mass on Didymos/Dimorphos after 15 days between $\sim$2.8$\times 10^6$~kg and $\sim$4.4$\times 10^6$~kg. Re-accreted mass increases in time, potentially leading to non-negligible modification of the mass/inertia properties of the asteroids. This affects the dynamics of the binary system, as discussed in \sect{hardening} below.

\subsubsection{Effect of impact site curvature/geometry}

Both the ejecta cone edge's mass and geometry play a role in determining $\beta$, as well as the dynamical evolution of the ejecta in the system. LICIACube's LUKE and Hubble Space Telescope (HST) images captured a relatively wide cone \citep{Dotto-Nature-2024,Li-Nature-2023}. The HST image analysis derived the opening angle as $125\pm10^\circ$ under the assumption that the ejecta cone was axisymmetric \citep{Li-Nature-2023}. We expand this earlier work to apply both LICIACube LUKE and HST images to further constrain the ejecta cone edge geometry using the different view planes of these observers. The cone axis was found to point towards $(\textrm{RA}, \textrm{DEC})=(141.0\pm4.0^\circ, 20.2\pm8.0^\circ)$ in the J2000 equatorial frame, which is consistent with the DART spacecraft's incoming direction to within $20^\circ$ \citep{Hirabayashi2024}. The analysis determined the ejecta cone was elliptical in opening angle, \ie its intersection with a plane perpendicular to its axis formed an ellipse. The narrower cone opening angle in the direction of the ellipse's shorter axis was $94.8\pm5.4^\circ$ and the wider cone opening angle in the direction of the ellipse's longer axis was $133.3\pm9.2^\circ$. The latter direction is favored along Dimorphos's north-south direction but was rotated by $26.0\pm16.0^\circ$ counterclockwise from the north about the cone axis in the Dimorphos-fixed frame. This condition results in ejecta from high latitudes tending to depart from Dimorphos more shallowly relative to a plane normal to the along-track direction than from lower latitudes. This ejection angle variation indicates the contribution of Dimorphos's higher curvature along its body frame north-south direction compared to its east-west direction \citep{Hirabayashi2024}. Further assessments applying the derived ejecta geometry analysis and numerical simulations of the DART impact reveal that the curvature of the target led to a reduction of the along-track momentum transfer enhancement \citep{raducan2024physical, Hirabayashi2024}.

\subsubsection{Binary hardening effect} \label{s:hardening}

Most of the dynamical modeling of the Didymos system after the DART impact has focused on understanding the system's excitation in the full two-body problem; however, the complete Didymos system immediately and for some time after the impact consisted of many interacting bodies. Indeed, pre-impact numerical simulations predicted that a large amount of debris could be generated from the impact \citep[e.g.,][]{Rossi-PSJ-2022,Ferrari-PSJ-2022,Fahnestock-PSJ-2022,Tancredi2023}. Observations from LICIACube and nearly simultaneous space-based and ground-based observations demonstrated that large amounts of debris were generated at the time of impact \citep[e.g.,][]{Graykowski-Nature-2023,Kareta_2023,Li-Nature-2023}. Follow-up observations from both space- and ground-based observatories have also seen a persistent tail that has lasted for over 250~d \citep[e.g.][]{Li-Nature-2023,Moreno-PSJ-2023}. Finally, dynamical models of the post-impact 2-body system indicate a reduction in the orbital period of $34 \pm 15$~s \citep{Naidu2024} and $19 \pm 6$~s over the first $\sim$15~d \citep{Scheirich2024} (1$\sigma$ uncertainties). \edit1{These aspects motivate considering whether the binary was ``hardened,'' \ie the binary's orbital energy decreased due to gravitational interactions with the debris, so that the lost orbital energy was transferred to either escaping unbound debris or transferred via collisions into the rotation states of Didymos and/or Dimorphos.}

\edit2{The bound ejecta would have formed a thick, broad, evolving annulus about the Didymos-Dimorphos system \citep{Ferrari-PSJ-2022}. Collisions between debris particles would only be important if the debris cloud were a very thin and narrow annulus, \eg $\sim$100~m wide \citep{Stewart1984} assuming a very thin annulus given an ejecta mass on the order of 10$^7$~kg (see Table \ref{t:ejecta_mass}) and a typical particle of about 1~mm diameter with a density of 3~g~cm$^{-3}$ \citep{Moreno-PSJ-2023}. Instead, the bound ejecta disk must have been very broad, extending from the binary orbit ($\sim$1.2~km) out to the Hill sphere of the system \citep{Li-Nature-2023,Kareta_2023,Jewitt-ApJ-2023}. Furthermore, the disk will hardly be thin since the debris was ejected in a cone geometry \citep{Li-Nature-2023,Deshapriya2023,Dotto-Nature-2024}, placing debris on a range of inclinations relative to the Didymos-Dimorphos orbital plane. Thus, the post-impact debris cloud was very diffuse, particle-particle collisions were rare, and the orbits of debris particles evolved due to other processes.}

Due to interactions both with the binary gravitational dynamics and solar-radiation pressure, ejecta are unlikely to settle onto stable orbits \citep{Ferrari-PSJ-2022,Rossi-PSJ-2022}. Instead, the particles will have strong interactions at their orbit pericenters with Didymos and Dimorphos. Each encounter will scatter the particles onto new orbit planes and create a random walk in the semi-major axis, eccentricity, and inclination of the ejecta orbits \citep{Jacobson-Icarus-2011,Rossi-PSJ-2022}. Furthermore, solar radiation pressure will strongly perturb particle orbits and will quickly unbind particles less than about 1~mm in diameter \citep{Ferrari-PSJ-2022, Tancredi2023}. However, larger particles are likely to persist for a longer time that depends on their size, consistent with the observed long-lasting tail \citep{Moreno-PSJ-2023}.

The population of larger particles in the Didymos system will decay with time due to loss onto heliocentric orbits and collision with Didymos and Dimorphos, as shown in \figr{plot_mass}. While particle-particle collisions are unlikely to play a significant role in the evolution of their orbits, a significant number fraction of the particles will likely collide with Didymos or Dimorphos (see \figr{plot_mass}). In other words, each particle's residence time in the Didymos system depends on its properties (radius, density, etc.) and ejection location and velocity, but not on the other particles. This means the particle loss rate is proportional to the total number of particles $N$ remaining in the system, \ie $\frac{dN}{dt} \propto - N$ which has the well-known solution of exponential decay, $N = N_0 e^{-(t-t_0)/\tau}$, where $\tau$ is the exponential decay timescale, $N_0$ is the initial number of particles, and $t_0$ is the time of the DART impact. Using this exponential decay model, \edit1{the timescale for clearing the Didymos system is proportional to the timescale associated with the scattering events that occur near the pericenter of each particle's orbit. Thus, $\tau$ is proportional to a characteristic ejecta orbital period $P_e$: $\tau \propto P_e$.}

\edit1{Ejecta will have a range of Didymos-Dimorphos system orbital periods $P_e$ depending on their ejection velocity after the DART spacecraft impact. All ejecta not lost immediately to heliocentric orbit have ejection speeds less than the escape speed from the Didymos system \citep[\ie $\lesssim24$~cm~s$^{-1}$;][]{Tancredi2023}. For ejecta launched with very low speeds relative to Dimorphos, the ejecta period is similar to the mutual orbital period of Didymos-Dimorphos $P_e \approx 0.5$~d. Other bound ejecta will have been launched on orbits with apocenters $Q_e$ very close to the Didymos-Dimorphos system Hill radius $r_H$ at the time of the DART impact:
\begin{equation}
    r_H = r_D \left( \frac{m}{3M_\odot} \right)^{1/3} = 73\ \text{km} ,
\end{equation}
where $r_D = 1.1$~au is the heliocentric distance of the Didymos system at the time of the DART impact \citep{Rivkin2021}, $m = 5.3 \times 10^{11}$~kg is the Didymos-Dimorphos system mass (\tbl{params}), and $M_\odot = 2.0 \times 10^{30}$~kg is the solar mass. Debris with apocenters closest to the Hill radius are very easily perturbed by solar-radiation pressure onto heliocentric orbits, contributing to the long-lasting tail, and so they do not interact much with the Didymos-Dimorphos system. However, stably bound debris \citep[$Q_e \lesssim 0.5 r_H$, \ie with apocenters well within the Hill radius, motivated in part by studies of the stability of irregular satellites;][]{Nesvorny2003} will have orbital periods of about $P_e \lesssim 15$~d assuming pericenters $q_e$ near the Didymos-Dimorphos separation distance $q_e \approx a = 1.2$~km. These estimates set a clearing timescale $\tau$ on the order of days to weeks, which is consistent with both the more detailed numerical experiments shown in \figr{plot_mass} and is consistent with the observed mutual orbit period change timescale \citep{Scheirich2024}.}

The effect of this exponentially decreasing population of debris on the dynamics of the Didymos-Dimorphos system is the dynamical hardening of the binary due to exchange of angular momentum of the binary system with the escaping ejecta, potential small impulses upon the binary components from accretion impact events and to a much-lesser extent an increase of the system mass due to that same accretion. \edit1{To test whether these processes may be responsible for the observed change in the Didymos-Dimorphos orbit period, we calculate an independent estimate of the amount of ejected mass needed to effect that change and compare it to observed estimates of the ejected mass.}

First, using a Keplerian model for the Didymos-Dimorphos system, we determine how much the orbital angular momentum of the binary changed given the observed period change:
\begin{equation}
    \Delta L = \frac{L}{3 P} \Delta P , \label{e:1}
\end{equation}
where $L = m_s \sqrt{G m a}$ is the initial angular momentum of the Didymos-Dimorphos circular orbit, $m = m_p + m_s$ is the combined mass of Didymos $m_p$ and Dimorphos $m_s$, and we assume the reduced mass can be approximated by the mass of Dimorphos. \edit1{The change in angular momentum of the binary $\Delta L$ is the opposite of the change in the angular momentum of the particle orbits $\Delta L_e$, which is the difference in angular momentum of the ejected particles on their final parabolic orbits and their elliptic orbits immediately after ejection. Thus, the change in the angular momentum of the binary is:
\begin{equation}
    \Delta L = - \Delta L_e = - m_e \left(\sqrt{2 G m a_e \left( 1 - e_e \right)} - \sqrt{G m a_e \left(1 - e_e^2 \right)} \right) , \label{e:2}
\end{equation}
where $m_e$ is the mass of ejected material that is not lost immediately to heliocentric orbit. To good approximation, the Didymos-Dimorphos orbit is circular and since most of the ejecta was launched off the leading face of Dimorphos, the ejecta would have an apocenter exterior to the Didymos-Dimorphos orbit and a pericenter similar to the semi-major axis of Dimorphos's orbit: $a_e (1-e_e) = a $. \eqntwo{1}{2} can then be solved for the initially bound but ejected mass,
\begin{equation}
m_e =  \frac{m_s}{ \sqrt{1 + e_e} - \sqrt{2}} \left(\frac{\Delta P}{3 P} \right) = \frac{ \left( 9.5 \pm 5.9 \right) \times 10^{5} \text{ kg} }{\sqrt{2} - \sqrt{1 + e_e}} ,
\end{equation}
where the mass of Dimorphos is $m_s = ( 4.5 \pm 0.4 ) \times10^9$~kg and the post-impact mutual period is $P = 11.3674 \pm 0.0010$ (see \tbl{params}). The two estimates of the post-impact period change are $\Delta P = -34\pm15$~s \citep{Naidu2024} and $\Delta P = -19\pm6$~s \citep{Scheirich2024}. \citet{Scheirich2024} cautions that their estimate is only a lower limit on the absolute magnitude of the change and that the real change may be slightly higher. If we take these estimates at face-value and average them, we estimate the period change as $\Delta P = -26 \pm 16$~s. A minimum ejected mass estimate $m_e \gtrsim \left( 2.3 \pm 1.4 \right) \times 10^6$~kg is obtained if the initial ejecta eccentricities are low ($e_e \approx 0$). If the ejecta that interact strongly with the Didymos-Dimorphos system (\ie have many pericenter passages) are initially spread out evenly on orbits with apocenters between the Didymos-Dimorphos orbit ($Q_e = a_e$ and so $e_e \approx 0$) and half the distance to the Hill radius ($Q_e = 0.5 r_H$ and so $e_e \approx 0.9$), then the estimated ejecta mass is $m_e = \left( 4.5 \pm 3.7 \right) \times 10^6$~kg.} This estimate of the unbound ejecta mass estimate is a similar order of magnitude as other estimates of the ejected mass, although most of those are for the unbound component (see \tbl{ejecta_mass}). Notably, it is very similar to the comparable the fallback mass of $5.4 \times 10^6$ kg, estimated from analysis of Hubble Space Telescope images \citep{Kim2023}. 

Next, we consider the change in angular momentum of the binary system due to re-impacting mass. \edit1{This is very similar to the original DART experiment except that the debris is not likely striking Dimorphos only in the along-track direction. Irrespective of the momentum transferred from collding ejecta to the binary components, in a Keplerian model, growth in the binary mass changes the binary orbital period according to $\Delta P = - P \Delta m / (2 m)$. Using the same values as above,} the change in the system mass would need to be about 0.13\% or $\Delta m = 6.6\times10^8$~kg to explain the observed period change. This is much more ejecta than other independent estimates suggest. Thus, mass re-accretion will play only a small role in any detected period change.

However, re-accretion will also deliver angular momentum to the orbit due to small torques associated with each impact event. \edit1{If these torques are truly random, then they will cancel out, but if not, then this re-impacting ejecta material may change the orbital period. In this case,} the change in the angular momentum due to re-impacting material can be estimated as
\begin{equation}
    \Delta L = m_r \sqrt{G m a_r (1-e_r^2)} , \label{e:3}
\end{equation}
where $a_r$ and $e_r$ are the average semi-major axis and eccentricity of ejecta material prior to re-impact and $m_r$ is the re-impacting mass that contributes to the final change in angular momentum of the binary \edit1{(note that this re-impacting mass is a lower limit since some mass might have re-impacted that did not change or even increased the angular momentum of the binary, canceling out yet other mass that did decrease the binary angular momentum).} \eqntwo{1}{3} can then be solved for an estimate of minimum re-accreted mass,\edit1{ then
\begin{equation}
    m_r \geq m_s \sqrt{1 + e_r} \left(\frac{\Delta P}{3 P} \right) = \sqrt{1 + e_r} \left( 9.5 \pm 5.9 \right) \times 10^{5} \text{ kg} ,
\end{equation}
where we have again asserted that the pericenter of the re-impacted debris is similar to the semi-major axis of the Didymos-Dimorphos system. If the debris is mostly circular ($e_r \approx 0$), then the minimum re-impacting mass is $m_r \geq \left( 9.5 \pm 5.9 \right) \times 10^{5}$ kg. Since much of the debris is likely placed on eccentric orbits and some of the debris is likely to collide with Didymos or Dimorphos to increase the binary orbital angular momentum, we emphasize that the required mass of re-accreted debris is likely much higher than this estimate. Many independent estimates of the ejected debris are on order $10^6$ kg, so it may be reasonable that a similar mass was lofted but ultimately re-accreted by Didymos and Dimorphos.  This calculation establishes that such a large mass of debris would be necessary to fully explain the observed period change of the binary, but whether re-accreted mass would ultimately add or remove angular momentum requires future detailed simulations. Thus, re-accretion and ejection likely play a role in the observed post-impact binary hardening of the system.}

From these simple considerations, we find that the cloud of non-escaping ejecta should dissipate on an exponential timescale and gravitational interactions with that cloud would lead to angular momentum transfer that would ultimately harden the orbit of the Didymos-Dimorphos system due to slowly escaping material. These considerations are consistent with both observations of a mutual orbital period change of the system as well as independent estimates of the amount of debris created and its evolution. Future work should examine the many-body dynamics of the ejecta cloud and binary system more fully. The angular momentum transfer between the bound ejecta cloud and the binary system appears measurable, and so this is an independent constraint on the amount of slow ejecta created by the DART impact.

The previous analysis has implications for $\beta$. The negative period change induced by the DART impact appears to be slightly increasing in magnitude over time (and therefore the measure of $\beta$ is also slightly increasing). That is, the ``immediate'' $\beta$ is slightly lower than the ``eventual'' $\beta$ for this system. This is likely to be true of any binary system when impacting the secondary more or less head-on. Although this effect is measurable in this instance, it has a very minimal effect on the final period change and eventual $\beta$. Consider the persistent particles in the system. Among the particles ejected by the DART impact, many were launched from the surface of Dimorphos at speeds less than the escape speed of the system. This resulted in particles orbiting in the system for some time after the impact. These persistent particles can reach a number of outcomes: they can re-impact Dimorphos, transfer impact onto Didymos, escape the system after close encounters with Dimorphos, or find a stable orbit within the system. Finding a stable orbit in the system is very unlikely for most of the particles but we include this possible outcome for completeness. Particles that have close encounters with Dimorphos are efficient at increasing the period reduction for Dimorphos and can explain the continuous period change. Such change appears to be a product of Dimorphos being in a binary system and it is possible that a single asteroid would not experience a similar phenomenon. This is one instance where the DART impact diverges from the case wherein a single asteroid is impacted. An impact into a single body would still result in a fraction of persistent particles but this would likely be a much smaller fraction than what we see at Dimorphos. The particles ejected from a single body below its escape speed would (mostly) be on re-impact trajectories with the body. Some may orbit the single body but they could not escape the system via close encounters with a secondary body. This is a fundamental difference between two-body systems and three-body systems, where three-body systems often eventually eject the smallest body and two-body systems have no ability to do so without external perturbation.


\section{Shape and Structure Effects} \label{s:shape}

Detailed images from DRACO, DART's camera, documented unique geologic features of both Didymos and Dimorphos \citep{Daly2023}. This binary system, or at least its satellite, is likely made of gravitational aggregates rather than monolithic bodies, given the observed unique surface morphology and shapes \citep{Daly2023}. The Didymos binary system is a member of the most common class of binary systems, in which a satellite is much smaller than its primary (a few \% of the system's mass) \citep{Pravec2006}. Binary systems should form and evolve due to mass transport between gravitationally bound bodies driven by events changing their dynamical and structural configurations such as impacts, tidal effects, and solar radiation. Didymos looks like a top shape but not exactly so; compared to Ryugu \citep{Watanabe2019} and Bennu \citep{Barnouin2019}, the very oblate shape likely resulted from various mechanisms related to its fast spin. The oblate shape of Dimorphos (at least prior to impact) is at odds with the expectation of a prolate shape \citep{Walsh2015} and is an important constraint on formation models. Structural modifications during the system's evolution also change the system's dynamical behavior. This section describes the current state-of-the-art understanding of the binary system's structural stability and how it contributes to dynamical evolution.

\subsection{Structural properties and stability of Didymos}

\subsubsection{Pre-encounter understanding constrained by shape and spin state}

Prior to the DART encounter, the binary system's physical characteristics were primarily inferred from ground-based observations. Photometric and radar observations estimated Didymos's spin period to be $2.2600\pm0.0001$~h, and the mutual orbit elements yielded bulk density estimates of $\rho_p=2170\pm350$~kg~m$^{-3}$ and $2370\pm300$~kg~m$^{-3}$ for Didymos (see Table \ref{t:params}). The radar-derived shape model gave Didymos's dimensions along the principal axes as $(832 \pm 6\%) \times (838 \pm 6\%) \times (786 \pm 10\%)$~m \citep{Naidu2020}. 
 
The pre-encounter understanding implied Didymos possesses mechanical strength, ensuring both interior and surface structural stability \citep{Hirabayashi2022}. Numerical simulations using discrete-element modeling provide two avenues to maintain Didymos's structural integrity at its rapid spin rate, \ie through either global weak cohesion or strong mechanical core strength \citep[e.g.,][]{zhang2021creep, Ferrari2022}. Assuming a homogeneous rubble-pile structure with various potential boulder size distributions, Didymos's minimum required bulk cohesive strength is estimated to be $\sim$11--17~Pa for $\rho_p=2170$~kg~m$^{-3}$ and $\sim$2--13~Pa for $\rho_p=2370$~kg~m$^{-3}$ \citep{zhang2021creep}. Alternatively, if Didymos's strength distribution is heterogeneous, a cohesionless external layer can suffice for structural stability, provided that over half of the interior possesses strong mechanical strength or rigidity \citep{Ferrari2022}.  

\subsubsection{Post-encounter understanding constrained by shape and spin state}
\label{s:postencounter}

By analyzing the close-encounter images obtained from DRACO, Didymos's dimensions were updated to $819 \pm 14 \times 801 \pm 14 \times 607 \pm 14$~m \citep{Barnouin2023}. This shape is remarkably smaller and more oblate than the radar shape model, yielding a significantly higher bulk density of $2790\pm140$~kg~m$^{-3}$ (see Table \ref{t:params}). Material cohesive strength is no longer a crucial requirement for maintaining Didymos's interior structural stability at the spin period of 2.26~h, provided its material internal angle of friction \edit1{(a measure of friction between particles that opposes shear and that is typically similar in value to the macroscopic ``angle of repose'')} is larger than $\sim$40$^\circ$ \citep{ZHANG201798, Agrusa2024}. For a lower friction angle of $\sim$35$^\circ$, interior cohesion of $\sim$10 Pa is still needed to prevent internal deformation \citep{Agrusa2024, Barnouin2023}. Some surface regions on Didymos have slopes exceeding $45^\circ$, indicating that a small cohesive strength of $\sim$1--2~Pa might be present on Didymos for maintaining the stability of these high-slope regions \citep{Barnouin2023}. The surface cohesive strength might need to exceed the static estimate in order to maintain the structural stability of the surface during the impact of ejecta resulting from the DART impact (see \sect{Didy_surface_ejecta}).

If the interior does not possess a proper level of mechanical strength, internal failure of a rubble-pile body may occur during a spin-up process driven by YORP torque \citep{Hirabayashi2015, Sanchez2018}. This could explain the pronounced oblateness observed in Didymos's current shape. Conversely, high internal strength coupled with low surface strength might induce surface mass transport, which is hinted at in the DRACO images \citep{Barnouin2023}. The possibility of shedding these mobile surface materials, followed by their gravitational accumulation in orbit around Didymos, offers a plausible explanation for the origin of Dimorphos (see \sect{Dimorphos_origin}). It is likely that Didymos's structural properties lie between these two extremes, allowing it to exhibit both interior- and surface-failure behaviors at fast spin \citep{Barnouin2023}. The internal failure could give rise to a density distribution heterogeneity with a less-dense core, akin to the case of Bennu \citep{Zhang2022inferring}. The upcoming gravity-field measurements and radio experiments to be conducted by the Hera mission (\sect{hera}) will provide the means to confirm the existence of such heterogeneity and quantify the interior structure's influence on the dynamics of the binary system.

\subsubsection{Reshaping of Didymos}\label{s:Didy_reshaping}

Due to Didymos's fast spin, it is possible that \edit1{the body is sensitive to structural failure at present \citep[e.g.,][]{Hirabayashi2015, nakano2022nasa}.} Small perturbations, such as ejecta from the impact site on Dimorphos striking Didymos at various speeds, could thus trigger a reshaping process, \edit1{wherein its equatorial radius increases while its polar radius decreases}, resulting in a more oblate shape \citep{Hirabayashi2022, nakano2022nasa}. Such reshaping would perturb the mutual gravitational fields and therefore the mutual dynamics between the bodies. Studies prior to the DART impact demonstrated that Didymos's reshaping under constant volume (\ie no mass change due to ejecta accretion) always reduces the orbit period of Dimorphos owing to an increased $J_2$ moment \citep{nakano2022nasa,Richardson2022}. It was found that \edit1{0.7~m of change in the polar radius} leads to an orbit period change of 7.8~s, which exceeds the DART Level 1 measurement requirement \citep[\ie 7.3~s;][]{Rivkin2021}. This implies that the $\beta$ estimation process likely needs to account for the effect of Didymos's reshaping when the magnitude of reshaping exceeds 0.7~m, as otherwise, the $\beta$ value could be overestimated \citep{nakano2022nasa, Richardson2022}.

Importantly, if Didymos's reshaping were to occur, its spin period would change as its moment of inertia is modified. This potentially provides a means to constrain the magnitude of reshaping, as the spin period change can be precisely measured. However, current measurements show that the spin period has been constant since pre-impact, with an uncertainty of 1~s at a 3$\sigma$ level (J. \v{D}urech \& P. Pravec 2024, in preparation). This suggests that Didymos's reshaping likely did not occur, possibly due to either the impact-related perturbations not being strong enough to trigger reshaping or the asteroid's structure being relatively resistant to reshaping, allowing it to withstand the perturbations. Therefore, it is reasonable to conclude that the $\beta$ estimation process currently does not need to account for the effect of Didymos's reshaping.

\subsubsection{Surface strength constrained by ejecta impacting on the surface}
\label{s:Didy_surface_ejecta}

\begin{figure}
    \centering
    \includegraphics[scale=0.3]{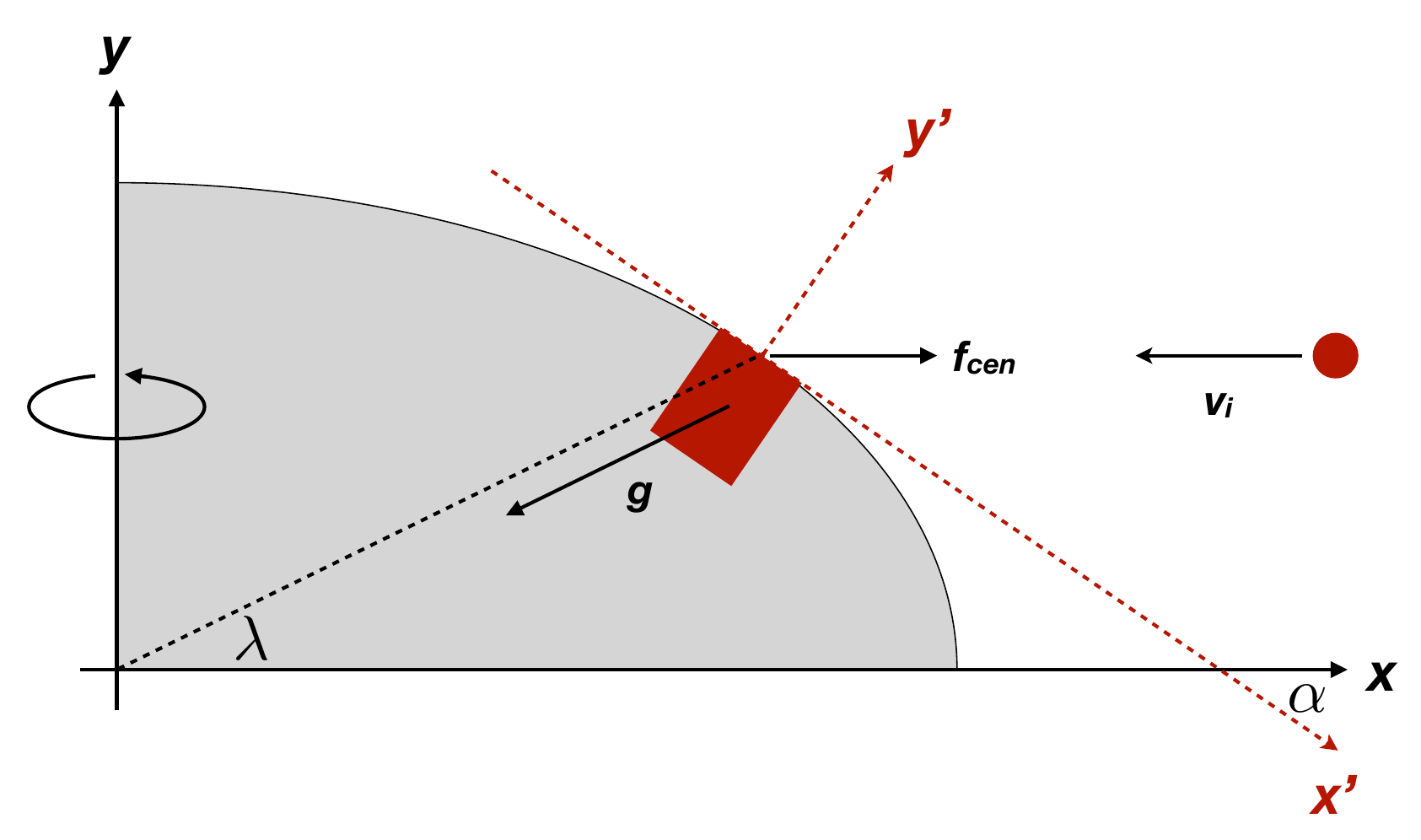}
    \caption{Scheme for positioning boxes on the surface of an ellipsoidal body. The x'--y' coordinates represent the rotating frame of reference of the simulation box. The angles $\lambda$ and $\alpha$ represent the latitude and the surface slope respectively in the inertial frame. The vectors $\vec{g}$, $\vec{v}_i$ and $\vec{f}_{\mathrm{cen}}$ are gravity, the projectile's impact velocity, and centrifugal force at the surface respectively.}
    \label{f:scheme}
\end{figure}

As noted earlier, the DART impact generated a lot of debris, so much so that Dimorphos acquired a long-lived tail that transformed it into an active asteroid. Part of this debris very likely landed on Didymos, but no appreciable change in either its shape or dynamics was observed (\sect{Didy_reshaping}). This implies Didymos's surface was strong enough to withstand such impacts. Here we simulate this process using an oblate spheroid of homogeneous density and strength rotating around its shortest axis to represent Didymos (\figr{scheme}), updating the work of \cite{Hirabayashi2022} who assumed a spherical shape. We  simulate impacts onto surface elements of Didymos to infer a lower limit for its cohesive strength. Impact simulations were carried out using GDC-i \edit1{(Granular Dynamics Code - Impacts)}, a soft-sphere \edit1{Discrete Element Method (DEM)} code for simulating impacts at various velocities \citep{Sanchez2018,Ballouz2021}. \edit1{Within the method, each particle is treated as an individual object that follows the laws of classical mechanics and moves under the influence of the forces imposed on it. Particles interact with each other through repulsive soft potentials and a collision is said to have occurred when the distance between the centre of any two particles is less than the sum of their radii. The normal repulsion force is provided by a spring (linear or Hertzian) and dashpot model. Friction forces are implemented through tangential springs whose response is truncated to satisfy to local Coulomb criterion. Rolling and twisting friction can also be implemented to mimic the behavior of non-spherical particles (for details see \cite{Ai-Chen2011,Sanchez2011,Sanchez2016,Schwartz2012}).}

Following \cite{Hirabayashi2022}, we place 1$\times$2$\times$1~m boxes, with periodic boundary conditions, on the surface of the oblate spheroid and fill them with 60,000 spherical grains with a uniform distribution of diameters between 2--3~cm to a height of $\approx$84~cm.  \edit1{We assume Didymos has a bulk density of 2950 kg/m$^3$; particle density is set to 4916.67 kg/m$^3$ so that the bulk density of the box is the same as that of Didymos.} Note that, as investigated by \cite{Sunday2022}, boundary effects could affect the outcomes of the simulated impact scenarios. In our case, the periodic boundary conditions eliminate the effects of the walls and emulate an infinite granular surface on which particles of the ejecta field will impact \edit1{\citep{Radjai2018}}.
\edit2{If we tried to eliminate boundary effects by increasing the size of the system so that the sound wave does not reach the border, we would need a granular bed of 4x4x2m (95\% of energy is dissipated every $\approx$2~m: \cite{Sanchez2022}); this would increment the simulation time by a factor of 32. If we add a reduction in the particle size to just half of what we have, that time would increase 4 fold.  Together they would increase the computation time by a factor of 128 ($\approx$1~yr per simulation) making them impractical with the current computational facilities and code.}
Furthermore, since it is the impact of many particles of an ejecta curtain that we seek to simulate, ours is an appropriate representation of a realistically much larger system as it would be unlikely that only one impact takes place in isolation of all others. 

The angle of friction of this granular bed is set to $\sim$35$^\circ$, a common value for geological regolith and also what was estimated for asteroid Dimorphos \citep{robin2023}. These boxes are placed in the northern hemisphere of the spheroid, along the longest meridian and at latitudes of 0$^\circ$, 15$^\circ$, 30$^\circ$, 45$^\circ$, 60$^\circ$, and 75$^\circ$. Cohesive strength was tested at 0, 1, 2, 3, 4, 5, 10, 15, and 20~Pa. Impact speed was kept at 1~m~s$^{-1}$ in the inertial frame \citep{Yu2018}, but it changes in the body frame depending on the latitude of impact due to the rotation of the asteroid, which provided a horizontal velocity component (up to 0.32~m/s). All impacts were set to be perpendicular to the axis of rotation in the inertial frame of reference (see \figr{scheme}). Each granular bed was impacted 5 times by identical projectiles at the same speed at 2~s intervals. The first projectile is always placed just above the surface of the granular bed and then shot. After 2~s, the same projectile is repositioned outside of the box and re-shot. The process is repeated 4 times to complete 5 impacts. In most cases, this 2~s time interval was enough for the projectiles to either stop due to the collision or simply pass through the granular material if the previous impact had already fluidised it.  \edit1{A granular material is said to be fluidised when the kinetic energy of the system is comparable to its potential energy.} The projectiles were aimed to a point in the centre of the box, 10 cm below the surface. Note that though the simulation has periodic boundary conditions, the projectile is impervious to them and simply sees the granular bed and not the containing box.  \edit2{We do this because of two reasons, one computational and one physical.  Computationally, given that 4 out of the 5 projectiles start their motion from outside the box, this would have brought complications for the algorithms. Physically, if the granular bed is already so fluidised that a projectile goes through the boundary after the impact, it will not have made a difference in our assessment of whether or not the bed was already fluidised.  If the granular bed is not fluidised and the projectile merely bounced on the surface, most of its energy would have been already dissipated in the first impact; a second impact after crossing the boundary would not have done anything significant.} 

\edit1{As the simulation boxes themselves change their orientation with respect to the body frame depending on their latitude, we need to calculate some angles and distances so that gravitational, Coriolis, and centrifugal forces are adequately calculated (see \figr{scheme}).} So, if $a$, $b$, and $c$ are the dimensions of the semi-axes of the spheroid, $a>b>c$. The distance $r$ between the body center and surface at a given latitude $\lambda$ is
\begin{equation}
    r(\lambda)=\frac{ac}{\sqrt{a^2 \sin^2{\lambda}+c^2 \cos^2{\lambda}}} ,
\end{equation}
which means that the radius of rotation of any box \edit1{(the distance between the box and the axis of rotation of the body)} is $r\cos{\lambda}$. Differentiating the ellipse equation (see \figr{scheme}), the slope of the curve (\ie the meridian of the ellipsoid) is
\begin{equation}
    \alpha=\arctan{\frac{-cr(\lambda)\cos{\lambda}}{a\sqrt{a^2-r^2(\lambda)\cos^2{\lambda}}}} ,
\end{equation}
which is a negative number. The angle of the gravity vector in the rotating frame of reference is $\pi+\lambda-\alpha$ and the angle of the centrifugal force is $-\alpha$, where $\alpha$ is the surface slope in the inertial frame. \edit1{Finally, the angle of the angular velocity vector, for the calculation of the Coriolis force, is $\pi-\alpha$.  These angles are used in the simulation for the calculation of all these forces on each grain.}

\begin{figure}
    \centering
    \includegraphics[scale=0.16,trim={4.5cm 2.5cm 4.5cm 2.5cm},clip]{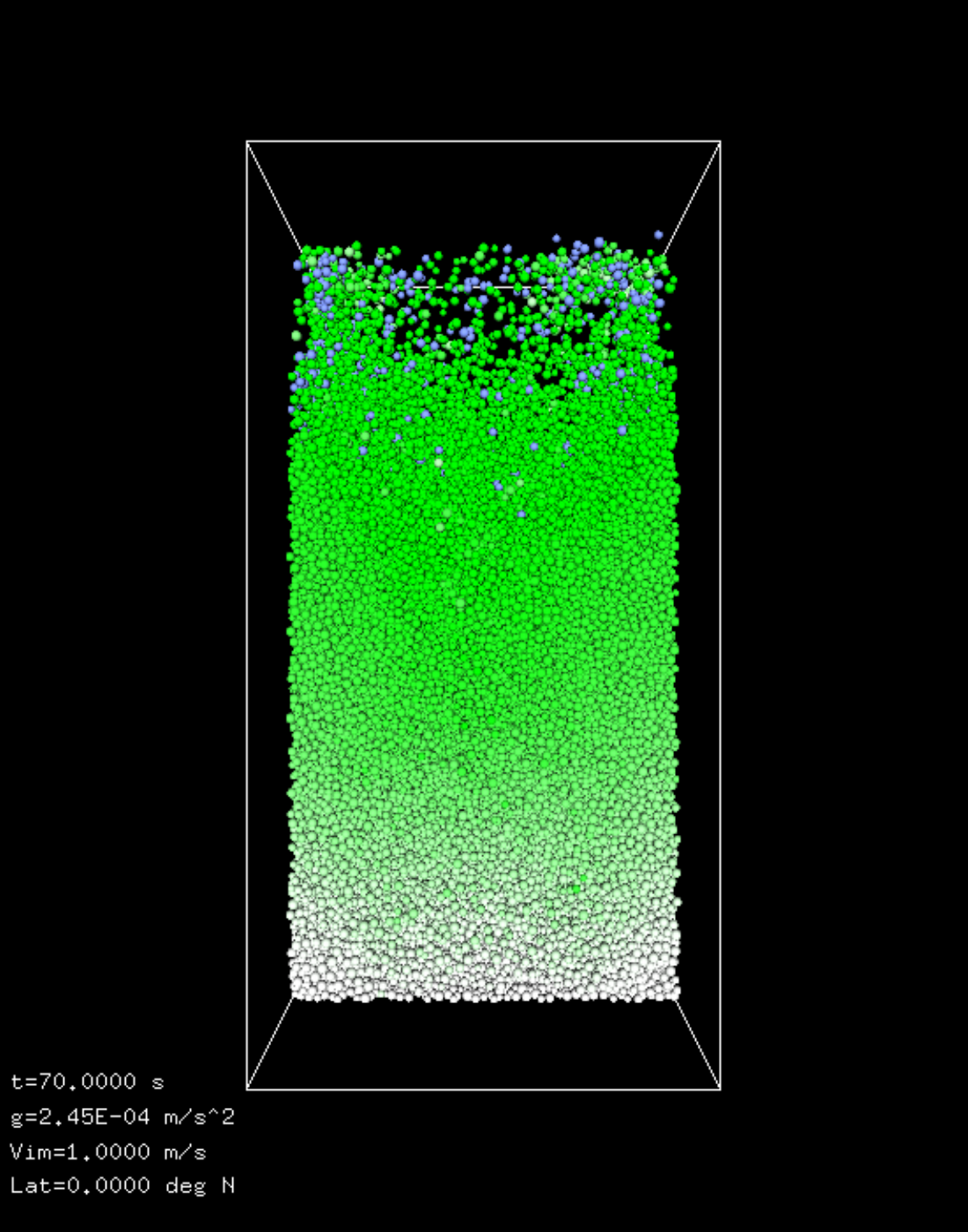}
    \includegraphics[scale=0.16,trim={4.5cm 2.5cm 4.5cm 2.5cm},clip]{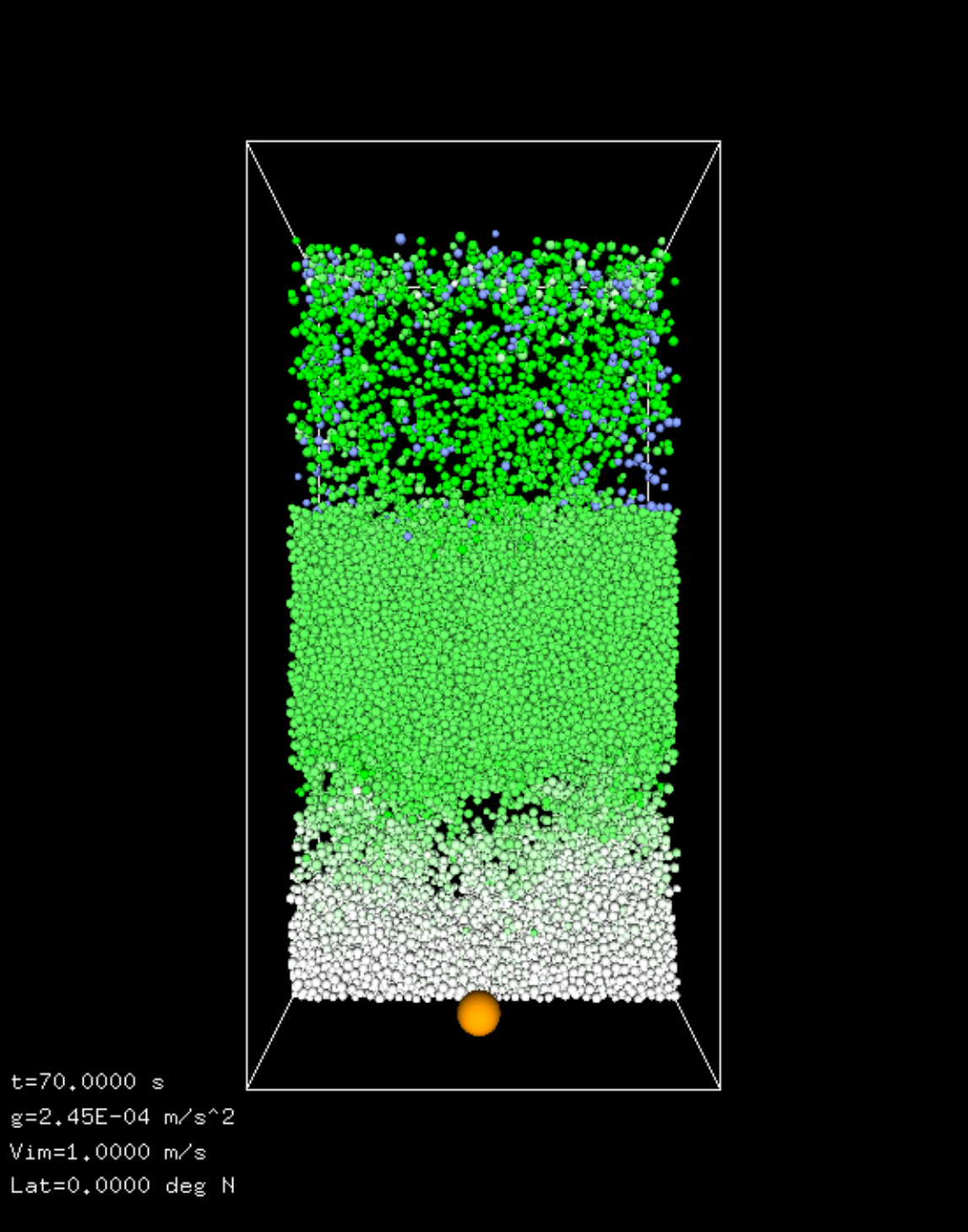}
    \includegraphics[scale=0.16,trim={4.5cm 2.5cm 4.5cm 2.5cm},clip]{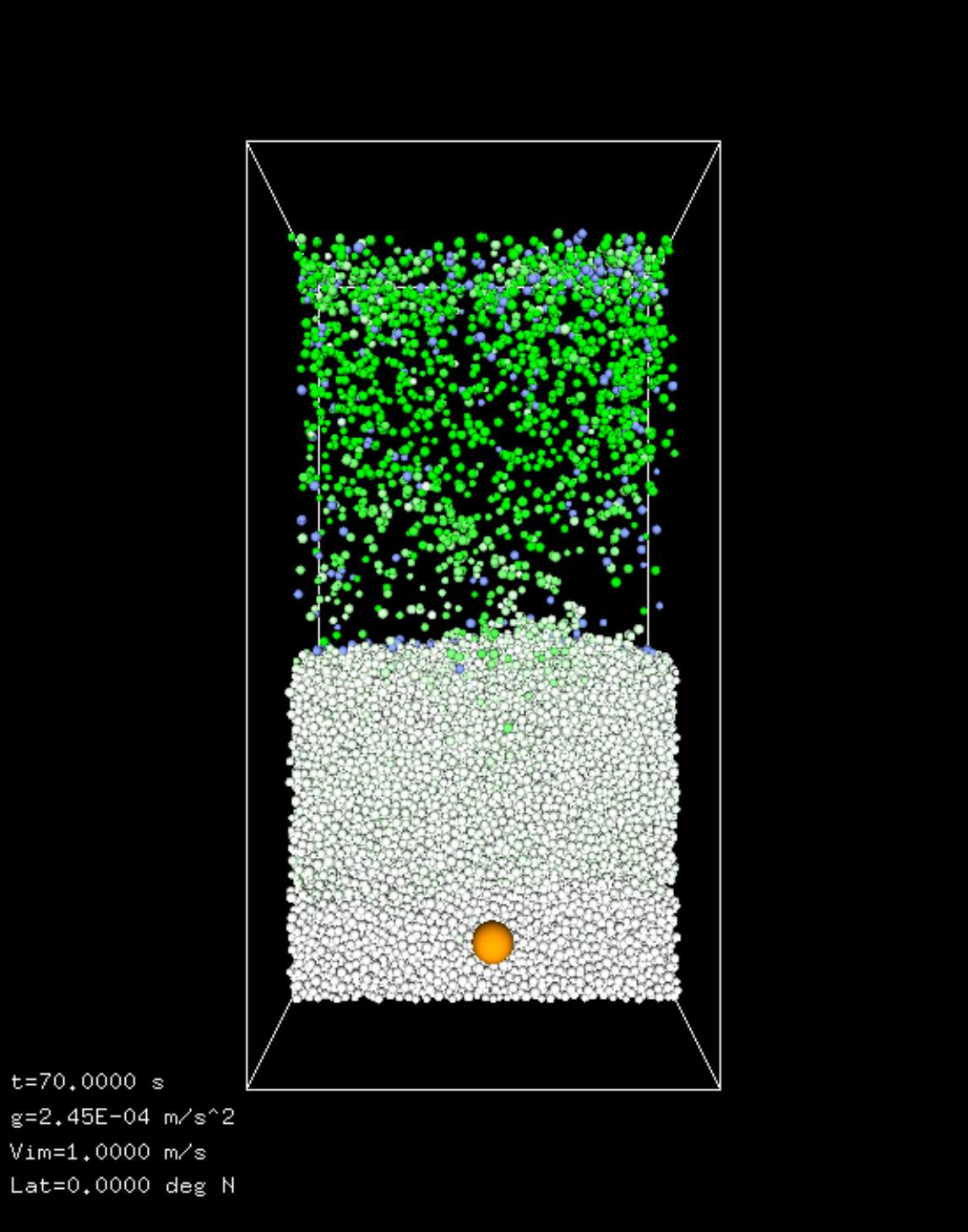}
    \includegraphics[scale=0.16,trim={4.5cm 2.5cm 4.5cm 2.5cm},clip]{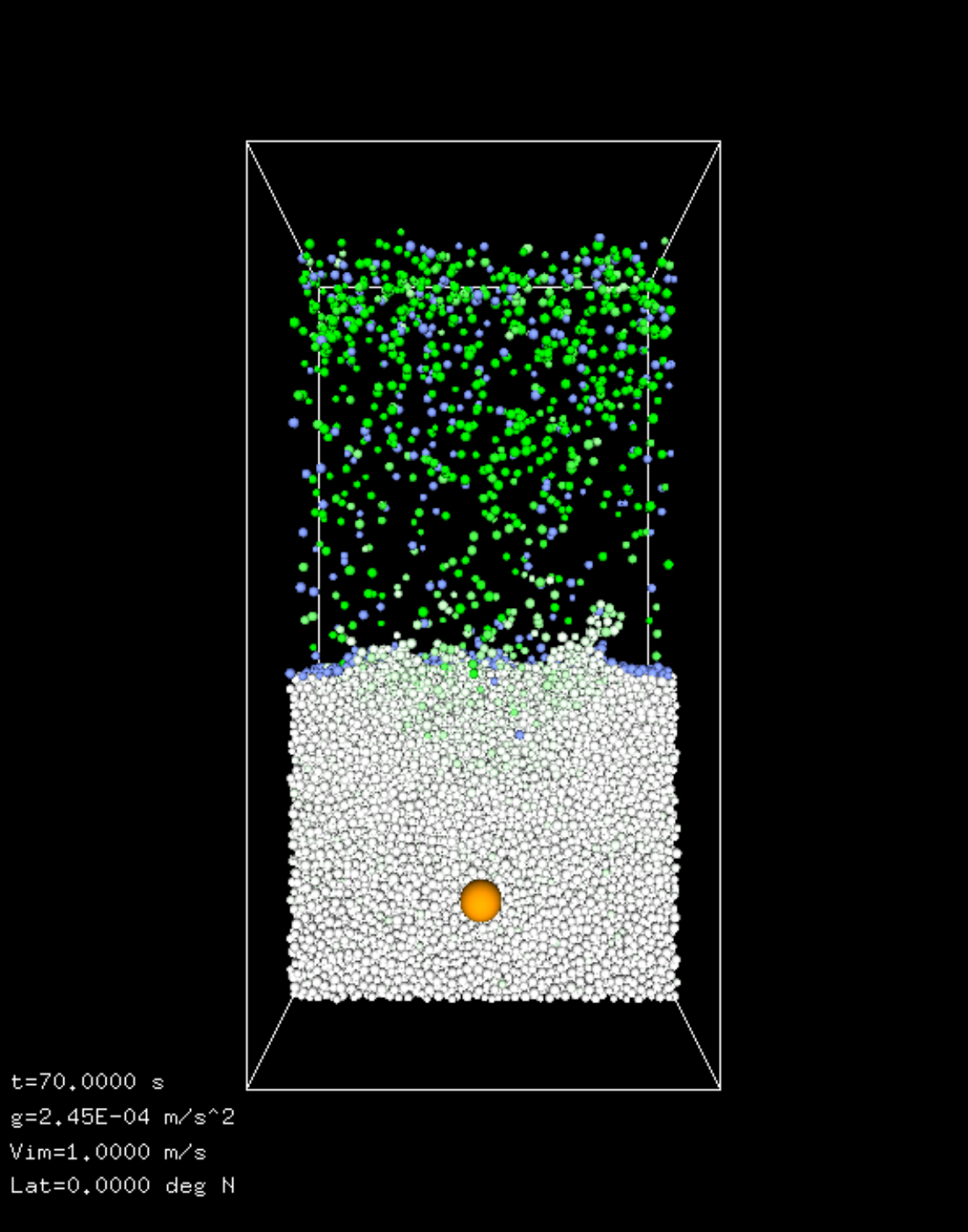}
    \includegraphics[scale=0.16,trim={4.5cm 2.5cm 4.5cm 2.5cm},clip]{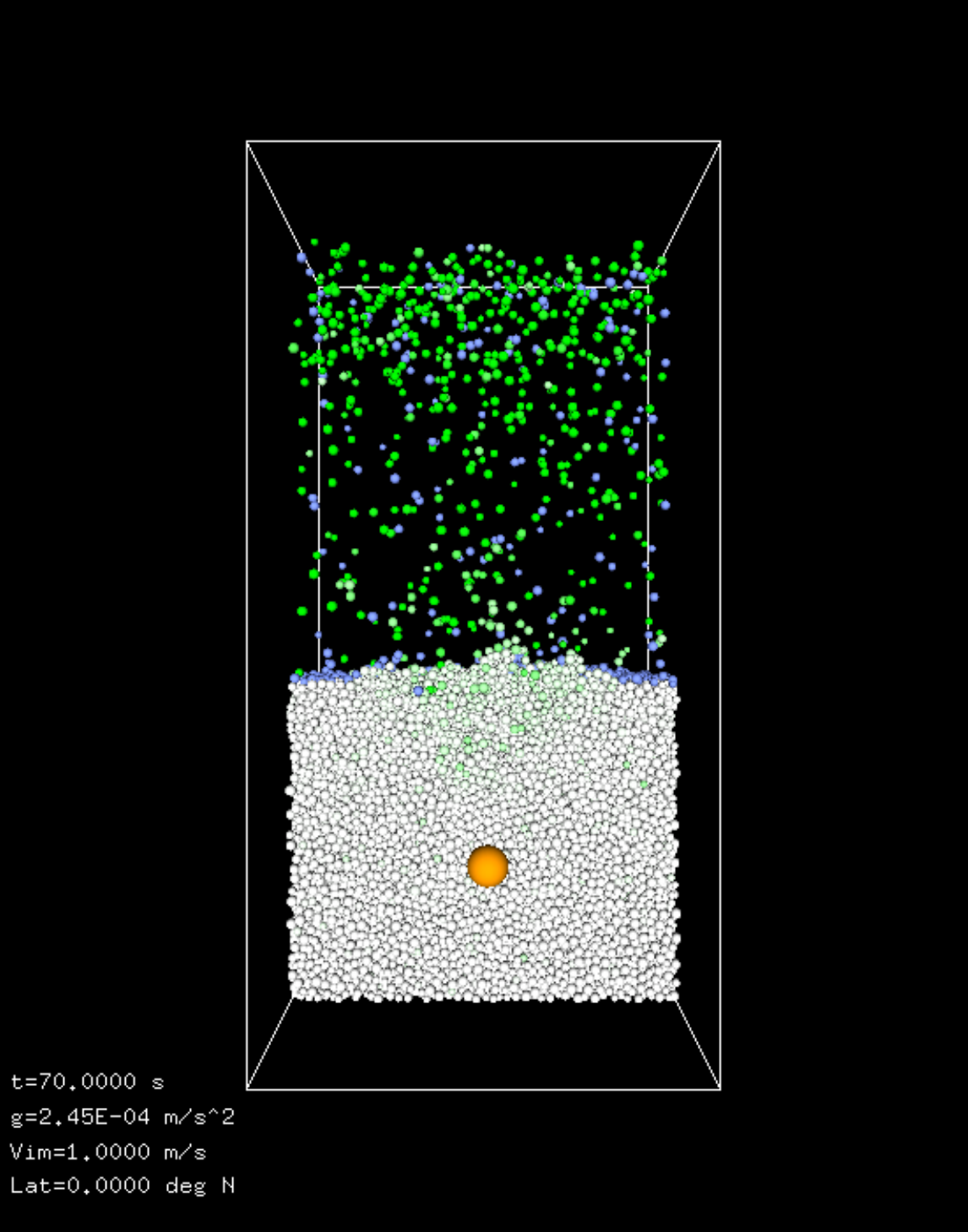}
    \includegraphics[scale=0.16,trim={4.5cm 2.5cm 4.5cm 2.5cm},clip]{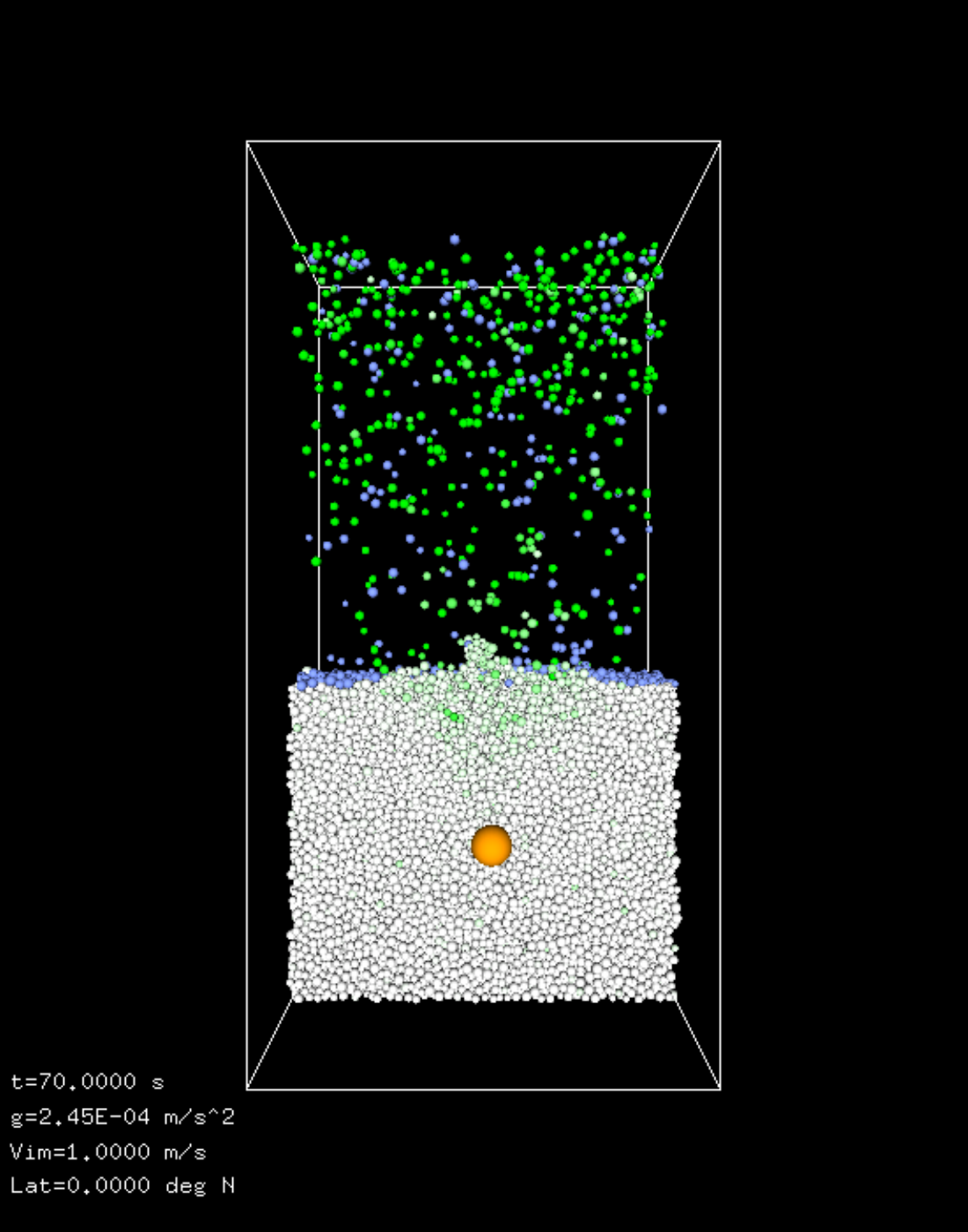}
    \includegraphics[scale=0.16,trim={4.5cm 2.5cm 4.5cm 2.5cm},clip]{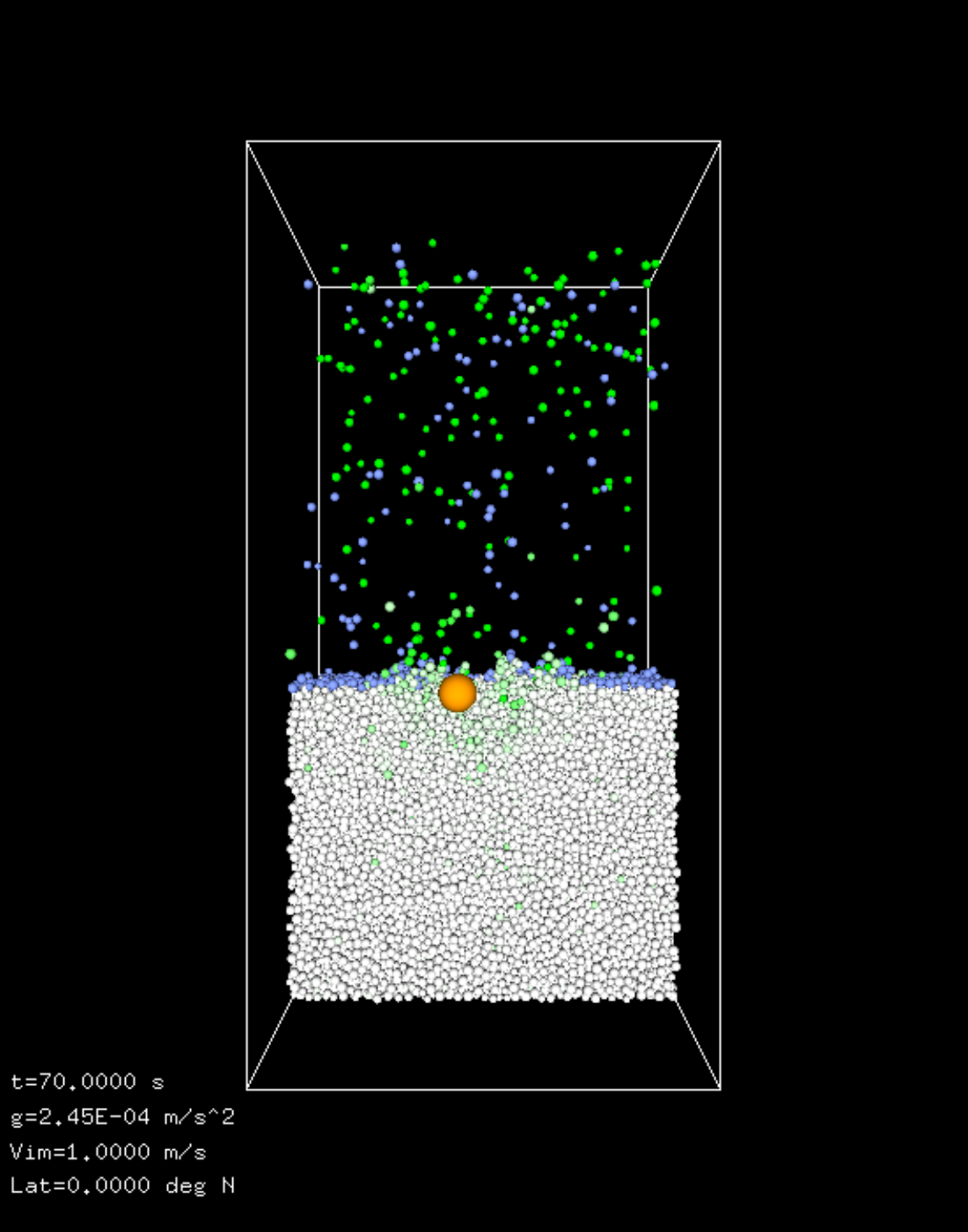}
    \includegraphics[scale=0.16,trim={4.5cm 2.5cm 4.5cm 2.5cm},clip]{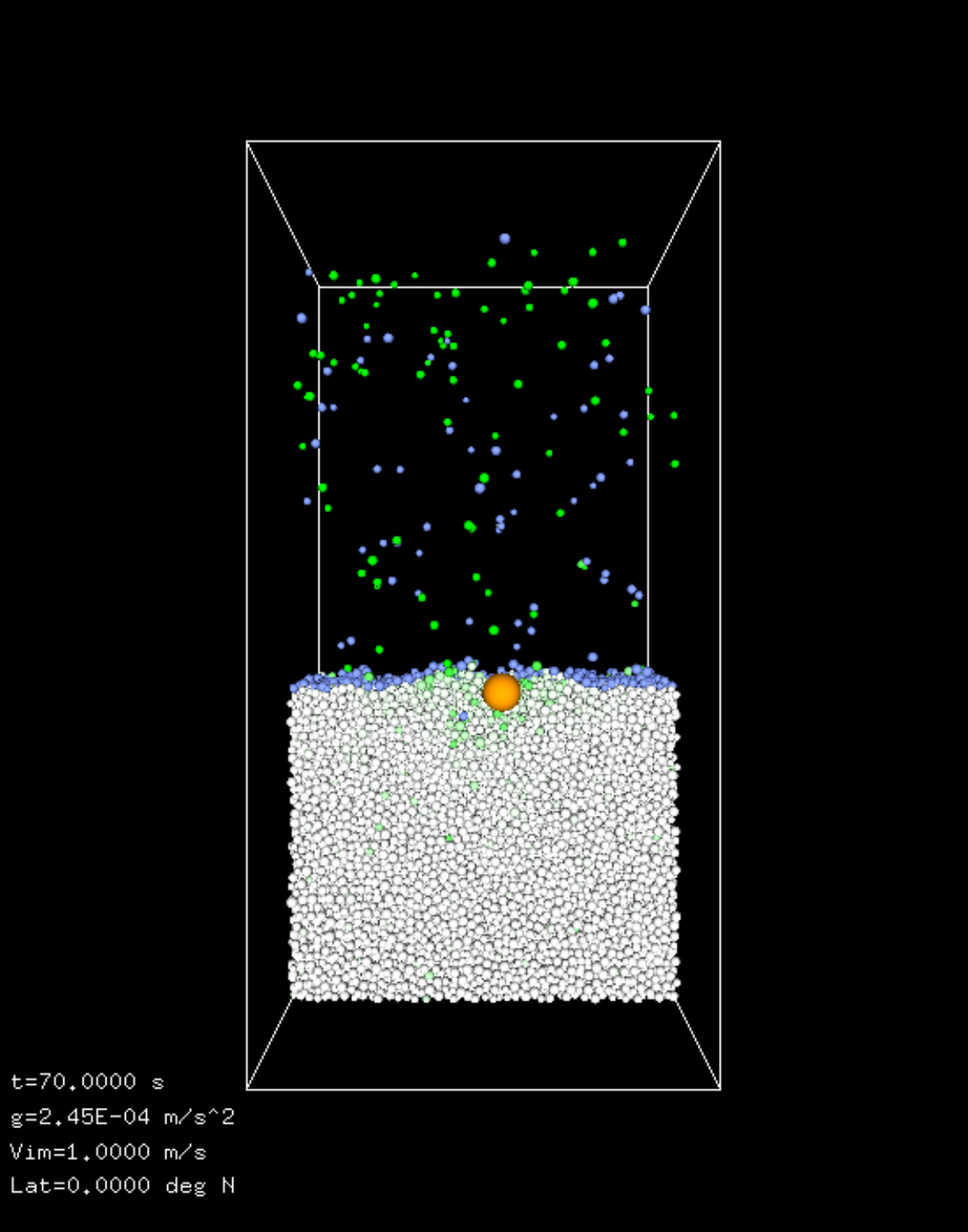}
    \includegraphics[scale=0.16,trim={4.5cm 2.5cm 4.5cm 2.5cm},clip]{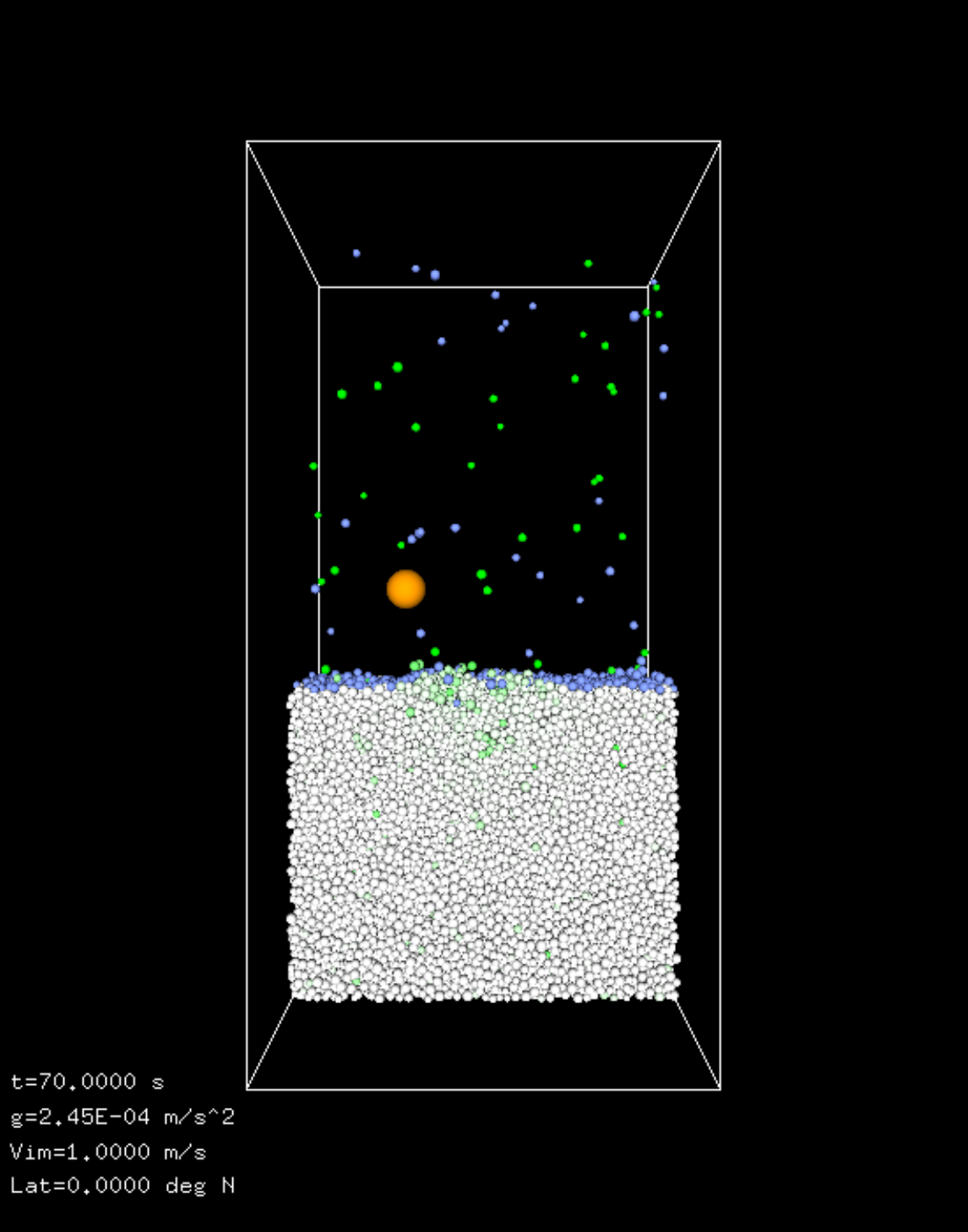}\\
    \includegraphics[scale=0.16,trim={4.5cm 2.5cm 4.5cm 2.5cm},clip]{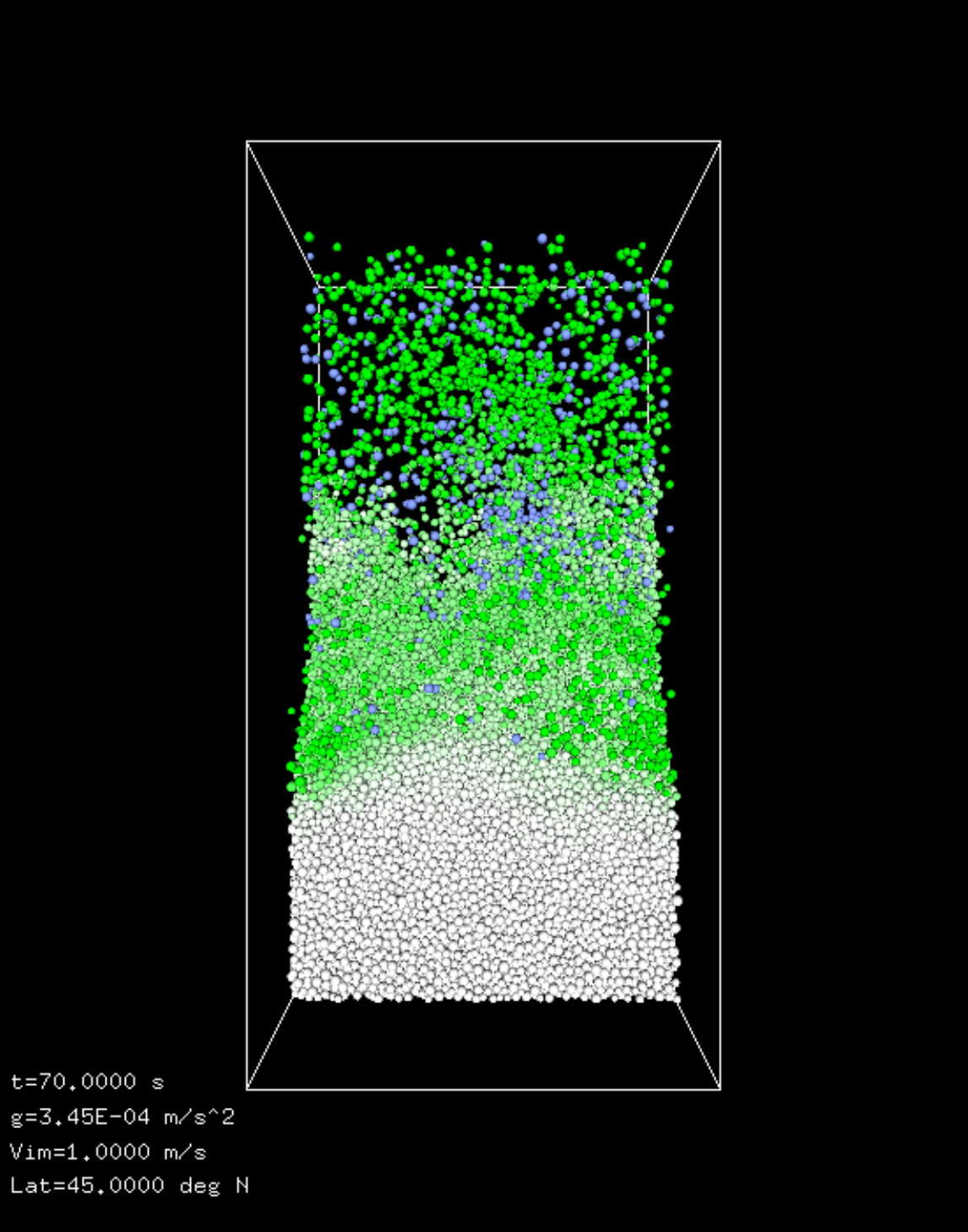}
    \includegraphics[scale=0.16,trim={4.5cm 2.5cm 4.5cm 2.5cm},clip]{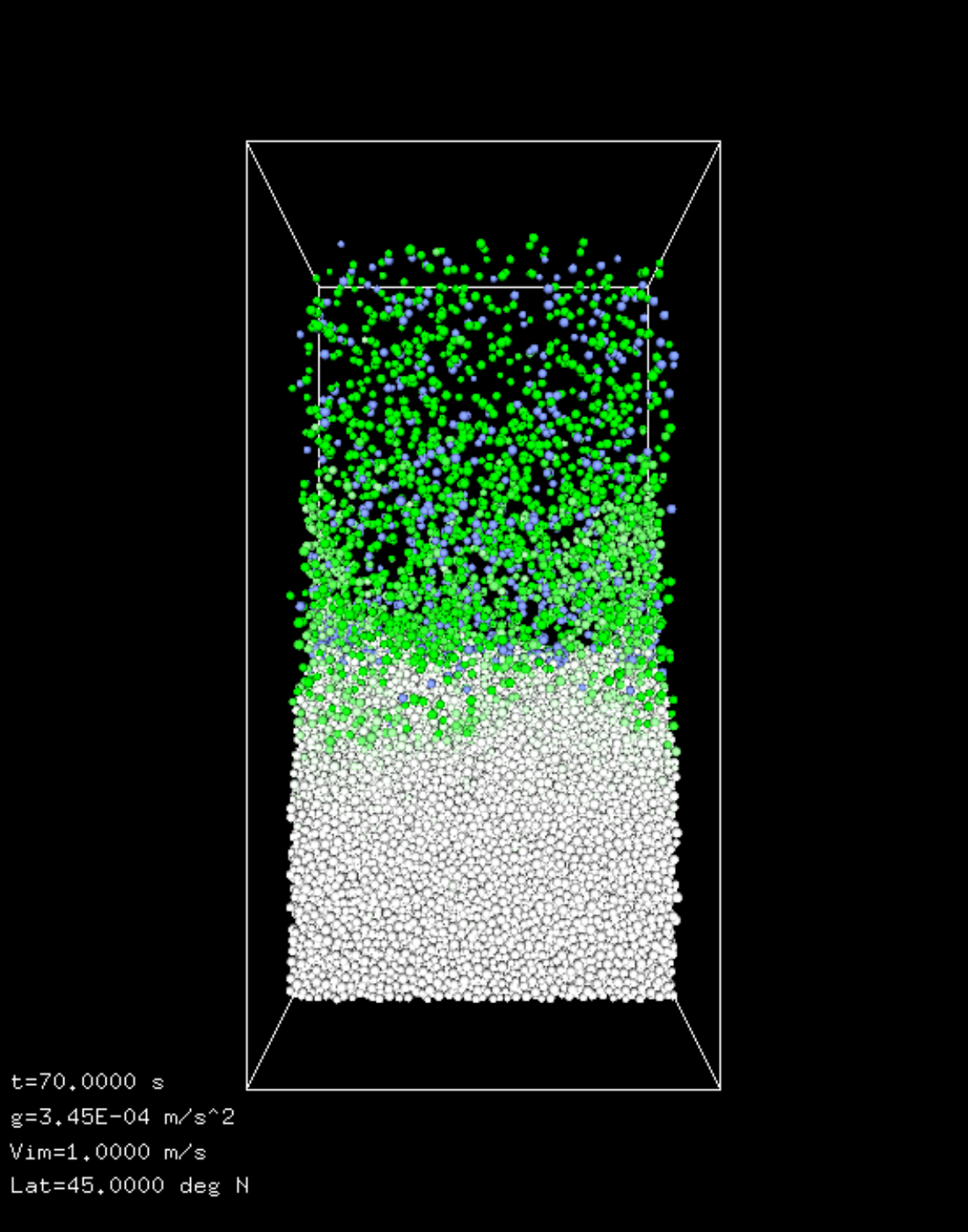}
    \includegraphics[scale=0.16,trim={4.5cm 2.5cm 4.5cm 2.5cm},clip]{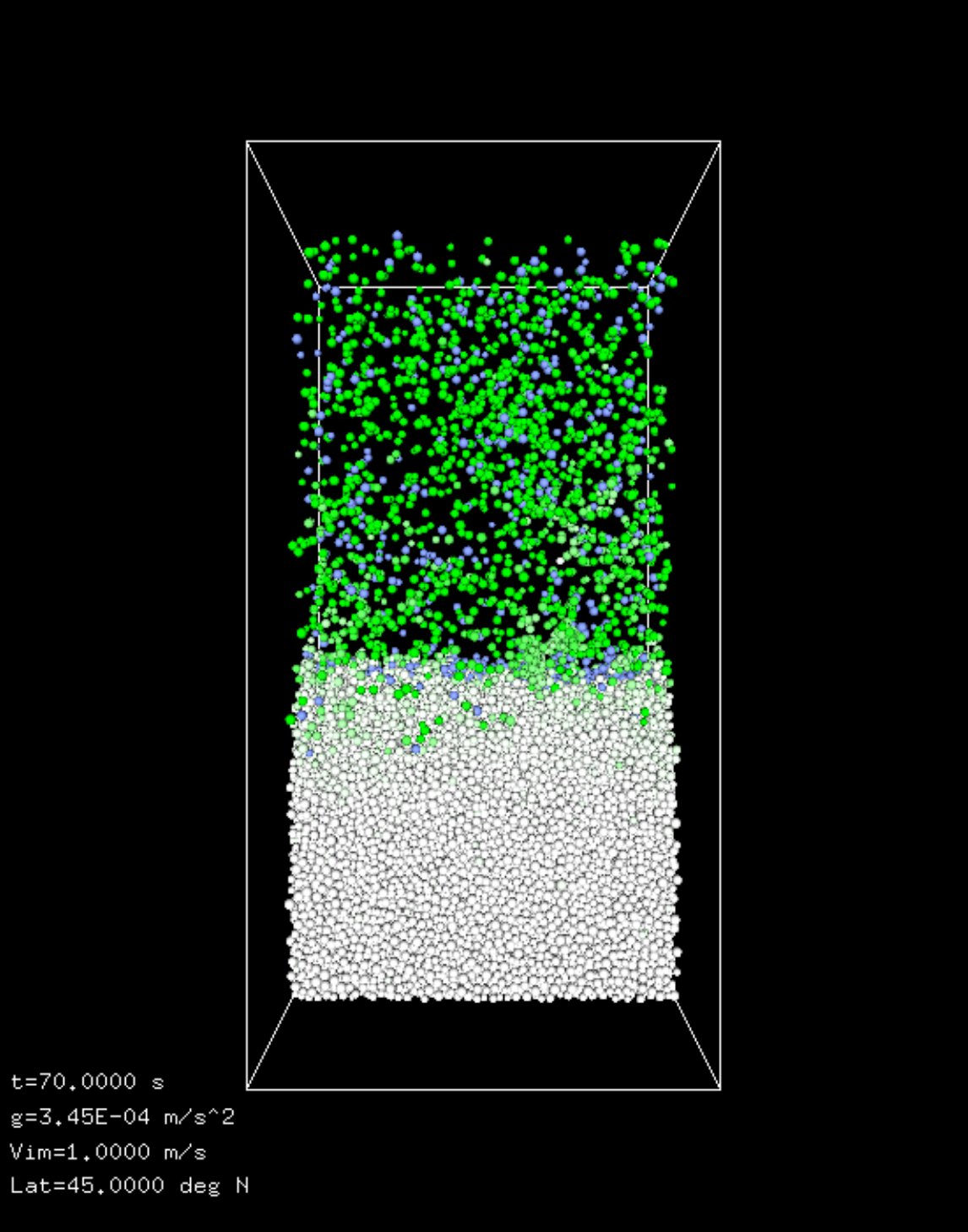}
    \includegraphics[scale=0.16,trim={4.5cm 2.5cm 4.5cm 2.5cm},clip]{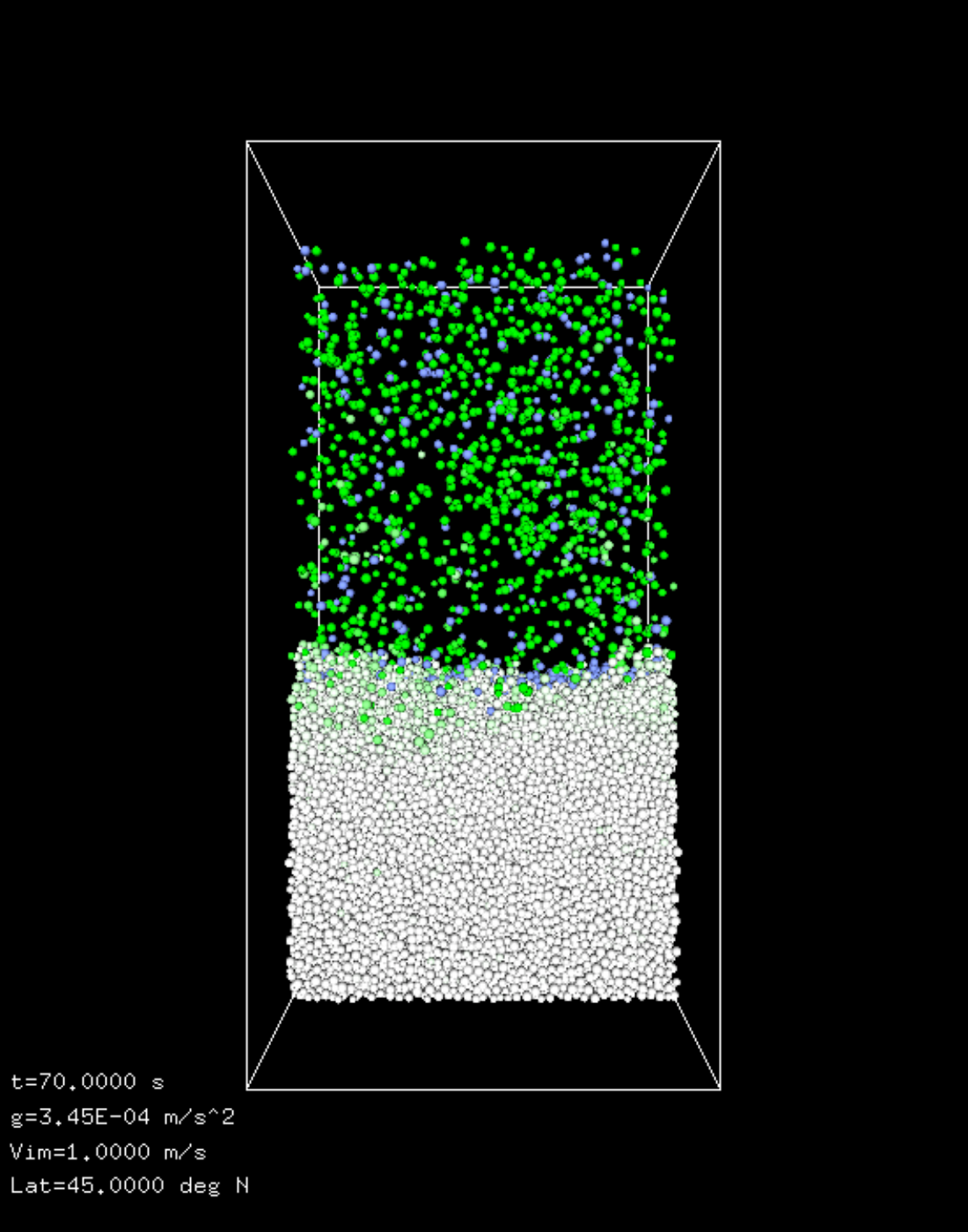}
    \includegraphics[scale=0.16,trim={4.5cm 2.5cm 4.5cm 2.5cm},clip]{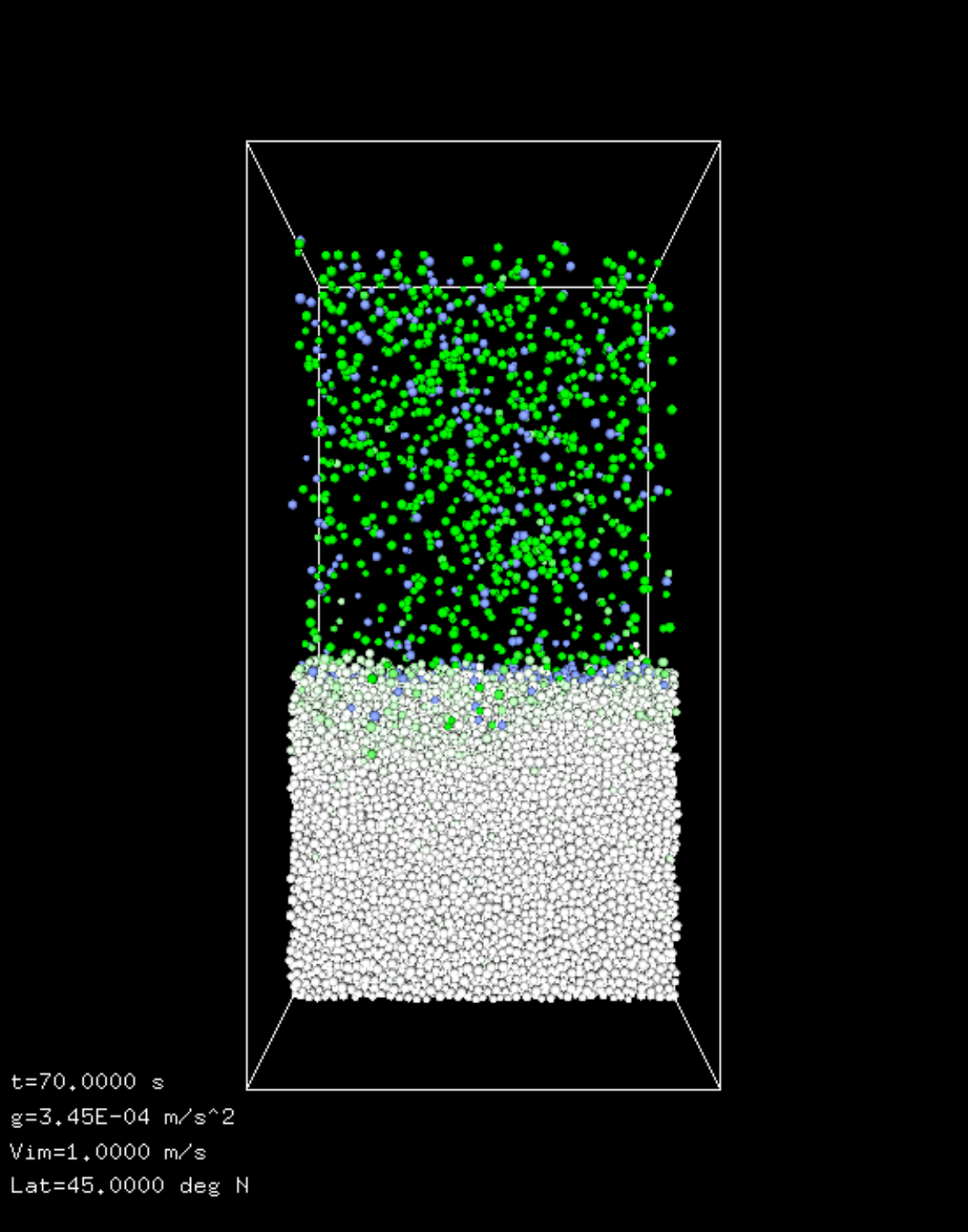}
    \includegraphics[scale=0.16,trim={4.5cm 2.5cm 4.5cm 2.5cm},clip]{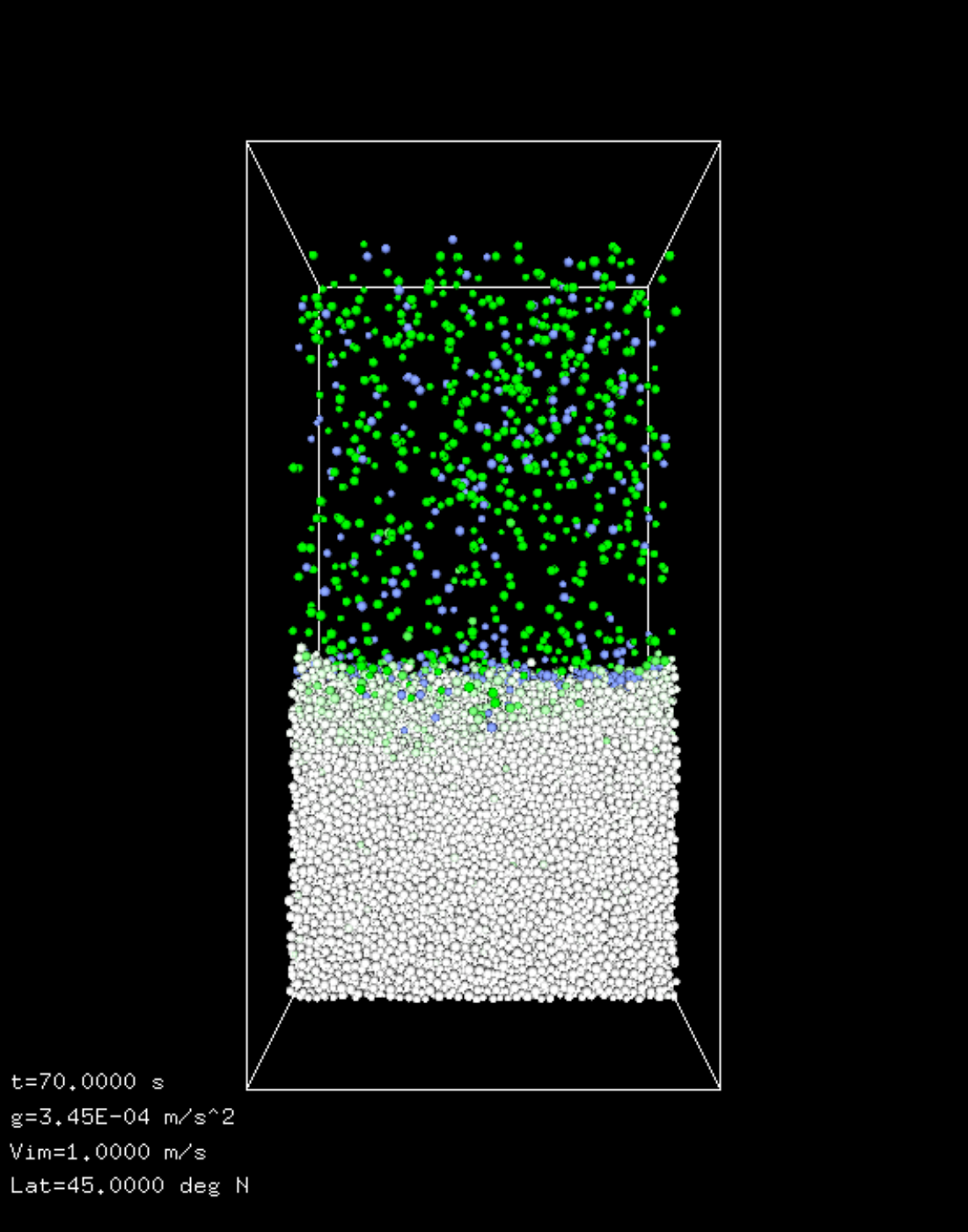}
    \includegraphics[scale=0.16,trim={4.5cm 2.5cm 4.5cm 2.5cm},clip]{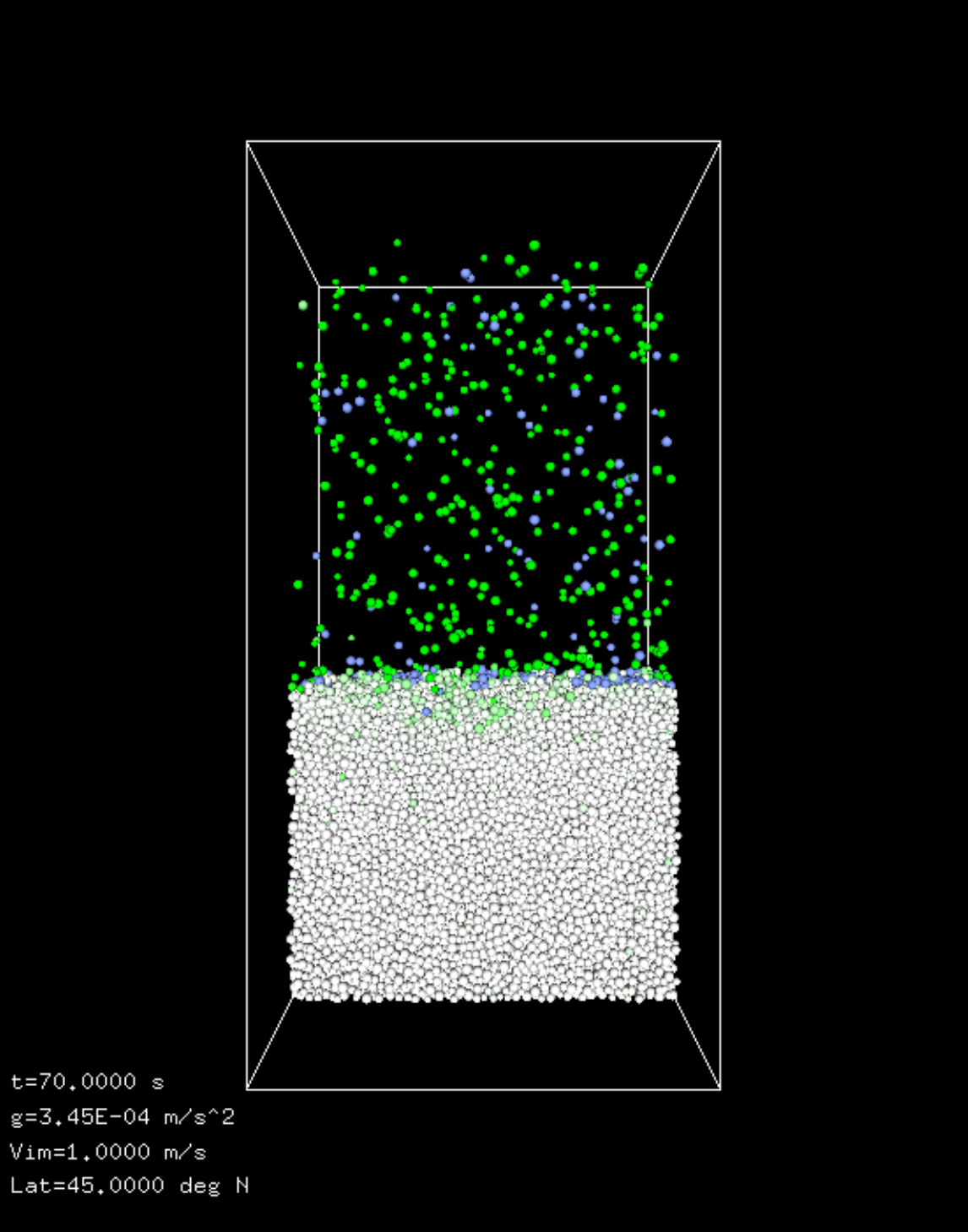}
    \includegraphics[scale=0.16,trim={4.5cm 2.5cm 4.5cm 2.5cm},clip]{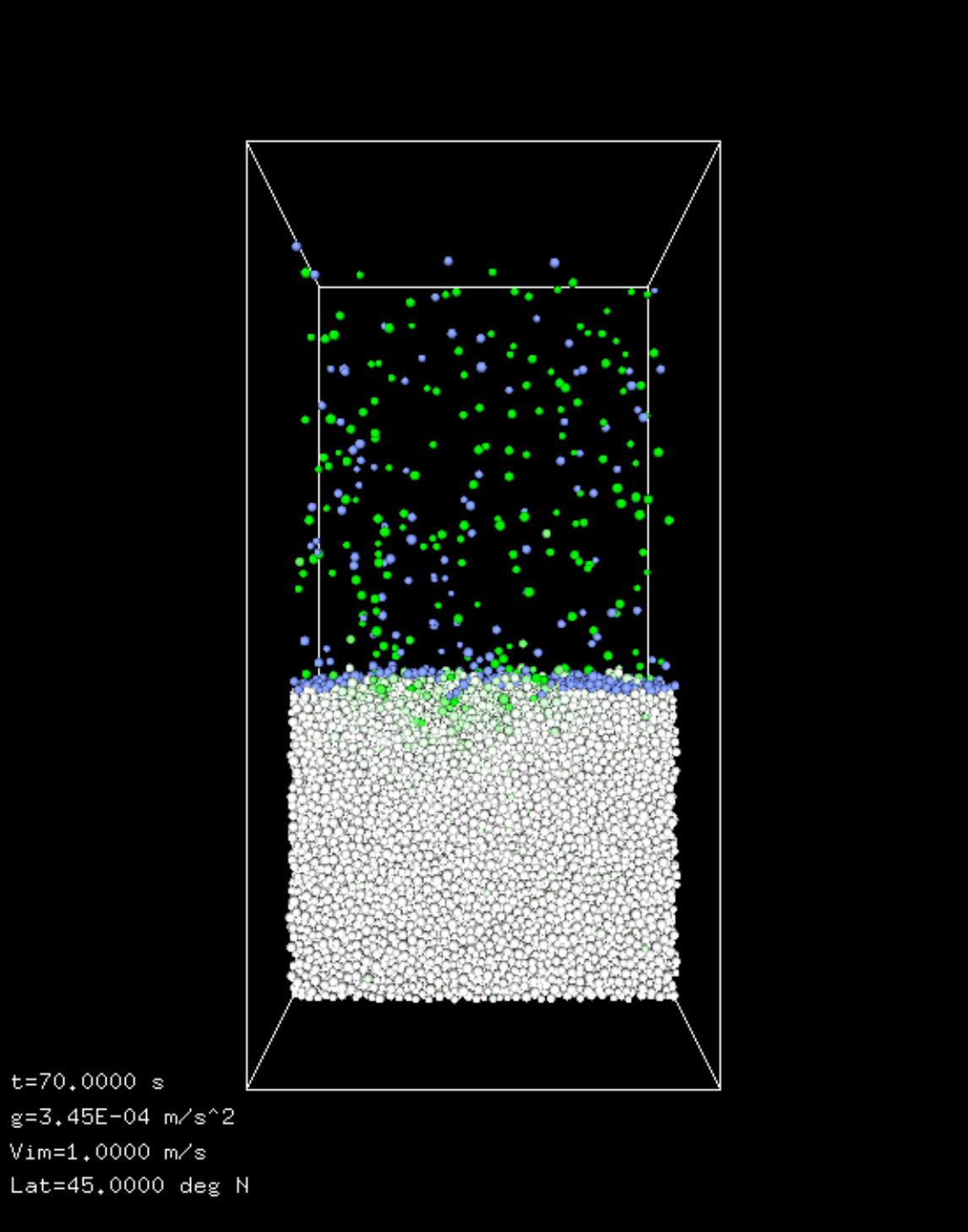}
    \includegraphics[scale=0.16,trim={4.5cm 2.5cm 4.5cm 2.5cm},clip]{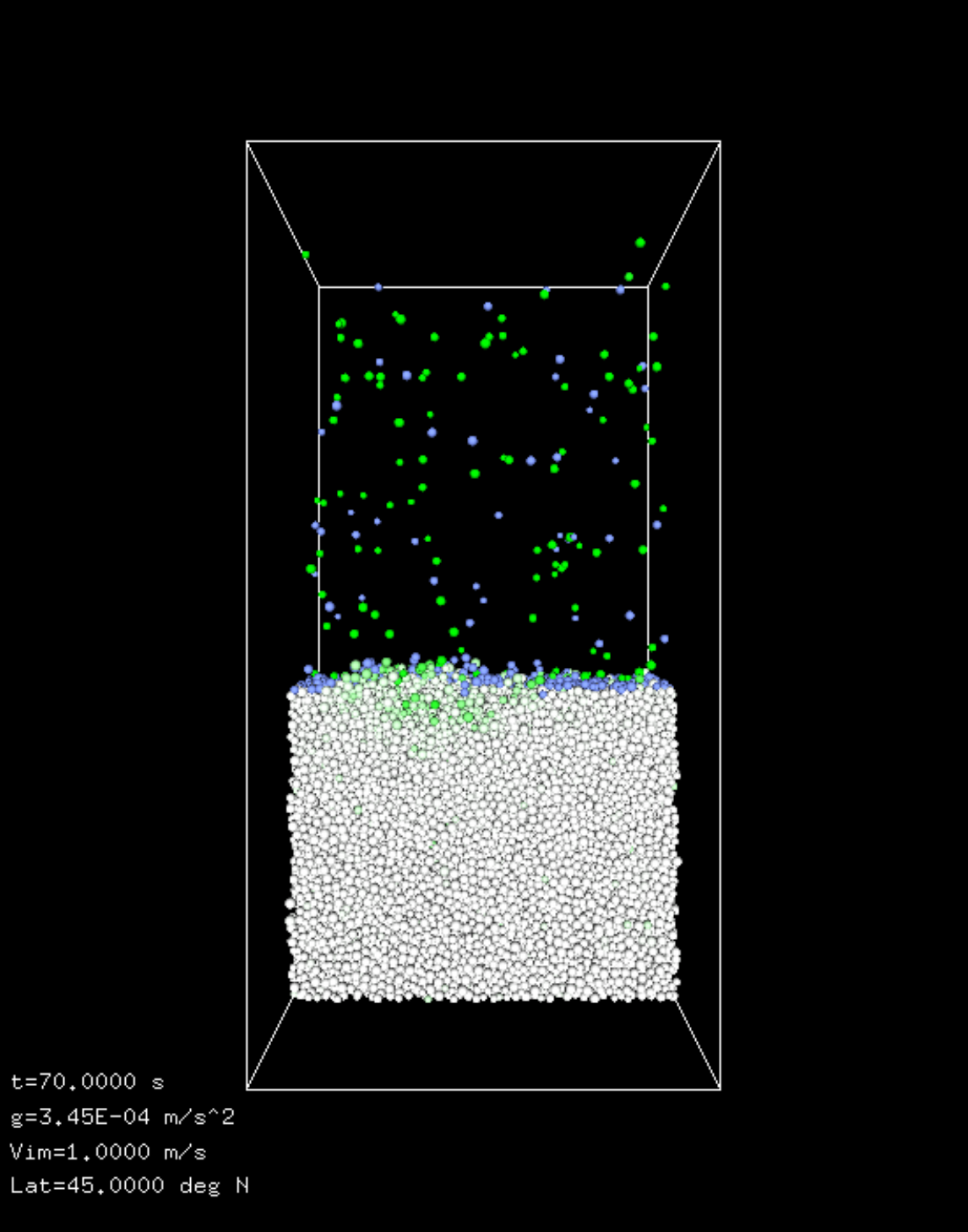}\\
    \includegraphics[scale=0.49]{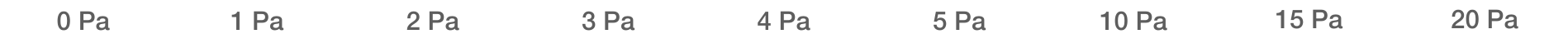}
    \caption{Snapshots of projectile impacts into a granular surface for different levels of cohesive strength. The images are taken 60~s after the last projectile was shot (top: $\lambda=0^{\rm{o}}$; bottom: $\lambda=45^{\rm{o}}$).  From left to right, cohesive strength is 0, 1, 2, 3, 4, 5, 10, 15, and 20~Pa. The orange sphere that appears in some of the images is the last projectile. Particles with speeds $>1\times10^{-4}$ m/s are turned green; blue particles were those on the very surface of the granular bed before the first impact.  The particles in the front half of the box have been made transparent to aid visualization.}
    \label{f:sim_imp}
\end{figure}

\figr{sim_imp} shows snapshots of the simulated granular bed after being impacted by the final projectile.  With the new constraints, this revised set of simulations shows that a bulk cohesive strength of 3~Pa would be more than sufficient to avoid material flow and any kind of global reshaping of Didymos. \edit1{This finding, along with those about the needed cohesive strength for structural stability under rotation (see above), show that a cohesionless structure is incompatible with the observations. Our results show that asteroid Didymos needs to have a non-zero cohesive strength to be able to maintain its structure intact under rotation and debris impacts.} Additionally, and as observed experimentally by \cite{Murdoch2017}, and \cite{Walsh2022-tagsam} during the OSIRIS-REx mission TAGSAM event, cohesionless systems are more easily penetrated and fluidized in weaker gravitational fields compared to Earth's \citep{Zacny2018}.

Given that we have used only spherical particles, the amount of geometrical interlocking (also called geometrical cohesion), and its contribution to shear strength, is diminished \citep{Govender2021}. Even if rolling friction is implemented, it is difficult to establish how similar its effect is to interlocking to avoid fluidization or granular flow. This implies then that the exact amount of cohesive strength needed to avoid granular flow on the surface of Didymos could be somewhat lower, though still necessary.

\subsection{Origin of Dimorphos and its pre-impact oblate shape}
\label{s:Dimorphos_origin}

Didymos's shape, fast rotation, and geophysical properties suggest that---at some point in the past---Didymos likely exceeded its spin limit, leading to the formation of Dimorphos \citep{Walsh2015,Barnouin2023,Agrusa2024}. Dimorphos's obliquity being near $180^\circ$ \citep{Rivkin2021} implies that this spin-up process was dominated by YORP \citep{Vokrouhlicky2015}, although natural impacts may have played some role as Didymos spends a significant fraction of time in the inner main belt \citep{Holsapple2022,CampoBagatin2023}.

If Dimorphos is indeed a rubble pile with little-to-no cohesion \citep{Barnouin2023,raducan2024physical}, then Dimorphos likely formed via gravitational reaccumulation outside the Roche limit of Didymos, as the body would not survive a fission event intact and would naturally be disrupted by tides \citep[e.g.,][]{Agrusa2022a}. However, Dimorphos's apparent oblate shape is not immediately explained by such a scenario, since gravitational accumulation in a strong tidal environment would tend to create a more prolate shape \edit1{\citep{Madeira2023,Agrusa2024,Wimarsson2024}}.

Although Dimorphos's unexpected oblate shape is structurally stable at its pre-impact orbit and assumed spin period \citep{Holsapple2006,Sharma2009}, this shape stands in contrast to the measured shapes of other secondaries---either by radar \citep{Ostro2006,Naidu2015b,Becker2015} or by lightcurve measurements \citep{pravec2016binary,Pravec2019}---which tend to be more prolate. Given that DART only saw one hemisphere of Dimorphos and the extent of the $b$ axis was inferred by the geometry of the terminator \citep{Daly2023}, it is possible that Dimorphos's true shape is prolate, with a small libration that shifts the illumination so as to mimic an oblate shape. If the true shape is oblate, then this means that Dimorphos either formed with such a shape or longer-term processes such as additional mass-shedding events or natural impacts could have led to an oblate shape. The issue of prolate/oblate shape and its implications on the formation and evolutionary processes at work are ongoing topics of study.

\subsection{Structural properties and potential post-impact reconfiguration of Dimorphos}

The high-resolution images of Dimorphos's surface returned by DART exhibited a predominantly blocky texture composed primarily of boulders with diameters in the 0.1--10~m range, indicating a low-cohesion surface structure \citep{Daly2023, Barnouin2023}. If Dimorphos's subsurface and interior share a comparable distribution of boulder sizes, its overall mechanical strength would be notably weak, and its shape and surface morphology could be susceptible to alteration due to the tidal forces exerted by Didymos or potential impacts from external objects. 

As a direct result of the DART impact, Dimorphos undoubtedly experienced significant resurfacing near the impact site \citep{raducan2024physical}. Additionally, indirect downstream effects would likely alter Dimorphos's post-impact morphology on a longer time scale. For example, the impact released significant amounts of ejecta \citep{Li-Nature-2023,Moreno-PSJ-2023}. Any material that was not on an escape trajectory or blown away by solar-radiation pressure can remain bound to the binary system for weeks or months, if not longer (\sect{hardening}). A significant portion of this material will re-impact Dimorphos at speeds on the order of a few cm/s, likely disturbing the surface upon impact \citep{Ferrari2024}. Another indirect mechanism for resurfacing is caused by the post-impact excited dynamical state of Dimorphos. Building upon the current understanding of Dimorphos's shape and structural properties, we analyze the likelihood of resurfacing and reshaping behaviors of Dimorphos after the DART impact.

\subsubsection{Tidal stress on Dimorphos, material displacement, and resurfacing possibility}
\label{s:tidal}

Periodic tidal stress studies done pre-encounter found that local material failure due to tides was only possible for a cohesionless Dimorphos with very low friction angle, \ie $\leq 10 ^\circ$, and failure was restricted to surface layers at the poles for homogeneous models and at the equator for layered internal-structure models \citep{Murdoch2017probing}. This result, however, was made using models based on monolithic rocks, did not take into account the rotation of the satellite, and underestimated the oblateness of Dimorphos \citep{Daly2023}. 
    
From the latest DART results, considering a weaker rubble-pile model instead and matching the estimated shape and rotation of Dimorphos would increase tidal stress inside Dimorphos up to a few tenths of Pa in amplitude. This is an order of magnitude lower than the upper bound of the best-fit results with the DART observables in \cite{raducan2024physical} where cohesion is estimated to be lower than a few Pa, consistent with the results of \cite{Cheng2023b} where cohesion was found to be lower than 500~Pa with no lower limit. If cohesion of $\geq\sim$0.1~Pa is present, tidal resurfacing and faults due to tidal stress are unlikely to occur at the surface of Dimorphos for synchronous rotation. Local resurfacing from non-tidal effects can still happen: due to structural re-adjustment of the whole asteroid, slight deformation about the impact antipodal site and moderate surface refreshment are still possible \citep{Liu2023}.

However, in addition to Dimorphos's excited post-impact orbital state, its rotation state could have been significantly excited (and possibly entered tumbling) \citep{Agrusa2021Excited}. As a result, the tidal, centrifugal, and Euler accelerations felt by boulders on Dimorphos's surface can undergo significant changes orbit-to-orbit. This can lead to surface slope changes on the order of tens of degrees \citep{Agrusa2022b}. Depending on the post-impact shape of Dimorphos (\ie the initial surface slopes), this may lead to motion on Dimorphos's surface long after the impact. However, this effect is difficult to quantify, as it depends strongly on Dimorphos's unknown post-impact rotation state and shape. These various sources of resurfacing suggest that Dimorphos may look very different upon Hera's arrival in late 2026 than it would have immediately following impact. Distinguishing between surface features (\ie crater morphology) that were created directly by the impact from these longer-term resurfacing processes will present a unique challenge to the Hera team (\sect{hera}).

\subsubsection{Large-scale reshaping of Dimorphos}
\label{s:reshaping}

Recent impact simulations have revealed that, if Dimorphos is indeed a structurally weak, low-cohesion body, the DART impact could have caused large-scale reshaping beyond just forming an impact crater \citep{Raducan2022reshaping, Raducan2022global, Raducan2024lessons, Stickle2022}. Material displacement and compaction around the impact site likely caused Dimorphos to become more elongated according to these simulations and Dimorphos's equatorial axis ratio $a_s/b_s$ could reach 1.2 \citep{raducan2024physical}, a significant difference from its pre-impact value of $1.06\pm0.03$.

Interestingly, ground-based observations also corroborate the elongated shape of the post-impact Dimorphos. While none were detected in the pre-impact lightcurve data, recent high-quality lightcurve data revealed secondary lightcurve variations corresponding to Dimorphos's rotation \citep{Pravec2024}. From these data, the upper bound of $a_s/b_s$ is estimated to be \edit1{$1.4$}. Though this estimate carries some uncertainties due to the unknown attitude state of Dimorphos, it aligns well with the estimate from the recent impact simulation \citep{raducan2024physical}. Additionally, post-impact orbit fitting using mutual-event observations over 5 months, accounting for the planar full two-body problem \citep{Naidu2024}, also leads to a similar $a_s/b_s$, $1.306 \pm 0.012$, as well as the polar axis ratio $b_s/c_s$ of $1.2 \pm 0.2$ (\tbl{params}).

Considering these numerical and observational constraints, the reshaping of Dimorphos appears to be highly plausible. Consequently, it is important to consider the effect of reshaping on the mutual dynamics of the Didymos system. The elongation of Dimorphos leads to a change in its orbit period \citep{nakano2022nasa}, which is critical to account for in the $\beta$ estimation process. Moreover, the post-impact shape is likely to have some degree of asymmetry, as the leading side of Dimorphos undergoes more significant reshaping than the opposite side. This asymmetry could influence Dimorphos's attitude dynamics and increase the likelihood of post-impact attitude instability. These effects of Dimorphos's reshaping will be discussed further in \sect{Dimorphos_reshaping}.

In addition, the DART impact caused a change in the shape of Dimorphos outside of a stable shape/dynamical configuration that may be providing a continued reservoir of material that maintains the tail. The resettling timescale is largely determined by the microgravity environment of the system, and thus a continued active release of particles up through the present time cannot be ruled out. Investigation into the timescale of how Dimorphos settles into a new shape and dynamical configuration is ongoing.

\subsection{Reshaping-induced perturbations on the mutual dynamics} \label{s:Dimorphos_reshaping}

\citet{nakano2022nasa} conducted a pre-impact statistical investigation into the effect of Dimorphos's reshaping on the mutual dynamics, specifically focusing on the orbit period---one of the key parameters in estimating $\beta$. The current estimate of $a_s/b_s \sim$1.3, however, exceeds the range explored in the pre-impact study. Here, we consider a range of $a_s/b_s$ from 1.1 to 1.5, which thoroughly encompasses the current estimates of $a_s/b_s$ with the uncertainties owing to Dimorphos's unknown attitude state and investigate the effect of significant reshaping on not only the orbit period but also the attitude state of Dimorphos. 

\edit1{Following \citet{nakano2022nasa}, we generate synthetic shape models of Dimorphos in its reshaped form (\figr{Dimo_reshaping_dP}(a)) and propagate the mutual dynamics using a finite-element full-two-body-problem model. In order to constrain specifically the effect of reshaping, the simulations do not account for the momentum and torque imparted by the DART spacecraft and only consider Dimorphos's reshaping. We assume that Dimorphos experiences instantaneous reshaping under constant volume. Assuming that the angular momentum is conserved before and after reshaping, the angular velocity of reshaped-Dimorphos is adjusted in each simulation based on the modified moment of inertia due to reshaping. This ensures that the system's initial state is dynamically consistent throughout the simulations. We compute the reshaping-induced orbit period change by comparing the orbit period of the reshaped-Dimorphos with that of the nominal, pre-impact state. Dimorphos's attitude state after reshaping is described by using the 1-2-3 roll, pitch, yaw Euler angles as used in \sect{Dimo_rotation}.}

\subsubsection{Orbit-period change and $\Delta V_T$}

\begin{figure}[ht!]
    \centering
    \begin{subfigure}
        \centering
        \includegraphics[width=0.35\linewidth]{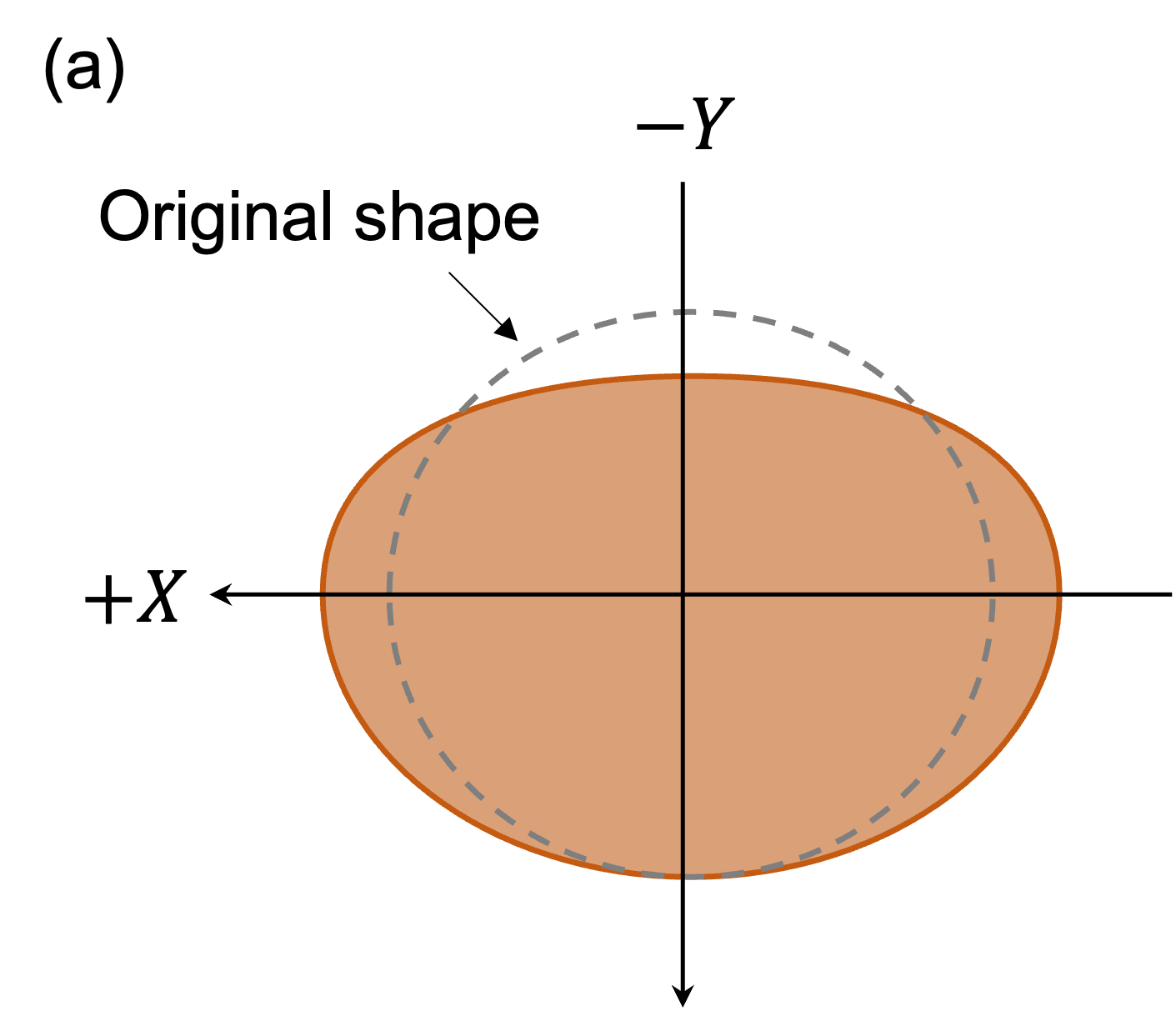}
    \end{subfigure}
    \hfill
        \begin{subfigure}
        \centering
        \includegraphics[width=0.5\linewidth]{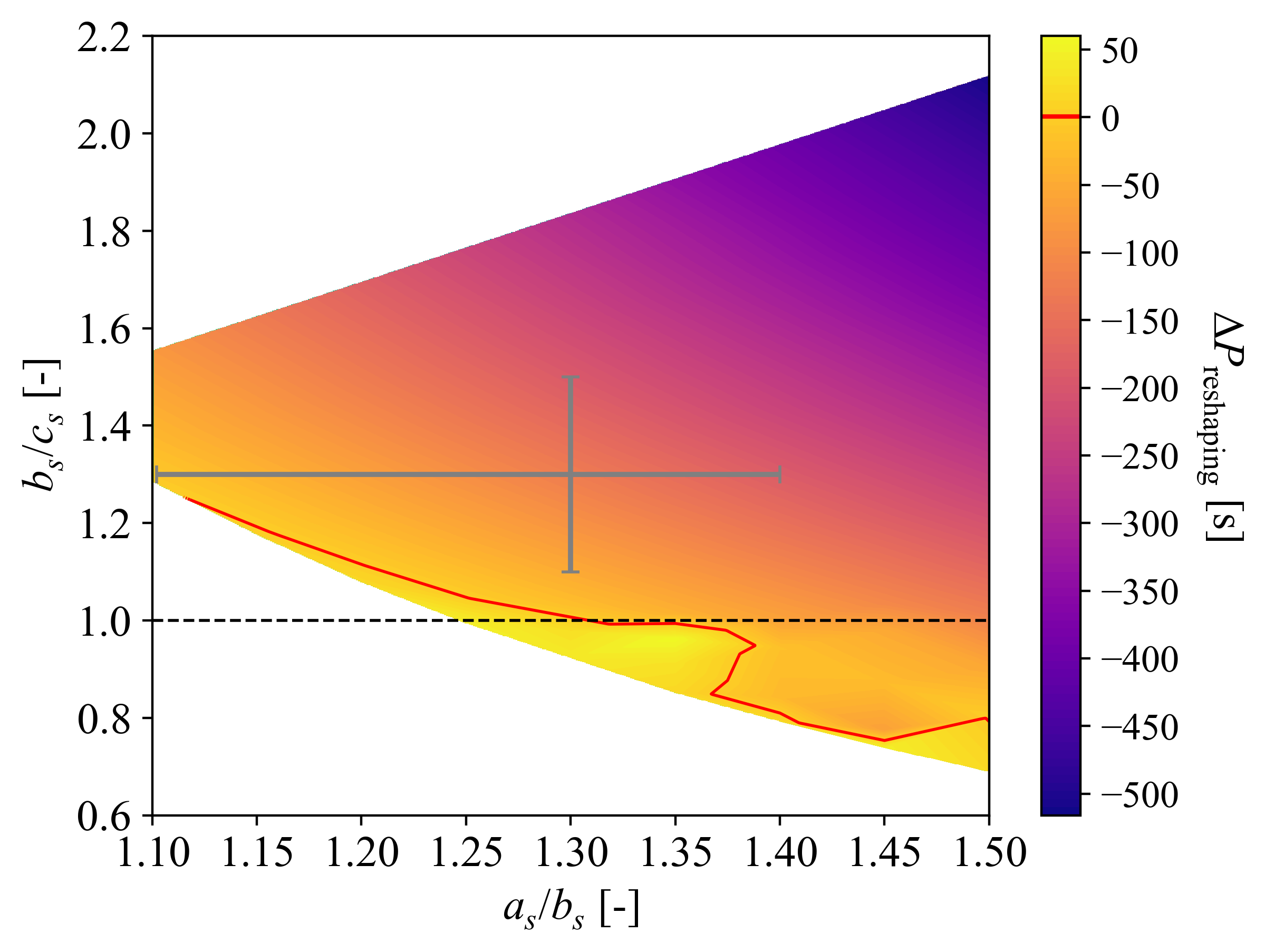}
    \end{subfigure}
    \caption{(a) Schematic diagram showing a synthetic shape model of a reshaped Dimorphos. The axes are in Dimorphos's body-fixed frame, with the X- and Y-axis corresponding to the long and intermediate axis, respectively. We consider a physically plausible reshaping condition inferred from recent impact simulations and describe such reshaping by changing the positive and negative sides of the X-, Y-, and Z-axis of the original Dimorphos shape model. We assume that the impact, which was approximately aligned with the Y-axis, instantaneously shortened the Y-axis and lengthened the X-axis. The Z-axis is modified such that the volume of Dimorphos before and after reshaping does not change. (b) Reshaping-induced orbit-period change, $\Delta P_\text{reshaping}$, as a function of the synthetic shape model's $a_s/b_s$ and $b_s/c_s$ that are permissible under the physically plausible reshaping condition. The red curve denotes $\Delta P_\text{reshaping} = 0$~s. \edit1{The current estimates of the axis ratios $a_s/b_s$ and $b_s/c_s$ are indicated by the horizontal and vertical error bars, respectively (i.e., $1.1 \le a_s/b_s \le 1.4$ from \edit1{\cite{Pravec2024}}; $a_s/b_s = 1.300 \pm 0.010$ and $b_s/c_s = 1.3\pm0.2$ from \citet{Naidu2024}).}}
    \label{f:Dimo_reshaping_dP}
\end{figure}

\figr{Dimo_reshaping_dP}(b) shows the reshaping-induced orbit period change, $\Delta P_\text{reshaping}$, as a function of the synthetic shape model's $a_s/b_s$ and $b_s/c_s$. We find that most of the shapes result in a reduction of the orbit period, but a subset of the shapes, characterized by $b_s/c_s$ close to or less than unity, can lead to an increase of the orbit period.\edit1{The red contour line indicates $\Delta P_\mathrm{reshaping}=0$~s. The un-smooth contour below $b_s/c_s < 1.0$ is because of the orbit-period variation owing to the unstable attitude of Dimorphos, as discussed in \sect{orbit_period}.} In the same figure, the horizontal and vertical error bars correspond to the current estimates for $a_s/b_s$ and $b_s/c_s$ ($1.1 \le a_s/b_s \le 1.4$ from \edit1{\cite{Pravec2024}}; $a_s/b_s = 1.300\pm0.010$ and $b_s/c_s = 1.3\pm0.2$ from \citet{Naidu2024}). Note that the synthetic shape models used here have symmetric configurations about the \edit1{Y-}axis (\figr{Dimo_reshaping_dP}(a)); our investigation revealed that minor asymmetries in shape do not exert any significant effect on $\Delta P_\text{reshaping}$ \citep{Nakano2023}.

Should post-impact Dimorphos fall within the uncertainty range defined by the error bars in \figr{Dimo_reshaping_dP}(b), the resulting $\Delta P_\text{reshaping}$ ranges from \edit1{$\sim –250$ to $\sim +10$~s}. This implies that within the observed 33-min orbit-period change, as much as \edit1{$\sim$4~min} could potentially be attributed to Dimorphos's reshaping, as opposed to being solely attributed to the momentum change caused by the DART impact and ejecta recoil.

The tangential component of Dimorphos's orbital-velocity change $\Delta V_T$ corresponding to $\Delta P_\text{reshaping}$ of \edit1{$–250$ to $+10$~s} can be computed; it ranges from \edit1{$-0.347$ to $0.014$~mm~s$^{-1}$} \citep{Nakano2023}. The initial estimate of $\Delta V_T$, $-2.70 \pm 0.10$~mm~s$^{-1}$, did not account for the effect of Dimorphos's reshaping \citep{Cheng2023, meyer2023a}. However, as the effects of reshaping and the DART impact are independent to first order \citep{nakano2022nasa}, adding this $\Delta V_T$ from reshaping to the earlier estimate of $\Delta V_T$ would lead to a good approximation of the true $\Delta V_T$ and thus $\beta$, effectively accounting for Dimorphos's reshaping. This adjustment can be made once the post-impact Dimorphos's shape is thoroughly characterized through the Hera mission.

\subsubsection{Attitude perturbation}

\begin{figure}[ht!]
    \centering
    \begin{subfigure}
        \centering
        \includegraphics[width=0.5\linewidth]{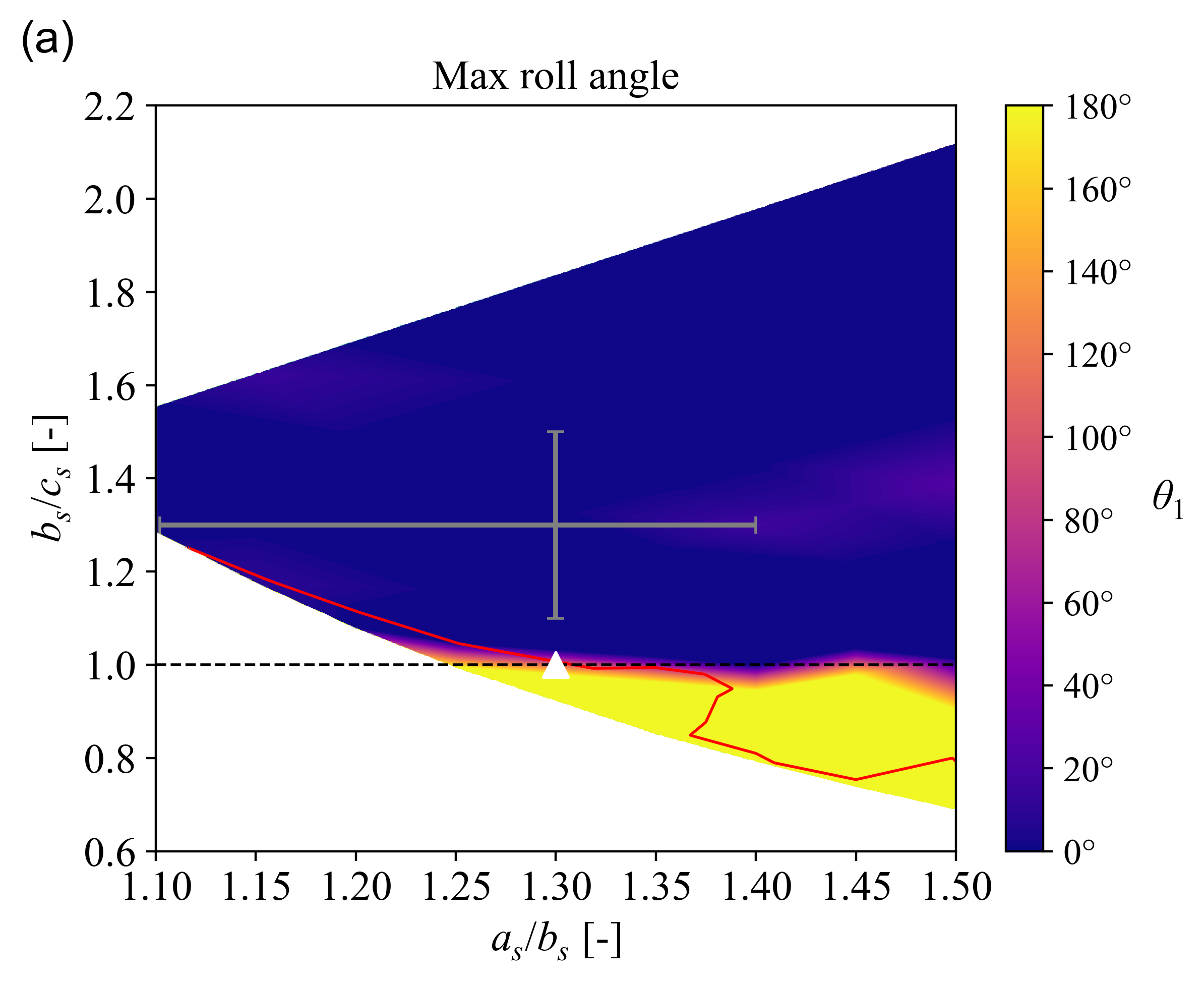}
    \end{subfigure}
    \par
    \hfill
    \begin{subfigure}
        \centering
        \includegraphics[width=0.325\linewidth]{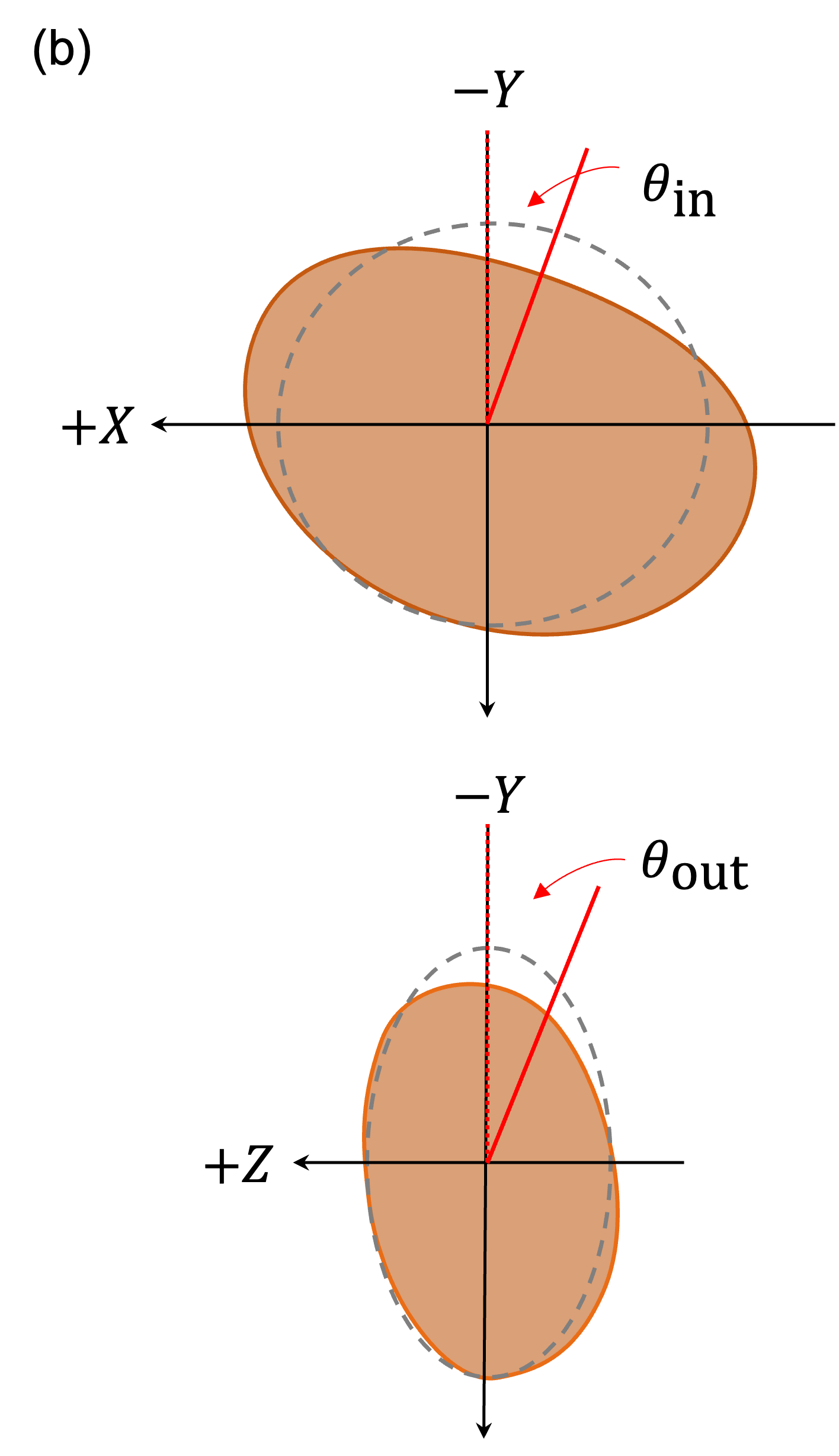}
    \end{subfigure}
    \begin{subfigure}
        \centering
        \includegraphics[width=0.5\linewidth]{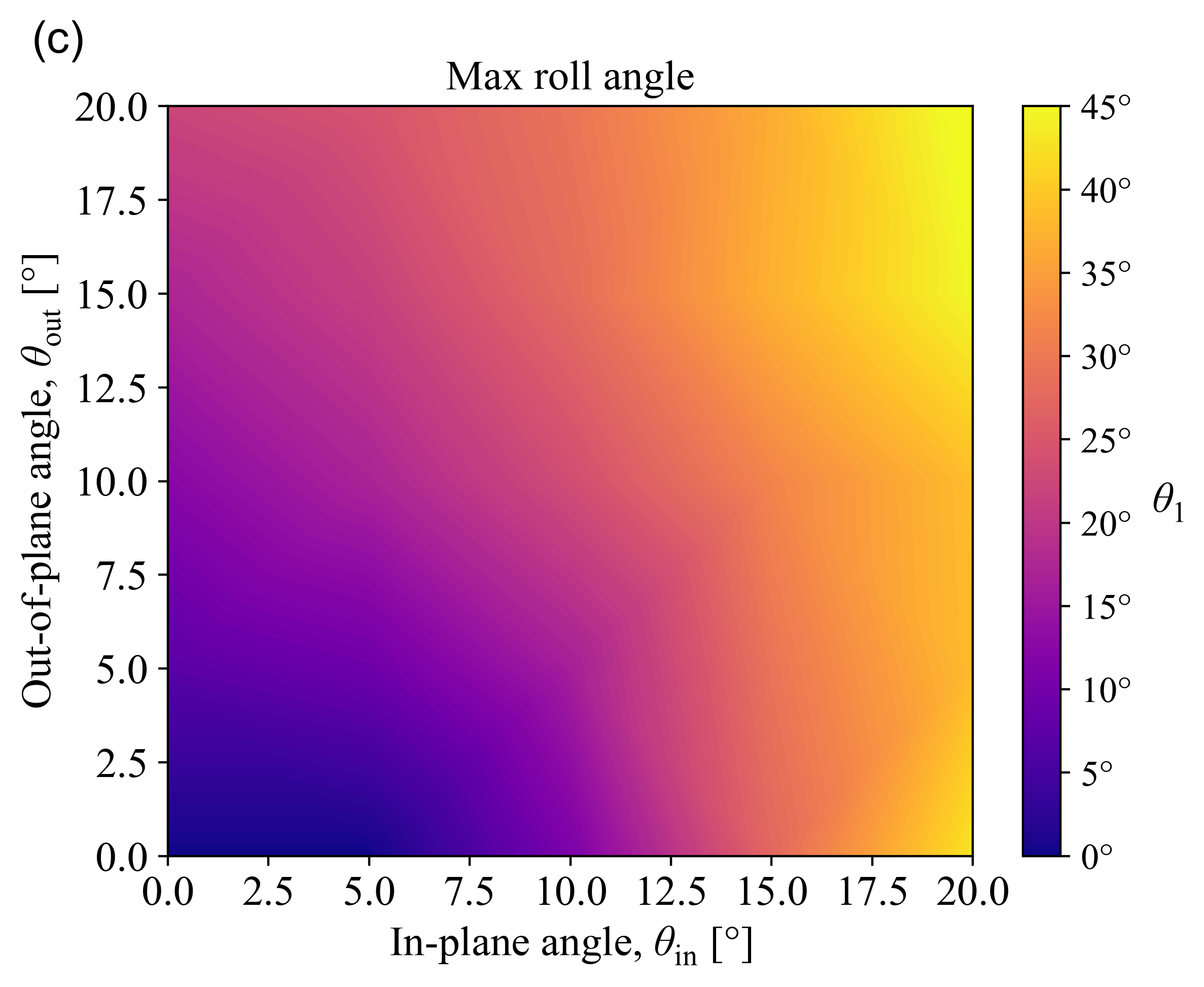}
    \end{subfigure}
    \caption{(a) Maximum roll angle amplitudes as a function of the synthetic shape model's axis ratios $a_s/b_s$ and $b_s/c_s$. The red curve denotes $\Delta P_\text{reshaping} = 0$~s from \figr{Dimo_reshaping_dP}(b). The error bars denote the post-impact axis ratios and their uncertainties (\tbl{params}). (b) A top and side view of a synthetic shape model of Dimorphos under off-axial reshaping. The in-plane reshaping angle $\theta_\text{in}$ and the out-of-plane reshaping angle $\theta_\text{out}$ are measured from Dimorphos's intermediate axis (Y-axis). (c) Maximum roll angle amplitude as a function of $\theta_\text{in}$ and $\theta_\text{out}$.}
    \label{f:Dimo_reshaping_attitude}
\end{figure}

Reshaping of Dimorphos, particularly for the magnitude considered in this study \edit1{(i.e., $a_s/b_s$ from 1.1 to 1.5)}, generally leads to an excitement of roll and pitch angles with an amplitude of less than 10$^\circ$ and yaw angles with an amplitude of about 20$^\circ$. However, as can be seen from \figr{Dimo_reshaping_attitude}(a), shapes with $b_s/c_s$ below $\sim$1.0 could experience higher amplitudes, particularly exceeding 90$^\circ$ in the roll angle. These shapes also experience high amplitudes in both pitch and yaw angles, although the amplitudes remain relatively small compared to the roll direction. This is likely attributed to the asymmetric configuration about the long axis (X-axis) caused by the impact (\figr{Dimo_reshaping_dP}(a)). The leading side of Dimorphos experiences more severe reshaping than the other side, inducing a stronger tendency for the body to roll about its long axis. Despite the instability induced by Dimorphos's reshaping, the body generally remains tidally locked with Didymos.

However, it is important to recognize that the orientation of the principal axes for the synthetic shape models considered thus far roughly aligns with that of the pre-impact Dimorphos. This alignment arises from the assumption that reshaping occurs along the original principal axes. If this is not the case, that is, if reshaping occurs along off-principal axes, Dimorphos's attitude becomes even more perturbed. \edit2{We have thus conducted an additional investigation accounting for the off-principal axes reshaping of Dimorphos. The off-principal axes reshaping was parameterized with two angles, the in-plane reshaping angle, $\theta_\text{in}$, and the out-of-plane reshaping angle, $\theta_\text{out}$, as depicted in \figr{Dimo_reshaping_attitude}(b), and the finite-element full-two-body problem model \citep{nakano2022nasa} was used to simulate the attitude dynamics of Dimorphos.} In \figr{Dimo_reshaping_attitude}(c), we show the roll angle amplitude of a synthetic shape model with $a_s/b_s = 1.3$ and $b_s/c_s = 1.1$ as a function of the in-plane reshaping angle $\theta_\text{in}$ and the out-of-plane reshaping angle $\theta_\text{out}$ measured from the intermediate axis (Y-axis). Notably, the attitude state can exhibit significant deviations from the principal-axes reshaping case (\ie $\theta_\text{in} = \theta_\text{out} = 0^\circ$). Similar sensitivity to $\theta_\text{in}$ and $\theta_\text{out}$ is observed in the pitch and yaw angles. This result highlights the difficulty of predicting Dimorphos's attitude state without precise knowledge of its post-impact shape. Even if the axis ratios of the post-impact shape are constrained through numerical and observational means, the attitude state could be significantly different, depending on how different the orientations of the new principal axes are from the original principal axes. 

Importantly, the discussed attitude instability of Dimorphos is solely a consequence of reshaping. As discussed in \sect{Dimo_rotation}, Dimorphos's attitude instability including non-principal-axis rotation is plausible, considering the pre-impact shape and $\beta$, even without reshaping. The findings presented here suggest that Dimorphos's reshaping further increases the likelihood of such attitude instability.

\subsection{Secular Evolutionary Effects}
\label{s:secular}

The long-term dynamical evolution of rubble-pile binary asteroid systems is driven by binary YORP (BYORP) and tides. A full discussion on these phenomena and how they affect binary asteroid evolution can be found in \citet{Richardson2022}. Note that the BYORP effect is dependent on the shape of the secondary, from which a set of Fourier coefficients can be computed \citep{McMahon2010}. \edit1{These coefficients are based solely on the shape model of the secondary asteroid.} The dominant along-track coefficient, denoted $B$, is the indicator for the direction of evolution. A positive $B$ corresponds to expansive BYORP, while a negative $B$ corresponds to contractive BYORP. Prior to DART, little was known about the BYORP coefficient or tidal parameters of Dimorphos. It is still difficult to constrain these quantities, but work has been done in \cite{Cueva2024} to bound predictions of how much $B$ could have changed from the DART impact, and the resulting implications for the dynamical evolution.

\begin{figure}[ht!]
    \centering
    \begin{subfigure}
        \centering
        \includegraphics[width=0.49\linewidth]{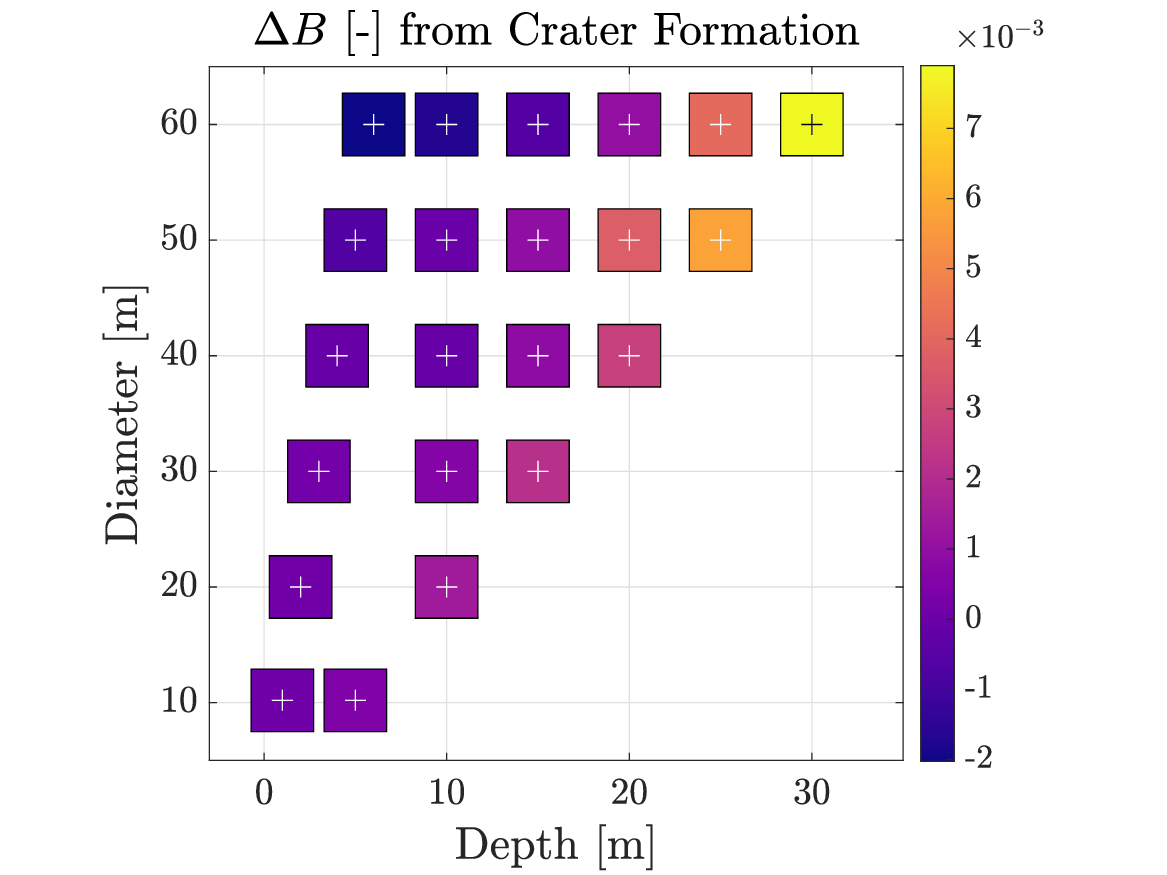}
    \end{subfigure}
    \hfill
        \begin{subfigure}
        \centering
        \includegraphics[width=0.49\linewidth]{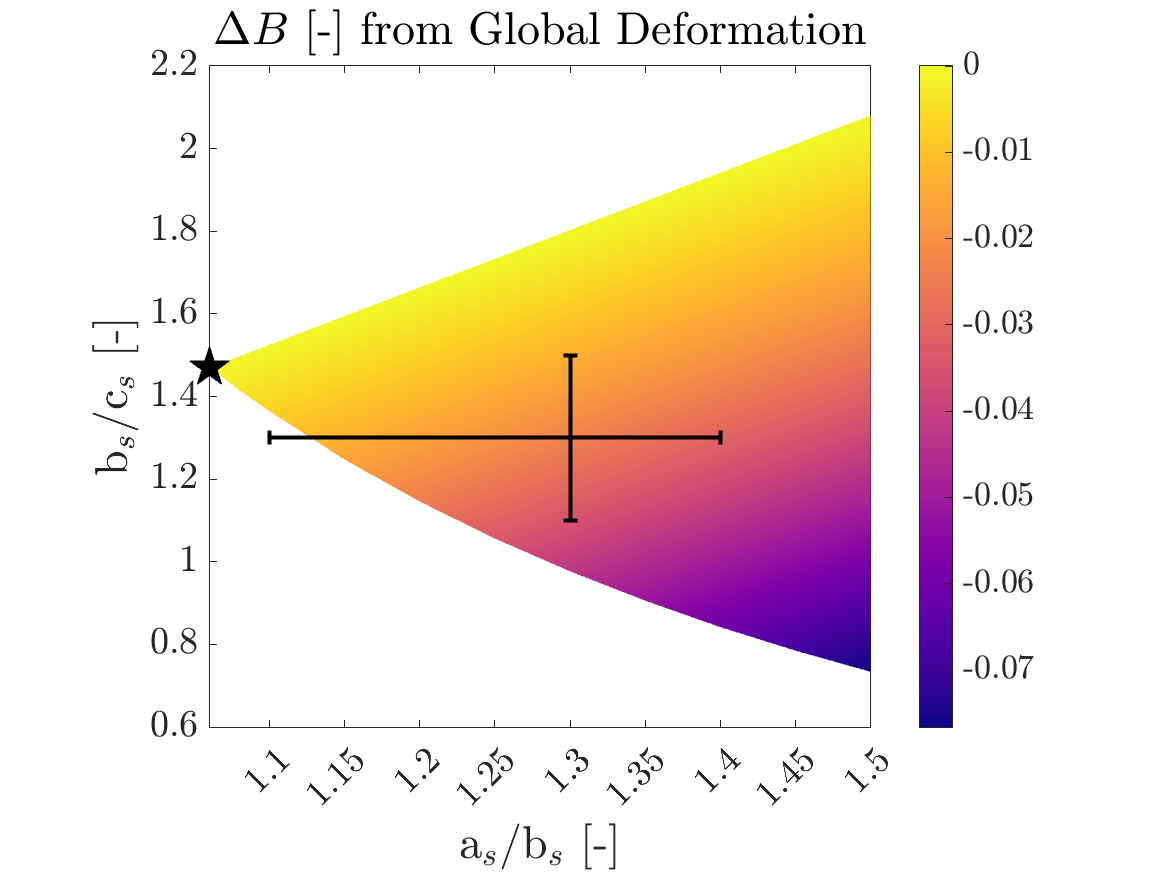}
    \end{subfigure}
    \caption{Changes in BYORP coefficient, $B$, as a function of crater depth and diameter (left) and post-impact elongation ratios $a_s/b_s$ and $b_s/c_s$ under physically plausible deformation (right). The white/black cross-hairs in the crater plot mark the specific depth and diameter values corresponding to each case. The colored boxes surrounding the cross-hairs indicate the magnitude of $\Delta B$. \edit1{The black error bars denote the current estimates of the axis ratios $a_s/b_s$ and $b_s/c_s$ ($a_s/b_s = 1.300 \pm 0.010$ and $b_s/c_s = 1.3\pm0.2$ from \citet{Naidu2024} with the $1.1 \le a_s/b_s \le 1.4$ range from \edit1{\cite{Pravec2024}}).} The black star represents the pre-impact elongation ratios provided in \tbl{params}.}
    \label{f:deltaB}
\end{figure}

We assume a pre-impact shape of a smooth oblate spheroid with nominal extents along the ($x$,$y$,$z$) axes of 179~m, 169~m, and 115~m from \cite{Daly2024} (corresponding to a near-zero BYORP coefficient) and tidal-BYORP equilibrium, as Didymos is suspected to be at or close to this state \citep{Richardson2022,Scheirich-2022,Scheirich2024}. \edit1{A nominal $B$ was computed for the assumed pre-impact shape model following the methodology in \citet{scheeres2007dynamical}.} We explore two avenues of reshaping, cratering and global deformation, and evaluate the shape's effect on $B$. \edit1{This is done by modifying the pre-impact shape model and recomputing $B$ to quantify how much $B$ deviates from the nominal value.} For the first regime, craters of various depths and diameters are placed at the estimated impact location of $8.84 \pm 0.45^{\circ}$S, $264.30 \pm 0.47^{\circ}$E \citep{Daly2023}. For global deformation, we use the methodology described in \citet{nakano2022nasa} and used in \sect{Dimorphos_reshaping} to explore deformed shapes ranging from $a_s/b_s = 1.06$ to $1.5$. We find that craters in general lead to a net positive change in $B$, but a large, shallow crater can cause a net negative change. All global deformation cases resulted in a net negative change in $B$ since flattening occurred on the leading hemisphere of Dimorphos. The change in magnitude becomes larger with more elongated secondaries and/or flatter deformed leading hemispheres. \edit1{The change in $B$ for the nominal post-impact elongations of $a_s/b_s = 1.3$ and $b_s/c_s = 1.3$ (\tbl{params}) is $\Delta B = - 2.33 \times 10^{-2}$.} \edit2{These elongation values are from the final orbit solution used throughout this paper \citep{Naidu2024}, therefore this is the latest estimation of $\Delta B$ for Dimorphos and is an update from the original value reported in \citet{Cueva2024}.} For reference, the pre-impact $B$ value for Dimorphos is expected to be on the order of $-5 \times 10^{-3}$, depending on the true strength of tides \citep{Cueva2024}. The change \edit1{in} magnitude of $B$ for both craters and global deformation are shown in \figr{deltaB}.

We ran dynamical simulations \edit1{of the attitude and mutual orbit dynamics} incorporating forces and torques from solar-radiation pressure, tidal dissipation on both the primary and secondary, and solar third-body effects. \edit1{The binary system's heliocentric orbit was also modeled in the simulations, which has a semimajor axis of 1.64 AU, eccentricity of 0.38, and inclination of $3.4^{\circ}$ with respect to the ecliptic \citep{Richardson2022}.} The $\Delta B$'s for both cratering and global deformation were added to three theoretical equilibrium $B$ values corresponding to three different tidal strengths. Cases were run for 10,000 simulated years in order to assess the initial result of the impact on the attitude of the secondary, as well as the long-term behavior of the system. Simulation results of the orbit and attitude evolution of Dimorphos are presented in \cite{Cueva2024}. Trends and observations from these results are summarized here. We find that the BYORP coefficient increases but remains negative for most crater cases, thus driving the semimajor axis to a new (higher) theoretical equilibrium orbit radius. Planar libration is excited for these cases. Large, deep crater morphologies can cause the true value of $B$ to become positive (depending on the tidal strength), resulting in joint expansive evolution with a theoretically unbounded semimajor axis. This ties in with the overall life cycle of rubble-pile binaries. Although expansive BYORP means it will work with tides to grow the orbit, there are a lot of factors that could affect the subsequent evolution and end state of the system. For example, Dimorphos could lose synchronicity with weakening tides from the expansion, stopping the evolution. This could result in a wide asynchronous binary system \citep{Jacobson2014}. Another possibility is that Dimorphos could escape the Hill radius of the orbit and form an asteroid pair, where the two bodies share a similar heliocentric orbit but are no longer bounded by a mutual orbit \citep{Jacobson-Icarus-2011}. Due to the chaotic nature of the dynamics, it is difficult to predict exactly what would happen if the DART impact caused expansive BYORP.

As for the global deformation cases, we found that the system can experience a wide range of evolutionary behavior depending on the magnitude of reshaping (and thus magnitude of inertia changes), tidal strength, and initial conditions. We observed cases with excitation of planar librations, tumbling, and barrel instability. Some cases with planar libration had a quick onset of nonplanar librations, likely due to resonances with the heliocentric orbit. Some cases recaptured in the 1:1 spin-orbit resonance, while some remained chaotic for the entire 10,000-yr simulation run. For cases that experience synchronous recapture after a period of chaos, we saw that the secondary can either recapture in its original orientation, or a different orientation where it has flipped $180^{\circ}$ about its minor axis, flipped $180^{\circ}$ about its major axis, or both---thus possibly leading to orbit expansion.

The range of results from \cite{Cueva2024} show that until the true post-impact spin state and shape of Dimorphos are resolved by Hera, it is difficult to assert exactly how the system's secular evolution changed from the impact. Detection of BYORP from Hera is unlikely due to the short time span between Hera and DART, but not impossible. BYORP will still act on the system if libration is excited, but tumbling or barrel instability will shut off BYORP\edit1{, and thus BYORP cannot be measured}. \edit1{If Hera observes that the secondary remained synchronous following the impact, then any measured drift in the orbit would be due to BYORP and tides. If Dimorphos is librating, then it gets more complicated.} For example, the range of secular drift rates surrounding the nominal $a_s/b_s$ and $b_s/c_s$ is approximately $-1.8$ to $-4.6$~cm~yr$^{-1}$ (corresponding to mean motion rates, $\dot{n}$, ranging from $5.5 \times 10^{-17}$ to $1.4 \times 10^{-16}$~rad~s$^{-2}$). This equates to an inward drift of about 9 to 23~cm in the 5-year gap between DART and Hera. These numbers are assuming only libration is excited from the impact. If Dimorphos is dynamically excited, \edit1{which Hera will measure}, there will be inevitable variations in the separation distance \edit1{due to inherent orbit-attitude coupling}. In the short timespan between DART and Hera, it will be difficult to distinguish any secular drift in these variations \edit1{since the expected secular drift in that timespan is so small.} However, Hera will be able to provide a full detailed shape model of post-impact Dimorphos. While DART supplied a high-resolution shape model, it is only one side of Dimorphos, so we are unable to fully compute a $B$ value for Dimorphos. Knowing the back side topography of Dimorphos would allow us to better resolve the pre-impact shape model of Dimorphos and use its $B$ value to constrain tidal strengths of Dimorphos from the pre-impact tidal-BYORP equilibrium. The detailed post-impact shape model will allow us to refine our predictions for how the secular evolution of the Didymos system will proceed. \edit1{Knowing the $B$ of the post-impact shape model acquired from Hera will also help us improve our computations for how much we would have expected the orbit to drift due to BYORP if the system is dynamically excited. If it in theory should drift more than we originally predicted, then it may be easier to detect BYORP.}


\section{Implications for Hera} \label{s:hera}

The Hera spacecraft will arrive at Didymos in October 2026 and will begin in December 2026 approximately 6 months of proximity operations. Details regarding mission objectives, profile, scenario, instrumentation, and measurements can be found in \citet{Michel2022}. The primary goal of Hera is to measure the mass of Dimorphos, which is the missing parameter for the precise estimate of $\beta$\edit1{, accounting for the fact that the ``eventual'' $\beta$ is slightly higher than the ``immediate'' $\beta$ (see \sect{hardening})}. This knowledge will be acquired through a series of measurements. Optical images will mainly be used to constrain the shape and dynamical state, and provide an early mass estimate during the distant flybys. Once Hera gets closer, its Radio Science Experiment (RSE), involving the main spacecraft and the two CubeSats, Juventas and Milani \citet{Gramigna2023a,Ferrari2021milani}, should obtain Dimorphos's mass to higher precision and measure the extended gravity fields and rotational states of both Didymos and Dimorphos. In this way, the RSE will also better constrain the interior structure and global properties (\eg density, porosity). The low-frequency radar called JuRa and the gravimeter called GRASS onboard the Juventas CubeSat will further characterize the interior. Additionally, Hera will also enable the accurate determination of the heliocentric $\beta$ by reconstructing the Didymos system ephemerides using pseudo range points (\sect{beta_helio}). 

Hera includes an ISL transceiver for ensuring correct communication with the CubeSats, relaying their data and commands to and from the operation centers on the ground. A dedicated radio-science mode of the ISL will also allow for collecting high-precision range and range-rate measurements, which will be used in combination with Earth-based radio tracking and optical navigation images provided by the Asteroid Framing Cameras (AFC) onboard Hera. Of particular importance are the ISL observables collected by Juventas and Milani, which orbit the system at closer distances than Hera \citet{Ferrari2021milani2}, and allows for obtaining higher accuracies in the parameters of interest. The AFC observations will also determine the rotational states of the asteroids by tracking their surface features (landmarks) and provide the necessary data to contribute to the mass measurement.

A covariance analysis has been performed to obtain the expected uncertainties in the scientific parameters of interest for the Hera RSE. The expected formal 1-$\sigma$ uncertainties for the $GM$ of Didymos and Dimorphos are on the order of $0.01\%$ and $0.1\%$, respectively. Additionally, the extended gravity fields of Didymos and Dimorphos can be estimated up to degree 3 and 2, with a $J_2$ accuracy of better than $0.1\%$ and $10\%$, respectively. \edit1{Regarding Didymos, such measurements will offer the possibility of detecting internal heterogeneity, with implications for understanding the structural behavior of the asteroid at fast spin and the formation of Dimorphos (\sect{postencounter})}.

Of utmost importance is getting an accurate estimate of the orbital parameters of Dimorphos, as well as its librations, since they are directly linked to the energy dissipation of the system and to the pre-impact state \citep{meyer2023a,meyer2023b}. \edit1{Measurements of those parameters may allow determination of secular effects, such as a possible orbital drift that can be linked to BYORP and tides, depending on Dimorphos's actual rotational state (\sect{secular})}. The RSE covariance analysis indicates that Dimorphos's relative position can be retrieved with sub-meter-level accuracy throughout the mission, while the semimajor axis and eccentricity uncertainties are on the order of $10^{-1}$~m and $10^{-4}$, respectively. Similarly, Dimorphos's spin pole can be constrained to less than 1$^\circ$, while its librations can be estimated with a relative accuracy of roughly $2 \times 10^{-2}$~deg for the libration amplitude, $8 \times 10^{-4}$~deg/h for the angular velocity, and 2$^\circ$ for the phase. (Note that a small perturbation in the rotational dynamics with respect to the equilibrium condition was assumed. If this is not the case, more analysis should be performed, and the results could vary.) All the details and results of the analysis can be found in \citet{Gramigna2023a}.

In addition, the LIDAR onboard Hera, called PALT (Planetary ALTimeter), will perform range measurements that will yield independent shape reconstructions with respect to the ones provided by the AFC images and will be used to help determine the wobble of the binary system, which will yield an independent estimate of the mass of Dimorphos. These altimetric measurements could also allow for a more precise reconstruction of Hera's trajectory, enhancing the RSE scientific return. Overall, when adding LIDAR altimetry crossover measurements, the expected improvement of Dimorphos's relative orbit is on the order of 60\% for the radial and tangential components, and up to 40\% for the orbit-normal one \citet{Gramigna2023b}. 

The current questions regarding possible shape change of Dimorphos \edit1{(\sect{reshaping})}, the actual $\beta$ value, and rotational properties will thus be answered by Hera thanks to a rendezvous with the binary system and its investigation at close proximity. \edit1{Regarding surface changes on Dimorphos resulting directly from the DART impact, the analysis will have to account for the long-term resurfacing processes that may contribute to the surface features (\ie crater morphology) that Hera will observe (\sect{tidal})}. 

JuRa onboard the Juventas CubeSat will reveal whether Dimorphos is a rubble pile made of boulders homogeneously distributed throughout, whether its interior contains a high level of heterogeneities, or whether it is simply made of a big core surrounded by a layer of pebbles and gravels. This has implications for binary-formation models and our interpretation of the response of the asteroid to the DART impact and to tidal forces. In turn, the possible presence of long-lived ejecta and the possibility that Dimorphos is tumbling have some implications for the operations of Hera and its two CubeSats, which need to be assessed. Operations for the Hera spacecraft that allow observing the entire surface of the asteroid may have to be adapted to this situation, as may the operations of the CubeSats that will come closer to the asteroid to perform their measurements. Furthermore, both CubeSats are expected to land on the smaller body. In particular, Juventas is planned to land on Dimorphos to perform additional measurements of the gravity field with its gravimeter, which requires that it remains stable on the surface. Landing on a tumbling asteroid increases the complexity of the operation and has never been attempted. Nevertheless, knowing in advance that the Hera mission may face such a situation is extremely useful to be best prepared for a more complex situation than originally expected. This is now taken into account in the development of operation plans at Didymos/Dimorphos.

The Hera data will also provide opportunities to constrain the material properties of both Dimorphos and Didymos thus feeding into the models discussed in this paper. The dynamics of the CubeSat landings on Dimorphos will be recorded with onboard accelerometers leading to direct measurements of the surface mechanical properties \citep{Sunday2022, murdoch2022}. The geology of the asteroids and the boulder morphology as observed with the Hera Asteroid Framing Camera or the CubeSat cameras can be used to constrain the mechanical properties such as the angle of internal friction \citep{Barnouin2023, robin2023}. High-resolution images of the surface of Didymos taken with different viewing and illumination geometries will constrain the currently unknown depth of the boulder tracks, allowing for an improved estimate of the bearing capacity of the surface of Didymos \citep{bigot2023}. Morphological mapping using, for example, surface-roughness measurements \citep{vincent2023} of the asteroid pair before and after the DART impact will highlight changes due to re-impacting ejecta, resurfacing, and mass-wasting events. The penetration depth of ejecta boulders that have re-impacted the surface of Didymos at low speed will provide additional constraints on the properties of the surface material \citep[friction, density, etc.;][]{Sunday2022}. 

In conclusion, Hera's mission will provide very accurate characterization of various dynamical aspects of the Didymos system, including its gravity field, orbits, dynamics, energy dissipation processes, surface, and interior structure. Thanks to the expected level of accuracy provided by the Hera mission to determine Dimorphos's important dynamical attributes (\eg mass, rotational state, and orbit), a more refined value of $\beta$ can be estimated, which in turn will significantly increase the understanding and validation of the kinetic impactor technique for deflecting potentially hazardous asteroids in the future.


\section{Conclusions} \label{s:concl}

Following the successful demonstration of kinetic impact provided by the DART mission, the dynamical state of the Didymos system before and after the event was assessed. The main conclusions are as follows:
\begin{itemize}
    \item The post-impact eccentricity is consistent with a circular pre-impact orbit and the post-impact precession rate strongly suggests impact-induced reshaping of Dimorphos. The possibility of an unstable rotation state of Dimorphos should be investigated further.
    \item The DART impact momentum transfer enhancement factor $\beta$ was found to be about 3.6 assuming a nominal bulk density of 2,400~kg~m$^{-3}$ for Dimorphos. Although some dynamical parameters have been revised since that determination, the uncertainty in $\beta$ remains dominated by the large uncertainty in the mass of Dimorphos. It is still too early for a heliocentric $\beta$ measurement, but future occultations and Hera's arrival should give strong constraints. It is evident that the impact generated a lot of ejecta ($\sim 10^7$~kg), despite the likely mitigating effect of target surface curvature. A possible secular decrease in orbital period in the days following impact may be evidence of coupling between the binary and persistent ejecta.
    \item Didymos's rapid spin, oblate shape, and comparatively high bulk density suggest global cohesion may not be required to maintain its structural stability at present. Given its shape and the likely formation of Dimorphos from accumulation of shed material from its parent, Didymos probably has an internal structure of moderate strength. High-slope surface features and the lack of any detectable spin change suggest a surface cohesion of a few pascals, consistent with granular impact simulations. The unexpected oblate shape of Dimorphos implies a more complex formation pathway than simply accretion at the Roche limit. Impact simulations suggest Dimorphos may have reshaped significantly, a possibility borne out by preliminary lightcurve measurements and dynamical modeling indicating a post-impact elongated shape. This in turn may have contributed to some of the measured $\beta$ and increases the probability of Dimorphos being in a tumbling state. All of these considerations are important for the long-term secular evolution of the system due to tides and BYORP, with a range of possible outcomes that will likely require Hera to resolve.
    \item Hera will greatly reduce the uncertainties on Dimorphos's mass and therefore on $\beta$, as well as on Dimorphos's shape, rotation state, and internal properties thanks to a wide array of measurement instruments. A possibly tumbling Dimorphos has some implications for the operation of Hera and its two CubeSats that needs to be assessed and is being taken into account in the development of operation plans.
\end{itemize}

The DART mission, together with the Didymos observing campaign, not only represented the first test at realistic scale of a hazard mitigation technique but also provided unprecedented measurements of dynamical effects in a non-ideal small solar system binary for testing theoretical models. Predictions prior to encounter were largely borne out, but there were some surprises and there remain unanswered questions. We look forward to revelations from the Hera mission, which promise to further refine our understanding of small bodies in general and the formation and evolution of binary asteroids in particular.

\newpage  


\begin{acknowledgments}

\section*{Acknowledgments}

We thank Paula Benavidez, \"Ozg\"ur Karatekin, and the anonymous reviewers who provided comments that helped improve this paper.

The work presented here was supported in part by the DART mission, NASA Contract \#80MSFC20D0004 to JHU/APL. Part of this work was supported by the Programme National de Plan\'etologie (PNP) of CNRS-INSU cofunded by CNES, by CNES itself, and by the BQR program of the Observatoire de la C\^ote d'Azur. The ACROSS project is supported under the OSIP ESA CONTRACT No.\ 4000135299/21/NL/GLC/ov. This study makes use of data obtained by the Observing Working Group of the DART Investigation Team. Some simulations were performed on the ASTRA cluster administered by the Center for Theory and Computation, part of the Department of Astronomy at the University of Maryland.

R.H.C. acknowledges that this material is based upon work supported by the National Science Foundation Graduate Research Fellowship Program under Grant No.\ DGE 2040434. Any opinions, findings, and conclusions or recommendations expressed in this material are those of the author(s) and do not necessarily reflect the views of the National Science Foundation. F.F. acknowledges funding from the European Research Council (ERC) under the European Union's Horizon Europe research and innovation programme (Grant agreement No.\ 101077758). R.M. acknowledges that this work was supported by a NASA Space Technology Graduate Research Opportunities (NSTGRO) award, NASA contract No. 80NSSC22K1173. P.M. acknowledges support from the French space agency CNES and ESA. R.N. acknowledges support from NASA/FINESST (NNH20ZDA001N).

A.C.B., E.G., M.J., P.M., R.L.M., S.D.R., P.T., K.T., and M.Z. acknowledge funding support from the European Union's Horizon 2020 research and innovation program under grant agreement No.\ 870377 (project NEO-MAPP). The work of S.R.C, E.G.F, and S.P.N was carried out at the Jet Propulsion Laboratory, California Institute of Technology, under a contract with the National Aeronautics and Space Administration (\#80NM0018D0004). E.G., R.L.M., M.Z., and P.T. wish to acknowledge Caltech and the NASA Jet Propulsion Laboratory for granting the University of Bologna a license to an executable version of MONTE Project Edition S/W. E.G., R.L.M., A.R., M.Z., and P.T. are grateful to the Italian Space Agency (ASI) for financial support through Agreement No.\ 2022-8-HH.0 in the context of ESA's Hera mission. M.J. and S.D.R. acknowledge support by the Swiss National Science Foundation (project number 200021 207359). J.M. acknowledges support from the DART Participating Scientist Program (\#80NSSC21K1048). F.M. acknowledges financial support from grants PID2021123370OB-I00 and CEX2021-001131-S funded by MCIN/AEI/10.13039/501100011033. N.M. acknowledges funding support from the European Commission's Horizon 2020 research and innovation programme under grant agreement No.\ 870377 (NEO-MAPP project) and support from the Centre National d'Etudes Spatiales (CNES), focused on the Hera space mission. L.P. was supported by an appointment to the NASA Postdoctoral Program at the NASA Jet Propulsion Laboratory, California Institute of Technology, administered by Oak Ridge Associated Universities under contract with NASA. P.P. and P.S. acknowledge support by the Grant Agency of the Czech Republic, grant \edit1{23-04946S}. S.R.S. acknowledges support from the DART Participating Scientist Program, grant no.\ 80NSSC22K0318. D.S. thanks Action F\'ed\'eratrice \emph{Gaia} of the Paris Observatory for financial support. G.T. acknowledges financial support from project FCE-1-2019-1-156451 of the Agencia Nacional de Investigaci\'on e Innovaci\'on ANII and Grupos I+D 2022 CSIC-Udelar (Uruguay). P.T., A.R., and M.Z. acknowledge financial support from Agenzia Spaziale Italiana (ASI, contract No.\ 2019-31-HH.0 CUP F84I190012600). J.M.T.-R. acknowledges financial support from the project PID2021-128062NB-I00453 funded by MCIN/AEI/10.13039/501100011033.

\end{acknowledgments}


\bibliography{revised}{}
\bibliographystyle{aasjournal}


\end{document}